\documentclass[aps,prd,amssymb,amsmath,amsfonts,superscriptaddress,nofootinbib,eqsecnum,reprint,showpacs,longbibliography]{revtex4-2}

\usepackage{graphicx}
\usepackage{lmodern}
\usepackage{amsmath,amssymb}
\usepackage{mathrsfs}
\usepackage{amsfonts}
\usepackage[utf8]{inputenc}
\usepackage{url}
\usepackage[colorlinks]{hyperref}
\usepackage{xcolor}
\usepackage[normalem]{ulem}
\usepackage{orcidlink}
\usepackage{placeins}
\usepackage{mathtools}
\usepackage{rotating}
\usepackage{tabularx}
\usepackage{xspace}
\usepackage{booktabs}

\definecolor{dodgerblue}{HTML}{1E90FF}
\definecolor{viennared}{HTML}{DA0A14}
\definecolor{ctorange}{HTML}{FF6C0C}
\definecolor{wales}{HTML}{ff0038}
\definecolor{benettongreen}{HTML}{009421}
\definecolor{ferrarired}{HTML}{ff2800}
\definecolor{austriawienpurple}{HTML}{441678}
\definecolor{steiermarkgruen}{HTML}{006747}

\hypersetup{
     colorlinks=true,
     linkcolor=dodgerblue,
     filecolor=dodgerblue,
     citecolor = dodgerblue,      
     urlcolor=dodgerblue,
     }

\def\phenom{\textsc{PhenomGSF}\xspace}
\def\teob{\textsc{TEOBResumS}\xspace}

\newcommand{\zeroGSF}{0\rm{GSF}}
\newcommand{\oneGSF}{1\rm{GSF}}
\newcommand{\twoGSF}{2\rm{GSF}}

\newcommand{\rlr}{r_{\rm LR}} 

\newcommand{\Birmingham}{School of Physics and Astronomy and Institute for Gravitational Wave Astronomy, University of Birmingham, Edgbaston, Birmingham, B15 2TT, United Kingdom}

\DeclarePairedDelimiterX{\norm}[1]{\lVert}{\rVert}{#1}

\begin{document}
%%%%%%%%%%%%%%%%%%%%%%%%%

\title{Phenomenological model of gravitational self-force enhanced tides in inspiralling binary neutron stars}

\author{Natalie Williams \orcidlink{0000-0002-5656-8119}} 
\email{nxw049@student.bham.ac.uk}
\affiliation{\Birmingham}

\author{Patricia Schmidt \orcidlink{0000-0003-1542-1791}} 
\email{P.Schmidt@bham.ac.uk}
\affiliation{\Birmingham}

\author{Geraint Pratten \orcidlink{0000-0003-4984-0775}} 
\email{G.Pratten@bham.ac.uk}
\affiliation{\Birmingham}

%\date{\today}

%%%%%%%%%%%%%%%%%%%%%%%%%%% ABSTRACT %%%%%%%%%%%%%%%%%%%%%%%%%%%
\begin{abstract}
Gravitational waves from inspiralling binary neutron stars provide unique access to ultra-dense nuclear matter and offer the ability to constrain the currently unknown neutron star equation-of-state through tidal measurements. This, however, requires the availability of accurate and efficient tidal waveform models. 
In this paper we present \phenom, a new phenomenological tidal phase model for the inspiral of neutron stars in the frequency-domain, which captures the gravitational self-force informed tidal contributions of the time-domain effective-one-body model \teob.
\phenom is highly faithful and computationally efficient, and by choosing a modular approach, it can be used in conjunction with \emph{any} frequency-domain binary black hole waveform model to generate the complete phase for a binary neutron star inspiral. 
\phenom is valid for neutron star binaries with unequal masses and mass ratios between $1$ and $3$, and dimensionless tidal deformabilities up to $5000$. Furthermore, \phenom does not assume universal relations or parameterised equations-of-state, hence allowing for exotic matter analyses and beyond standard model physics investigations. 
We demonstrate the efficacy and accuracy of our model through comparisons against \teob, numerical relativity waveforms and full Bayesian inference, including a reanalysis of the binary neutron star observation GW170817. 
\end{abstract}

\maketitle

%%%%%%%%%%%%%%%%%%%%%%%%%%%%
\section{Introduction}
\label{sec:intro}
%%%%%%%%%%%%%%%%%%%%%%%%%%%%
Detections of gravitational waves (GWs) from binary neutron stars (BNS) hold the potential to probe nuclear matter at densities which cannot easily be achieved under laboratory conditions, and shed light on the currently unknown neutron star equation-of-state (EOS)~\cite{Flanagan:2007ix, Wade:2014vqa}. The observation of the inspiralling BNS GW170817~\cite{TheLIGOScientific:2017qsa, GW170817-PE} provided the first constraints on the EOS obtained from GWs. Complementary EOS constraints for nuclear matter can be obtained from X-ray observations of pulsars~\cite{Raaijmakers:2021uju, Miller:2021qha} or the measurement of the thickness of the neutron skin in laboratory experiments~\cite{PREX:2021umo, Chatziioannou:2024tjq}. Notable differences in inferred constraints on the neutron star radius between different astrophysical and laboratory-based measurements~\cite{Reed:2021nqk}, however, highlight the crucial role of more observations, e.g. from GWs and NICER to resolve the tension. 

Matter leaves characteristic imprints on the GW signal, making the signal distinct from that of a binary black hole (BBH). The most prominent GW signature arises from the excitation of a neutron star's fundamental oscillation modes or $f$-modes, which, at leading order, is characterised by the quadrupolar tidal deformability $\lambda_2$~\cite{Flanagan:2007ix}. Its measurement in GW observations allows us to constrain the EOS~\cite{GW170817-EOS}. 
However, in order to do so, waveform models that accurately capture the relevant physics are needed. 
Current state-of-the-art tidal waveform models are obtained either within post-Newtonian (PN) theory~\cite{Flanagan:2007ix, Vines:2011ud, Damour:2012yf, Abdelsalhin:2018reg, Banihashemi:2018xfb, Landry:2018bil, Schmidt:2019wrl, Henry:2019xhg, Henry:2020ski}, 
the effective-one-body (EOB) framework~\cite{Damour:2009wj, Bini:2012gu, Damour:2012yf, Hinderer:2016eia, Steinhoff:2016rfi, Bohe:2016gbl, Nagar2018, Akcay:2018yyh, Steinhoff:2021dsn, Gamba:2023mww}, or by utilising a phenomenological approach~\cite{Dietrich:2017aum, Dietrich:2019kaq, Abac:2023ujg, Bernuzzi:2014owa, Kawaguchi:2018gvj}. One particular advantage of phenomenological waveform models is their computational efficiency, which is highly desirable when performing Bayesian inference which often requires $\sim 10^6-10^9$ model evaluations to accurately sample the posterior distributions. The computational efficiency of phenomenological models is achieved by directly describing the waveform as piecewise closed-form expressions in the frequency domain by fitting ansatzes for the amplitude and phase to hybrid waveforms, which combine state-of-the-art analytical knowledge with numerical relativity (NR) waveforms. This approach has been successfully applied to obtain fast waveform models for binary black holes~\cite{Ajith:2009bn, Santamaria:2010yb, Hannam:2013oca, Pratten:2020fqn, Pratten:2020ceb, Garcia-Quiros:2020qpx, Thompson:2023ase}, neutron star -- black hole binaries~\cite{Thompson:2020nei, Matas:2020wab} and BNSs~\cite{Dietrich:2019kaq, Abac:2023ujg}. Alternative approaches to decreasing the waveform generation cost include the stationary phase approximation (SPA)~\cite{Finn:1992xs} and reduced-order modelling~\cite{Field:2013cfa}. Parameter estimation itself can also be sped up by techniques such as relative binning~\cite{Cornish:2010kf, Zackay:2018qdy, Finstad:2020sok, Cornish:2021lje, Leslie:2021ssu}, multi-banding~\cite{Vinciguerra:2017ngf,Morisaki:2021ngj} and reduced-order quadratures~\cite{Antil:2012wf, Canizares:2013ywa, Canizares:2014fya, Smith:2016qas}.

Current BNS waveform models do not only differ in their computational speed but also in their physics content. In particular, the phenomenological models of the \textsc{NRTidal} series~\cite{Dietrich:2017aum, Dietrich:2019kaq, Abac:2023ujg} are calibrated to a limited range of BNS simulations. While this allows for the incorporation of non-perturbative information, it also comes with some caveats: firstly, the NR data are themselves limited in their physics content due to the complexity of matter simulations. Secondly, different tidal effects that arise in the strong-field regime such as higher-order multipoles or other oscillation modes including pressure ($p$-) and gravity ($g$-) modes, are difficult to separate in NR simulations and hence when modelled phenomenlogically are entangled in the parameterisation. Thirdly, NR simulations make specific assumptions on the nuclear EOS such a hadronic composition. Finally, to reduce the dimensionality of the parameter space, tidal models also often invoke quasi-EOS independent universal relations (URs)~\cite{Yagi:2016qmr} between different tidal parameters. While this dimensional reduction may be desirable, any such assumptions will limit the applicability of the model and exclude them from being used for exploring e.g. the possibility of the presence of exotic matter such as hyperons or deconfined quarks or boson stars~\cite{Raithel:2022orm}.

In this work we present \phenom, a new phenomenological tidal phase model in the frequency-domain based on the tidal EOB model \teob~\cite{Nagar2018}, which can readily be added to \emph{any} BBH baseline to provide a complete BNS model. We restrict our model to the dominant tidal effects associated with the quadrupolar gravitoelectric static Love number $k_2$ including tidal contributions informed by gravitational self-force (GSF), and only model the phase of the $\ell = |m|=2$ mode. \phenom is calibrated to both equal and unequal mass systems and freely fits the component tidal deformabilities without assuming URs, hardonic matter or specific parameterisations of the EOS. 

The paper is organised as follows: Section~\ref{sec:prelim} details some preliminaries, the used conventions and parameterisations. 
In Sec.~\ref{sec:inputs} we specify the input waveforms used to build the model, the parameter space of the model and detail the construction of \teob-NR hybrid waveforms. Then, in Sec.~\ref{sec:inspiral} we detail the hierarchical phenomenological fitting procedure and give details of the model construction. 
In Sec.~\ref{sec:validation} we validate our model against the input waveforms, a set of independent waveforms as well as a selection of NR-hybrid waveforms. We then perform Bayesian inference with \phenom on model injections, a hybrid waveform and reanalyse the data of GW170817.
We then conclude with a discussion in Sec.~\ref{sec:discussion}. 
The complete \phenom fit can be found in App.~\ref{sec:FitsAppendix}, while the full parameter estimation results for GW170817 are shown in App.~\ref{sec:GW170817appx}.
Throughout the paper we use geometric units by setting $G=c=1$. 

%%%%%%%%%%%%%%%%%%%%%%%%%%%%
%\section{Methodology}
%\label{sec:methods}
%%%%%%%%%%%%%%%%%%%%%%%%%%%%

%%%%%%%%%%%%%%%%%%%%%%%%%%%%
\section{Preliminaries}
\label{sec:prelim}

%%%%%%%%%%%%%%%%%%%%%%%%%%%%%%%
\subsection{Tidal Parameterisations}
\label{sec:tidal}
%%%%%%%%%%%%%%%%%%%%%%%%%%%%%%%
As opposed to BBH systems, the presence of matter in a BNS system induces an additional quadrupole moment and hence accelerates the inspiral. Physically, this results in a multipolar deformation of each neutron star, which, at leading-order, is characterised by the static quadrupolar $\ell=2$ (dimensionless) tidal deformability $\Lambda_{i,2} = \lambda_{i,2}/m_i^5$ of the $i$-th star with mass $m_i$. The tidal deformability strongly depends on the internal structure of the neutron star and hence is connected to the EOS. Its measurement can therefore be used to map macroscopic properties to a constraint on the microscopic EOS. As we only consider the $\ell=2$ multipole, we drop the multipolar index henceforth for convenience. We also define the total mass $M = m_1 + m_2$, mass ratio $q = m_1/m_2 \geq 1$ and symmetric mass ratio $\eta = q /(1 + q)^2$. 

In PN theory, the leading-order tidal effect in the GW phase occurs at 5PN~\cite{Flanagan:2007ix} and is governed by the parameter~\cite{Favata:2013rwa}
\begin{align}
    \tilde{\Lambda} =& \frac{8}{13}\Bigg[(1+ 7 \eta - 31\eta^{2})(\Lambda_1 +\Lambda_2) \nonumber \\
    &+  \sqrt{1-4\eta}(1+9\eta -11\eta^2)(\Lambda_1 -\Lambda_2)\Bigg].
\end{align}

The coefficient of the next-to-leading correction is given by~\cite{Favata:2013rwa, Wade:2014vqa}
\begin{align}
    \delta\tilde{\Lambda} =& \frac{1}{2}\Bigg[\sqrt{1-4\eta} \bigg(1-\frac{13272}{1319}\eta-\frac{8944}{1319}\eta^2\bigg)(\Lambda_1 + \Lambda_2) \nonumber \\
    &+ \bigg(1-\frac{15910}{1319}\eta-\frac{32850}{1319}\eta^2+\frac{3380}{1319}\eta^3\bigg)(\Lambda_1 -\Lambda_2)\Bigg].
\end{align}

While we will be sampling the binary parameter space in $ \{q, \Lambda_1, \Lambda_2\}$ to build our training dataset, we will then transform to $\{\eta, \tilde{\Lambda}, \delta\tilde{\Lambda}\}$ to construct the \phenom ~model.

%%%%%%%%%%%%%%%%%%%%%%%%%%%%%%%%
\subsection{Waveform Conventions}
\label{sec:conventions}
%%%%%%%%%%%%%%%%%%%%%%%%%%%%%%%%

\phenom ~models the phase of the $\ell=|m|=2$ spin-weighted spherical harmonic modes of the GW signal of nonspinning neutron star inspirals. The complex GW strain as a function of GW frequency $f$ is given by  
\begin{align}
    \tilde{h} &\equiv \tilde{h}_+(f, \Vec{\theta}, \vartheta, \varphi) - i\tilde{h}_\times(f, \Vec{\theta}, \vartheta, \varphi) \\ 
    &= \sum_{m=\{-2,2\}}\tilde{h}_{2m}(f, \Vec{\theta}){}^{-2}Y_{2m}(\vartheta, \varphi), 
\end{align}
where $\Vec{\theta}$ denotes the intrinsic source parameters, $(\vartheta, \varphi)$ the orientation of the binary with respect to an observer, and ${}^{-2}Y_{2m}$ the spherical harmonic of spin weight $s=-2$ given by
\begin{equation}
{}^{-2}Y_{2\pm2}(\vartheta, \varphi) = \sqrt{\frac{5}{64\pi}}(1\pm \cos \vartheta)^2 e^{\pm 2i\varphi}.
\end{equation}

We utilise the orbital plane reflection symmetry of aligned spin systems to map between the $(m = 2)$-mode and $(m=-2)$-mode via
\begin{equation}
\tilde{h}_{22}(f, \vec{\theta}) = \tilde{h}^*_{2,-2}(-f, \vec{\theta}),
\end{equation}
where ${}^*$ denotes complex conjugation.

Each mode can be further decomposed into an amplitude $A_{\ell m}(f, \vec{\theta})$ and a phase $\psi_{\ell m}(f, \vec{\theta})$. The presence of tides predominantly affects the phase, although there are small amplitude corrections \cite{Hotokezaka:2016bzh}, which we do not take into account here, yielding
\begin{equation}
\tilde{h}_{22}(f, \vec{\theta}) = A^{\rm{BBH}}_{22}(f, \Vec{\theta})e^{-i\psi_{22}(f, \Vec{\theta})},
\end{equation}
where $\psi_{\ell m}$ denotes the complete Fourier phase given by the point particle phase plus the tidal phase 
\begin{equation}
\psi_{\ell m} = \psi_{\ell m}^{\rm{BBH}} + \psi_{\ell m}^{\phenom} + \psi_{\ell m}^{\rm{SS}},
\label{eq:phase}
\end{equation}
where $\psi_{\ell m}^{\mathrm{SS}}$ denotes the tidal self-spin PN corrections \cite{Dietrich:2018uni, Nagar:2018plt} described in more detail in Sec.~\ref{sec:spin}.

%%%%%%%%%%%%%%%%%%%%%%%%%%%%%%%
\section{Input Waveforms}
\label{sec:inputs}
%%%%%%%%%%%%%%%%%%%%%%%%%%%%%%%

%%%%%%%%%%%%%%%%%%%%%%%%%%%%%%%
\subsection{\teob Waveforms}
\label{sec:wfphysics}
%%%%%%%%%%%%%%%%%%%%%%%%%%%%%%%

The EOB framework provides a map from the general relativistic two-body problem to the motion of a test particle in a deformed effective spacetime. The EOB formalism allows one to calculate the full orbital dynamics and the complete waveform through merger. The EOB Hamiltonian describes the conservative dynamics with dissipative effects entering through the radiation reaction force in the equations of motion. Here we provide a brief overview of how tidal effects enter the \teob~\cite{Nagar2018}, with a particular focus on the GSF-resummed tidal potentials in the Hamiltonian introduced in~\cite{Bini:2014zxa} and extended in~\cite{Bernuzzi:2014owa,Akcay:2018yyh}. We emphasise that no NR information is included in the tidal sector. Tidal contributions also enter the radiation reaction force within the waveform multipoles as PN corrections to the amplitude (see~\cite{Gamba:2023mww} for more details). The EOB tidal model outlined here sets the baseline for the tidal content included in the \phenom calibration dataset.

Tidal forces enter the Hamiltonian through a modification of the EOB radial potential
\begin{align}
    A(u) &= A_0(u) + A_T (u),
\end{align}
where $u = 1/r$ is the Newtonian potential, $A_0 (u)$ denotes the point-particle potential, and $A_T(u)$ encodes the tidal interactions \cite{Damour:2009vw}
\begin{align}
     A_T (u) &= \displaystyle\sum_{\ell} \Big[ A_A^{(\ell+) \rm{LO}} (u) \hat{A}^{(\ell+)} (u) \nonumber \\ &\qquad + A^{(\ell-) \rm{LO}} (u) \hat{A}^{(\ell-)} (u) \Big] 
     + (A \leftrightarrow B),
\end{align}
where $A,B$ label the stars, $(+)$ denotes the gravitoelectric sector, $(-)$ the gravitomagnetic sector, and $\hat{A}$ denotes terms beyond leading-order (LO). 

The gravitoelectric tidal coefficients broadly relate to mass multipole moments induced in a star by an external gravitoelectric tidal field and the gravitomagnetic tidal coefficients to the current multipoles moments induced in a star by an external gravitomagnetic field \cite{Damour:2009vw}. The gravitomagnetic tides are therefore typically associated to frame-dragging effects which in turn excite inertial modes in the NS \cite{Provost:1981lfo,Ho:1998hq,Flanagan:2006sb}, including those that are associated to the Coriolis effect. As we predominantly focus on non-spinning binaries, we choose not to implement the gravitomagnetic terms at this time. The following therefore pertains only to the gravitoelectric tides. Likewise, we restrict ourselves to the dominant $\ell = 2$ quadrupole and disregard all higher multipole moments unless otherwise stated. 

Under the above restrictions, the Newtonian tidal potential can be written as
\begin{align}
    A^{\rm{LO}} (u) &= -\kappa_A u^{6}, 
\end{align}
where $\kappa_A$ denotes the electric tidal coupling constant, which can be expressed in terms of the dimensionless tidal love number $k_A^{(\ell)}$ as 
\begin{align}
    \kappa_A &= 2 k^{(2)}_A \frac{X_B}{X_A} \left(\frac{X_A}{\mathcal{C}_A} \right)^5, \\
    &= 3 X_B X_A^4 \Lambda_A,
\end{align}
where $\mathcal{C}_A = m_A / R_A$ denotes the compactness parameter, and the dimensionless tidal deformability is related to the Love number via
\begin{align}
\Lambda_A &= \frac{2}{3} k_A^{(2)} \mathcal{C}^{-5}_A.
\end{align}
The EOB radial potential subsequently simplifies to
\begin{align}
    A_T &= - [\kappa_Au^6\hat{A}_A(u) + \kappa_Bu^6\hat{A}_B(u)] .
\label{eq:EOBpotential}
\end{align}

Following \cite{Bini:2014zxa,Akcay:2018yyh}, the terms beyond LO admit a series expansion in terms of the expansion parameter $X_A = m_A/M \ll 1$, informed by results at first-order GSF
\begin{align}
\hat{A}_A(u) &= \hat{A}^{\mathrm{0GSF}} + X_A \hat{A}^{\mathrm{1GSF}} + X_A^2 \hat{A}^{\mathrm{2GSF}}.
\label{eq:GSFexpansion}
\end{align}

The $\zeroGSF$ potential is given by \cite{Bini:2014zxa}
\begin{align}
    \hat{A}^{\zeroGSF} (u) &= 1 + \frac{3 u^2}{1 - \rlr u},
\end{align}
where $r_{\rm LR}$ denotes the EOB light-ring (LR). 

The $\oneGSF$ potential is expressed in terms of the LR singularity together with a fit against strong-field GSF information \cite{Akcay:2018yyh} 
\begin{equation}
\begin{split}
\hat{A}^{\mathrm{1GSF}} = \frac{5u}{2(1-3u)^{7/2}} (1-a_1 u)(1-a_2 u)\frac{1 + n_1 u}{1+d_2 u^2}
\end{split}
\end{equation}
with coefficients
\begin{equation}
\begin{split}
&a_1 = 8.53352 \hspace{10pt} a_2 = 3.04309 \\
&n_1 =  0.84006 \hspace{10pt} d_2 = 17.7324.
\end{split}
\end{equation}

In contrast, the $\twoGSF$ potential only incorporates partial knowledge of the second-order GSF result \cite{Bini:2012gu} 
\begin{align}
    X^2_A \hat{A}_A^{\twoGSF} (u) &= \frac{337}{28} X^2_A u^2 \left( 1 + \mathcal{O}(u)\right),
\end{align}
such that near the LR, the quadrupolar electric tidal potential must blow up as \cite{Bini:2014zxa}
\begin{align}
    \frac{c_2 X^2_A}{(1 - \rlr u)^p}, \qquad p \geq 4.
\end{align}

The $\twoGSF$ tidal potentials can therefore be written as 
\begin{align}
    \hat{A}^{\twoGSF} &= \frac{337}{28} \frac{u^2}{(1-\rlr u)^p},
\end{align}
where the value of $p$ depends on as-of-yet-unknown $\twoGSF$ information. It was argued in \cite{Bini:2014zxa} that $p \geq 4$ together with a further argument that $p \leq 6$. Following \cite{Bernuzzi:2014owa,Akcay:2018yyh}, the unknown exponent is taken to be $p = 4$, though both coefficients may generically have mass ratio and EOS dependence. 

We choose to not include dynamical tidal effects within \phenom{}, which would further increase the dimensionality of the parameter space, corresponding to each stars fundamental mode frequency $\omega_2$, unless we use universal relations to relate the values of $\omega_2$ to $\Lambda_2$. 
We note that due to the modularity of our model, it is in principle straight forward to incorporate the PN dynamical tides using the \textsc{fmtidal} model \cite{Schmidt:2019wrl} .

As \teob is natively defined in the time-domain, we need to transform the model to frequency-domain in order to calibrate \phenom{}. We use the numerical SPA implemented in \teob{} \cite{Gamba:2020ljo}
\begin{align*}
\frac{d^2\psi(f)}{d\omega} &= \frac{1}{\omega}\frac{d\psi(t)}{d\omega},
\end{align*}
where $\omega = 2\pi f = \dot{\psi}(t)$. We use the SPA over a full Fourier transform due to small dephasing that was arising from spectral leakage and tapering in the binary black hole limit. We independently verified that the SPA agrees with the Fourier transform on a sub-radian scale over the parameter space under consideration.

%%%%%%%%%%%%%%%%%%%%%%%%%%%%%
\subsection{Parameter Space}
\label{sec:params}
%%%%%%%%%%%%%%%%%%%%%%%%%%%%%
\phenom~ is calibrated across the parameter space $q \in [1,3]$ and $\Lambda_{1,2} \in [0,5000]$ using 8446 \teob~ waveforms. The total mass $M$ scales out of the tidal phase when we consider the phase as a function of geometric frequency $M f$. This allows us to reduce the problem to a $3$-dimensional parameter space governed by a single mass parameter, taken to be either $q$ or $\eta$. 

\begin{figure*}
    \centering
    \includegraphics[width = \linewidth]{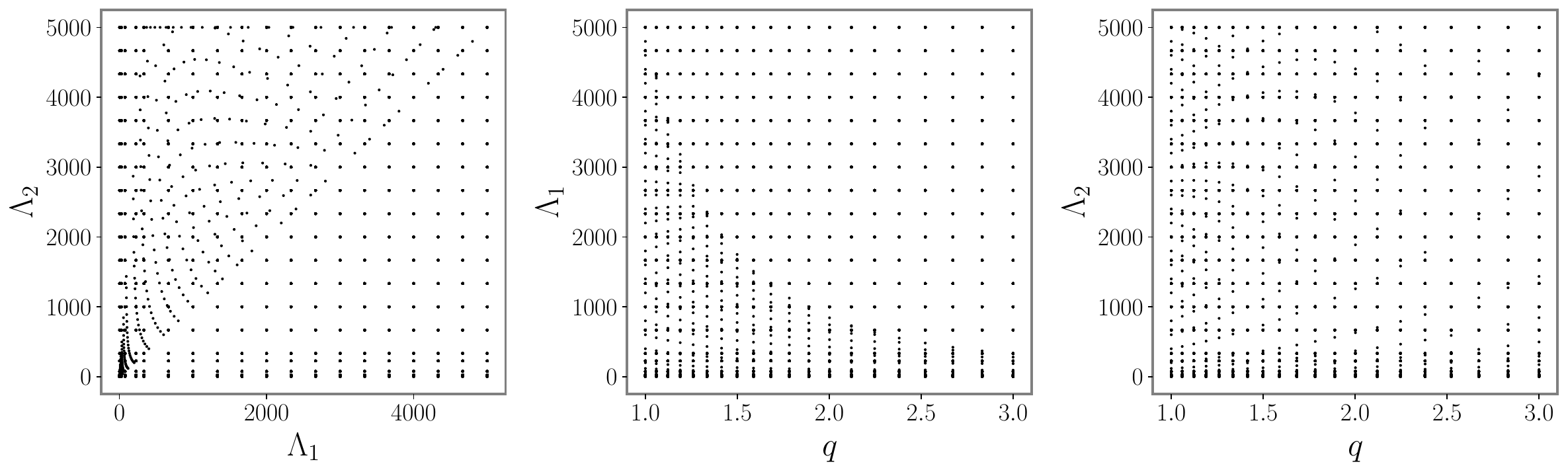}
    \caption{Visualisation of the parameter space covered by the \teob waveforms used in the calibration of \phenom.}
    \label{fig:parameterspace}
\end{figure*}

For sampling the mass ratio, we choose a log-uniform spacing in order to have higher resolution near the equal-mass limit $q=1$. Due to the hierarchical fitting process employed later on and explained in Sec.~\ref{sec:inspiral}, a subset of $446$ of these waveforms was specifically generated such that $\delta\tilde{\Lambda} = 0$. Within this subset, $\tilde{\Lambda}$ is uniformly sampled between the limits defined by a given $q$ for $\Lambda_{1,2} \in [0,5000]$. 
An additional layer of log-uniform spaced waveforms with $\tilde{\Lambda}\in [0,200]$ is also added in order to accurately capture the approach to the black hole limit, i.e. $\tilde{\Lambda}=0$. The remaining 8000 waveforms are sampled uniformly directly in $\Lambda_{1,2}$, including another log-uniform layer inserted between $\Lambda_{1,2}=0$ and the first uniformly chosen point. A visualisation of the calibration region is shown in Fig.~\ref{fig:parameterspace}. 

An important point to note is that we do not assume hadronic physics, i.e. we do not enforce for $q\neq 1$, $\Lambda_2>\Lambda_1$, and for $q=1$, $\Lambda_2=\Lambda_1$ when sampling the training space. 
This is done deliberately to ensure that the model is well-behaved without imposing such limiting assumptions and that no pathologies occur. 

This is a particular strength of \phenom ~as it allows for more flexibility 
and separates it from models that are informed by NR data that assume a hadronic EOS. Consequently, this also allows the consideration of objects such as boson stars and quark stars and other exotic models which are not comprised of hadronic matter.

When generating the \teob waveforms, we choose a sampling rate of $16384$ Hz and a starting frequency of $10$ Hz below the equivalent of $0.0005Mf$ in Hz for a given total mass $M$ to allow for a buffer between the start of the waveform and the fitting regime. We stop the EOB evolution at an orbital separation of $r=4M$. We also explicitly turn off gravitomagnetic tides, dynamical tides, the use of universal relations, and set the octupolar and hexadecapolar tidal deformabilities to $10^{-8}$, the lowest possible value we found.

%%%%%%%%%%%%%%%%%%%%%%%%%%%%%%%%%%%%%%%%
\subsection{\teob-NR Hybrids} 
\label{sec:hybrids}
%%%%%%%%%%%%%%%%%%%%%%%%%%%%%%%%%%%%%%%%
For the model validation of \phenom we perform comparisons against a set of six \teob-NR hybrid waveforms. The NR waveforms are taken from the CoRe database~\cite{Dietrich:2018phi, Gonzalez:2022mgo, Core} and were produced with the \textsc{BAM} code (see Refs. in~\cite{Gonzalez:2022mgo} for details), which we hybridise with \teob waveforms using identical settings to those used for the calibration of \phenom and generated from $30$ Hz. Details of the six aligned-spin NR waveforms used in this paper are given in Tab.~\ref{tab:hybrids}. The choice of these simulations is motivated by simulation length requirements and minimal residual eccentricity. 

\begin{table*}[]
\begin{tabularx}{\textwidth}{l@{\extracolsep{\fill}}l@{\extracolsep{\fill}}l@{\extracolsep{\fill}}l@{\extracolsep{\fill}}l@{\extracolsep{\fill}}l@{\extracolsep{\fill}}l@{\extracolsep{\fill}}l@{\extracolsep{\fill}}l@{\extracolsep{\fill}}l@{\extracolsep{\fill}}l@{\extracolsep{\fill}}} \toprule[1pt]\midrule[0.3pt]
CoRE ID & EOS  & $M_{\rm ADM}\,[M_\odot]$    & $q$   & $\Lambda_1$     & $\Lambda_2$     & $\chi_{1z}$ & $\chi_{2z}$ & $e$      & $f_{\mathrm{NR}}$ [Hz] & Ref.\\ \hline
\textsc{BAM 0001} & 2B   & 3.05 & 1.0 & 126.7  & 126.7  & 0.0   & 0.0   & 7.0$\times 10^{-3}$ & 454.5  & \cite{Bernuzzi:2014owa}       \\
\textsc{BAM 0120} & SLy  & 3.05 & 1.0 & 346.1  & 346.1  & 0.0   & 0.0   & 1.5$\times 10^{-2}$ & 424.4  &  \cite{Dietrich:2017feu}       \\
\textsc{BAM 0127} & SLy  & 3.07 & 1.5 & 93.1   & 1367.9 & 0.0   & 0.0   & 8.0$\times 10^{-3}$ & 426.3  & \cite{Dietrich:2017feu}         \\
\textsc{BAM 0124} & SLy  & 2.76 & 1.5 & 191.0  & 2313.7 & 0.0   & 0.0   & 1.2$\times 10^{-2}$ & 407.8  &  \cite{Dietrich:2017feu}        \\
\textsc{BAM 0104} & SLy  & 2.99 & 1.0 & 388.2  & 388.2  & 0.19  & 0.19  & 7.4$\times 10^{-4}$ & 453.1  &  \cite{Dietrich:2017aum}        \\
\textsc{BAM 0066} & MS1b & 2.94 & 1.0 & 1531.5 & 1531.5 & 0.18  & 0.18  & 1.9$\times 10^{-3}$ & 427.2 &  \cite{Dietrich:2017aum}\\
\textsc{BAM 0095} & SLy & 2.7 & 1.0 & 390.1 & 390.1 & 0.0  & 0.0 & 4.0$\times 10^{-4}$ & 453.3  &  \cite{Dietrich:2017aum} \\
\midrule[0.3pt]\bottomrule[1pt]
\end{tabularx}
\caption{Summary of the parameters of the aligned-spin NR waveforms from the CoRe database used to create the \teob-NR hybrids. 
The columns denote the CoRE ID number, the EOS~\cite{Read:2008iy} used for the simulations, the ADM mass $M_{\rm ADM}$, mass ratio $q$, tidal deformabilities $\Lambda_1, \Lambda_2$, component spins $\chi_{1}, \chi_{2}$, the residual eccentricity $e$, the NR starting frequency $f_{\mathrm{NR}}$, and relevant references. All quantities are taken from the CoRE metadata.}
\label{tab:hybrids}
\end{table*}

Our $(\ell,m)=(2,2)$ hybrids are generated as follows:
We take the Newman-Penrose scalar $\psi_{4,22}$ at the largest fixed extraction radius available $r_0$ the simulation and extrapolate it to infinity via~\cite{Nakano:2015rda,Nakano:2015pta}
\begin{align}
\psi_{4,22}^{\infty}(t,r_0)  =& \left( 1-\frac{2M_{\rm ADM}}{r_A} \right) \nonumber \\
\times & \left[ \psi_{4,22}^{r_0}(t,r_0)  - \frac{2}{r_A} \int^t_0 \psi_{4,22}^{r_0}(t',r_0) dt' \right],
\label{eq:exttoinf}
\end{align}
where $r_A = r_0[1+M_{\rm ADM}/(2r_0)]^2$ and $M_{\rm ADM}$ is the ADM mass of the system~\cite{Arnowitt:1962hi} obtained from the CoRE metadata. 

The Newman-Penrose scalar is related to gravitational-wave strain $h_{22}(t)$ via 
\begin{equation}
\psi_{4,22} (t) = \frac{d^2h_{22}(t)}{dt^2}.
\label{eq:FFI}
\end{equation}
We implement this double time-integration via the fixed frequency integration (FFI), where the waveform is windowed such that it smoothly tapers to zero below $t=200M$ to remove any residual ``junk radiation'' at the start of the NR waveform due to imperfect initial data. We also taper to zero at an arbitrary time just after merger for each waveform to avoid edge effects in the FFI. The waveform is then zero-padded to reduce any spectral leakage. We implement the FFI following Ref.~\cite{Reisswig:2010di} choosing the NR starting frequency $f_{\mathrm{NR}}$ given in Tab.~\ref{tab:hybrids} as the fixed-frequency to perform the integration. 

The NR waveforms are then hybridised with \teob waveforms, where all settings are the same as those detailed in Sec.~\ref{sec:inputs} for the \teob generation with the exception of using the time-domain rather than the SPA implementation. The \teob waveforms are initialised with the same intrinsic parameters as the NR simulations. We note here that we used the non-eccentric version of \teob, however, the NR waveforms have non-negligible eccentricity as shown in Tab.~\ref{tab:hybrids}, which constitutes a source of error in the hybridisation procedure. 

\begin{figure*}
\centering
\includegraphics[width = \linewidth]{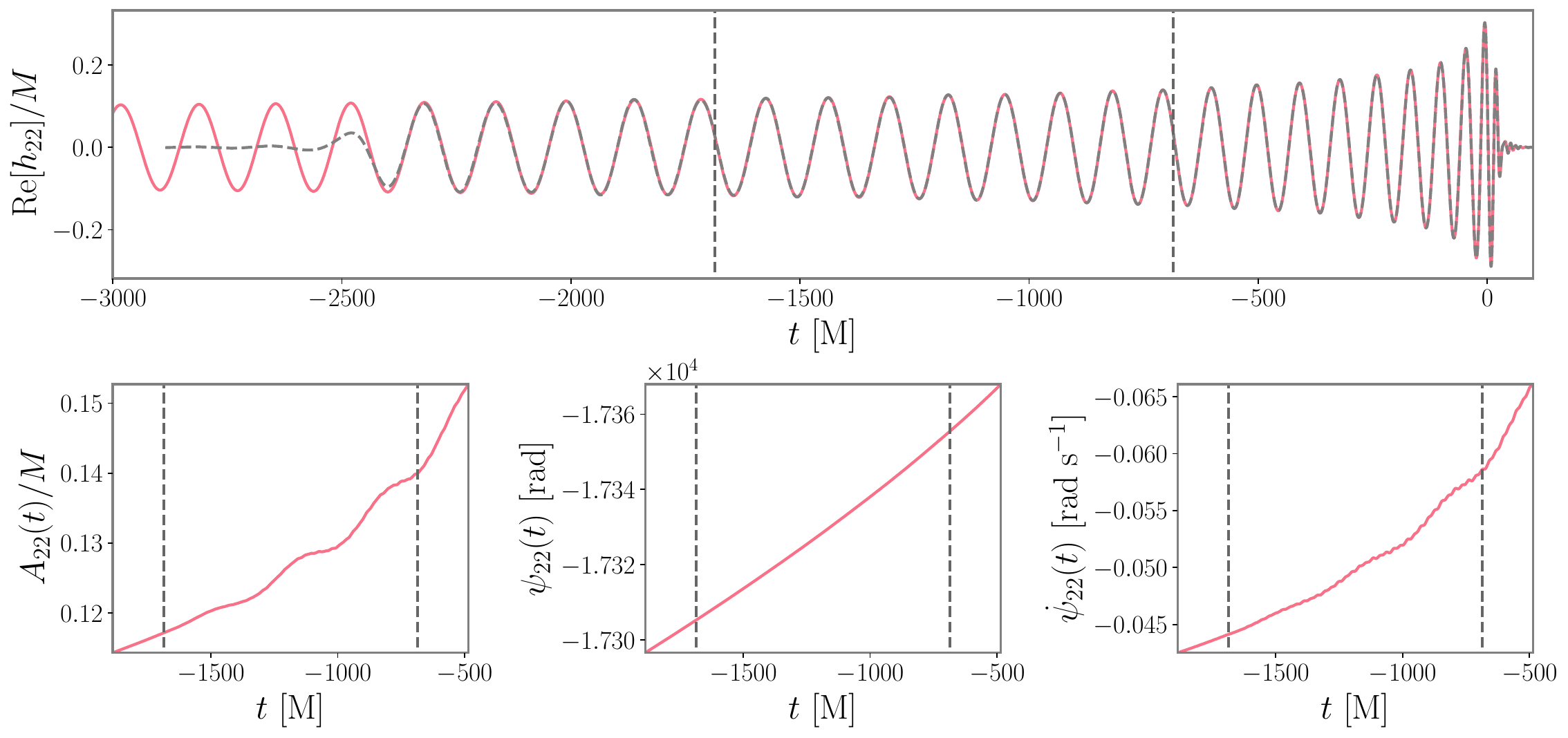}
\caption{\teob-NR hybrid example for BAM\_0001. \emph{Top panel}: Real part of $h_{22}$ showing the NR data (dashed grey), the aligned \teob ~waveform (pink) and hybridisation region marked by the two dashed vertical lines. \emph{Bottom panel}: Close ups of the hybridisation region for the resulting hybrid waveform for the amplitude (left), the phase (middle), and the first time derivative of the phase (right).}
\label{fig:Hybrids}
\end{figure*}

For the hybridisation between the NR and \teob waveforms, we first ensure that the amplitudes of both waveforms peak at $t=0$, and then choose a Planck window $\sigma(t)$~\cite{McKechan:2010kp} across a time interval $[t_1,t_2]$ to smoothly join the two waveforms such that 
\begin{equation}
  \sigma(t) =
    \begin{cases}
      0 & \text{if $t < t_1$},\\
      \frac{1}{1+e^z} , z=\frac{1}{t} - \frac{1}{1-t}& \text{if $t_1 \leq t \leq t_2$},\\
      1 & \text{if $t > t_2$},
    \end{cases}   
\label{eq:planck}
\end{equation}
where we find that $t_1=1200M$ (with the exception of \textsc{BAM 0095}, where $t_1=1000M$) after the start of the NR waveform and a window length of $1000M$ are sufficient. Across this window a time domain alignment is performed by minimising
\begin{equation}
    \min\left[ \int^{t_2}_{t_1} \Big|\psi^{\mathrm{TEOB}}(t) - \psi^{\mathrm{NR}}(t+t_0) - 2\psi_0\Big|^2 dt\right],
\end{equation}
and a visual inspection to check for the continuity of amplitude, phase and phase derivative is done as shown in Fig.~\ref{fig:Hybrids}.

%%%%%%%%%%%%%%%%%%%%%%%%%%%%%%%
\section{Tidal Phase Model}
\label{sec:inspiral}
%%%%%%%%%%%%%%%%%%%%%%%%%%%%%%%

%%%%%%%%%%%%%%%%%%%%%%%%%%%%%%%%%%%%%%%%%%
\subsection{Phenomenological Modelling}
\label{sec:phenom}
%%%%%%%%%%%%%%%%%%%%%%%%%%%%%%%%%%%%%%%%%%
A key motivation for building phenomenological (Phenom) waveform models is that they can yield closed-form frequency-domain expressions, making them extremely computationally efficient for GW data analysis whilst retaining a high level of accuracy. As a result, Phenom models have become a very popular choice in many aspects of gravitational-wave data analysis, including matched-filter searches, e.g.~\cite{Harry:2016ijz,Venumadhav:2019tad,McIsaac:2023ijd,Schmidt:2023gzj,Wadekar:2023kym,Wadekar:2024zdq}, and Bayesian inference, e.g.~\cite{LIGOScientific:2016vlm,LIGOScientific:2017vwq,LIGOScientific:2018mvr,LIGOScientific:2020ibl,LIGOScientific:2021qlt,KAGRA:2021vkt,LIGOScientific:2020zkf,LIGOScientific:2020ufj,LIGOScientific:2024elc}. We note that an alternative phenomenological model of~\teob using a different approach was developed in Ref.~\cite{Gamba:2023mww}.

In \phenom, we follow the framework previously described for the BBH waveform model \textsc{IMRPhenomXAS}~\cite{Pratten:2020fqn}. We will first briefly outline the main elements in the construction of a phenomenological phase model before presenting the detailed fits in the remainder of this section. 

We recall that the goal is to build a closed-form expression of the tidal phase of binary neutron stars in the frequency domain. The first step is to choose an appropriate mathematical ansatz for the tidal phase for each section of the different stages of the binary evolution. As we are only considering the inspiral phase here with a known PN expansion, a natural choice is a parameterised continuation of the finite-order PN series, schematically given by
\begin{align}
    \psi_T(\Vec{\theta},Mf) &= \psi_T^{\rm PN}(\Vec{\theta},Mf) + \sum_i a_i(\vec{\theta})(Mf)^i,
    \label{eq:exampleansatz}
\end{align}
where $\Vec{\theta}$ denotes the model parameters and $Mf$ is the dimensionless GW frequency. 

Note that the powers of the frequency are for illustrative purposes only. 
In the latest generation of phenomenological models, the phenomenological coefficients $a_i(\vec{\theta})$ are not directly calibrated due to numerical instabilities and poor numerical conditioning~\cite{Pratten:2020fqn}. 
Instead, the phenomenological coefficients are reconstructed by solving a linear system of equations expressed in terms of a finite number of collocation points that are placed at prescribed frequency nodes, $\lbrace M f_j \rbrace$.

In order for this to be a deterministic system, the number of constraints, e.g. boundaries, \emph{and} collocation points must equal the total number of free parameters. 
At each collocation point, a fit is constructed such that
\begin{equation}
\lambda_j(\Vec{\theta}) \equiv \psi_T(\Vec{\theta},Mf_j).
    \label{eq:examplefit}
\end{equation}
The resulting linear system of equations can then be solved for the phenomenological coefficients using standard matrix methods, e.g. via an LU decomposition.

%%%%%%%%%%%%%%%%%%%%%%%%%%%%%%%%%%
\subsection{Tidal Residual Fit}
\label{sec:fit}
%%%%%%%%%%%%%%%%%%%%%%%%%%%%%%%%%%
\begin{figure}
    \centering
    \includegraphics[width = \linewidth]{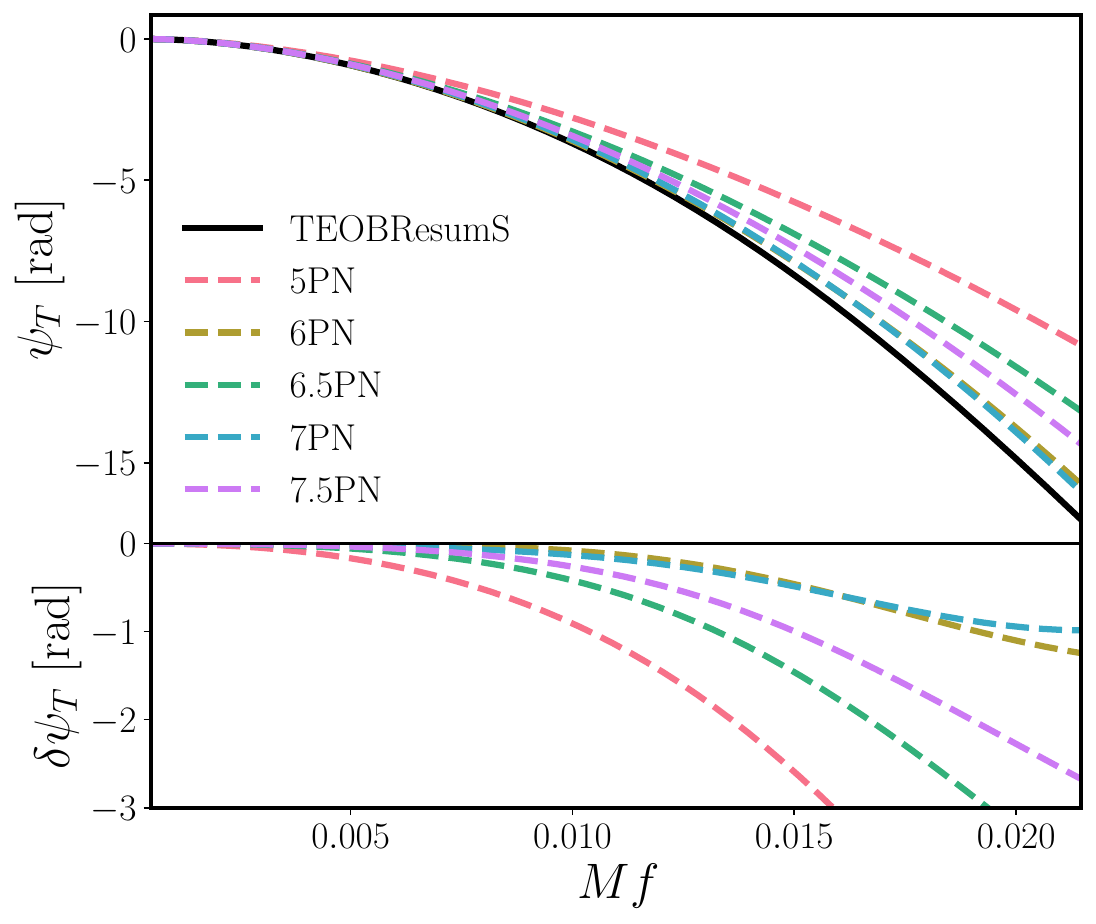}
    \caption{\emph{Top}: Tidal phase $\psi_{T}$ as a function of frequency for a binary with parameters $\{q=1,\Lambda_{1}=\Lambda_2=600\}$. We show the \teob ~data (solid black) alongside several TaylorF2 approximants of different PN orders (dashed lines) aligned as per Eq.~\eqref{eq:align}. \emph{Bottom}: Phase difference $\delta\psi_{T} = \psi^{\rm TEOB}_{T} - \psi_T^{\rm{xPN}}$ between between \teob ~and each Taylor approximant.}
    \label{fig:TaylorComp}
\end{figure}

For the PN baseline, we use \textsc{TaylorF2} with quadrupolar adiabatic tidal contributions up to $7.5$PN ~\cite{Bini:2012gu,Damour:2012yf,Henry:2020ski}. 
We add five pseudo-PN coefficients $a^i$ that are calibrated to the \teob dataset
\begin{align}
    \psi_{\ell m}^{\phenom} &= \psi^{\mathrm{7.5PN}}_T + \sum^{5}_{i=1}{a^i \left( \eta, \tilde{\Lambda}.\delta\tilde{\Lambda} \right) \left( Mf \right)^{ (10+i)/3 } }.
    \label{eq:PhenomGSF}
\end{align}
Instead of directly fitting the phase, we fit the residual between \teob and TaylorF2, 
\begin{align}
    r \equiv \psi^{\mathrm{TEOB}}_T-\psi^{\mathrm{7.5PN}}_T - 2\pi Mft_0 + \phi_0,
\end{align}
where $\{t_0, \phi_0\}$ are two gauge degrees of freedom corresponding to an overall time and phase shift and are determined by minimising the residuals over a frequency interval
\begin{equation}
    \min\left[ \int^{0.001M}_{0.0005M} |r|^2 dMf\right].
    \label{eq:align}
\end{equation}

A choice that must be made is the PN baseline for the phenomenological model. 
It is well known that the PN expansion is slowly converging, and higher-order PN terms do not necessarily lead to smaller residuals. In fact, successively higher-order PN terms often lead to partial cancellations. Whilst the 6PN and 7PN terms yield a smaller residual relative to \teob, they also display a turning point at high frequencies for large tidal deformabilities. This introduces additional complexity into the function which can be difficult to accurately fit, often requiring more phenomenological coefficients in the \phenom ansatz. Therefore, we adopt a 7.5PN baseline as it is the highest-order PN series and yields a monotonic residual across the parameter space, see Fig.~\ref{fig:TaylorComp}.

\begin{figure}
    \centering
    \includegraphics[width = \linewidth]{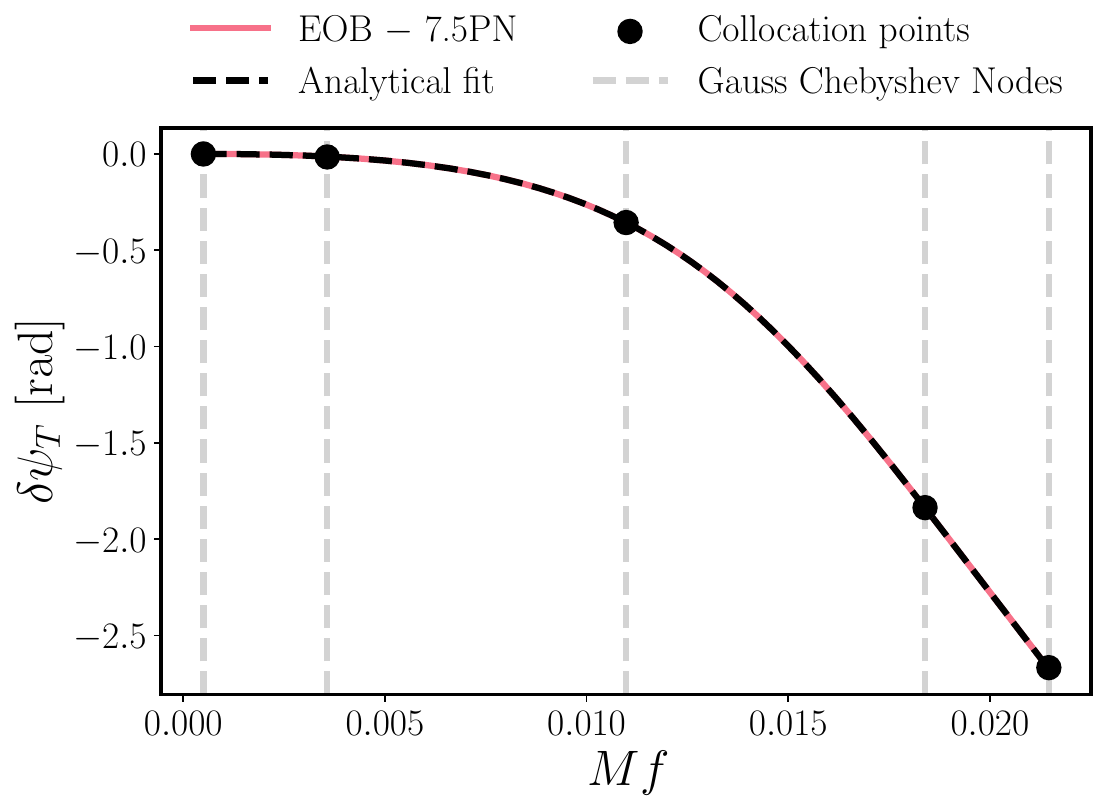}
    \caption{Example of a tidal residual $\{q=1,\Lambda_{1,2}=600\}$ as a function of frequency (pink solid line) with the corresponding analytical fit from a $5^{\mathrm{th}}$ order polynomial (black dashed line). The Gauss-Chebyshev nodes (grey dashed line) are shown alongside their corresponding collocation points (black circles).}
    \label{fig:CLpoints}
\end{figure}

We choose collocation points at Gauss-Chebyshev nodes to reduce fitting errors in comparison to equidistant nodes~\cite{Pratten:2020fqn}, in which the $i^{\rm th}$ node is placed at
\begin{align}
    f_i &= f_{\mathrm{min}} + \frac{f_{\mathrm{max}}-f_{\mathrm{min}}}{2} \left(1+\cos\left[\frac{i\pi}{n}\right]\right),
    \label{eq:GC}
\end{align}
where $\{f_{\mathrm{min}},f_{\mathrm{max}}\}$ is the frequency range over which the nodes are placed and $n$ is the number of collocation points.
We find that five phenomenological coefficients provides an optimal balance between goodness-of-fit and the subsequent computational efficiency of the model. An example of this procedure is shown in Fig~\ref{fig:CLpoints}.

A practical consideration is that the choice of termination frequency dictates the frequency spacing of the collocation points, and the concomitant structure of each phenomenological fit. 
A choice of a constant termination frequency would correspond to each collocation point lying at the same geometric frequency at every point in the parameter space. 
This leads to problems for binaries with extremely large and extremely small tidal deformabilities, with the relevant morphological features being poorly captured. 
For example, a termination frequency that is too high can result in information beyond $f_{\rm MECO}$ being included, where the PN expansion has formally broken down. 

To resolve this, one may try to adopt a parameter-space dependent termination frequency, though this often results in a more complex structure to the global fit of the collocation values. One such example is the NR informed contact frequency $M f_{\rm contact}$~\cite{Gonzalez:2022mgo}, though we find that this leads to a number of subtle issues. First, $M f_{\rm contact}$ spans a large dynamical range across the parameter space, introducing significant substructure to the fits. Second, we find issues in constructing a fit across the parameter space related to how the frequency smoothly transitions to the black-hole limit and exotic equations-of-state that are outside the calibration regime of $M f_{\rm contact}$, especially when $\Lambda_2 > \Lambda_1$. 

Overall, we find that the best compromise is the non-spinning hybrid minimum energy circular orbit (MECO) frequency, $M f_{\rm MECO}$, introduced in~\cite{Cabero:2016ayq}. This spans a comparatively small range of frequencies that are well-behaved and simplify the resulting phenomenological fits. In practice, we use the phenomenological fit for $M f_{\rm MECO}$ provided in Ref.~\cite{Pratten:2020fqn},
\begin{align}
    M f_{\mathrm{MECO}}(\eta) &= \frac{a_0 + a_1 \eta + a_2 \eta^2 + a_3 \eta^3}{1 + b \, \eta},
\end{align}
with coefficients
\begin{align*}
a_0 &= \phantom{-}0.018744, & a_1 &= \phantom{-}0.007790,\\
a_2 &= \phantom{-}0.003940, & a_3 &= -0.000067, \\
b   &= -0.104233.
\end{align*}
The \teob waveforms are then generated over the frequency interval $\{Mf_{\mathrm{min}},Mf_{\mathrm{max}}\}$ = $\{0.0005M, Mf_{\mathrm{MECO}}\}$. For a $2 M_{\odot}$ binary, the lower frequency limit corresponds to $50$Hz, which is significantly below the GW frequency ($\sim 400$Hz) where tidal effects introduce a measurable phase difference relative to the BBH case~\cite{Flanagan:2007ix}. As we do not recalibrate the point-particle sector, we find no practical benefit in going to lower starting-frequencies.  

\begin{figure}
    \centering
    \includegraphics[width = \linewidth]{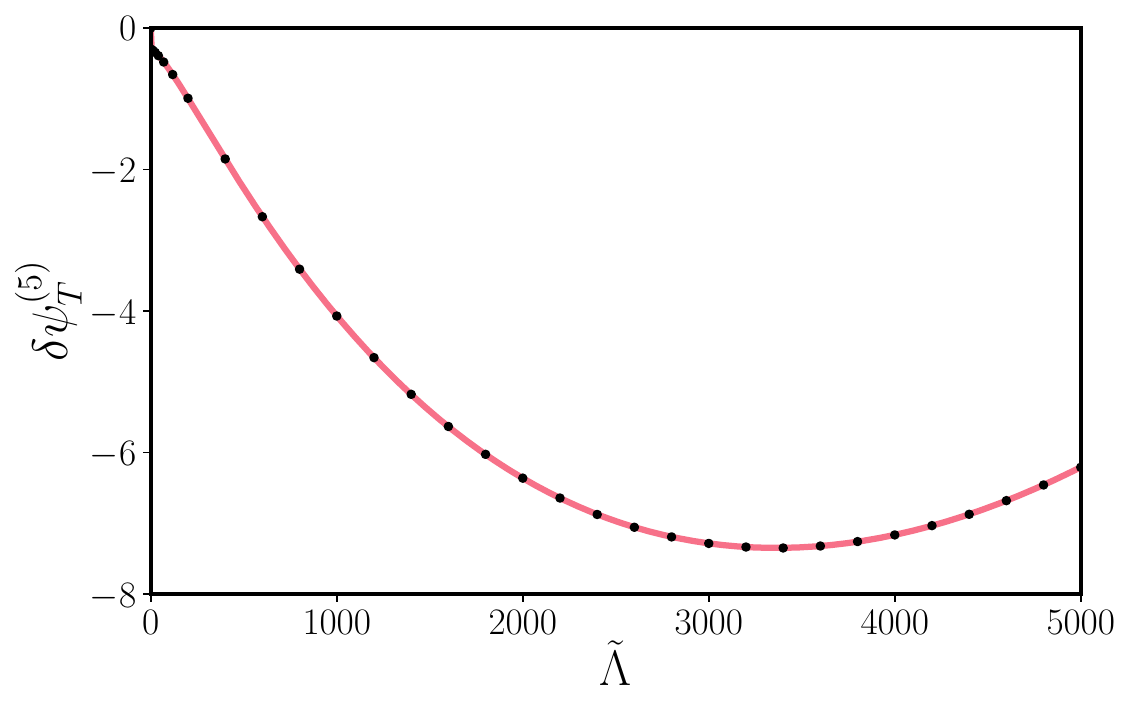}
    \caption{1D fit for equal mass ($\eta=0.25$) systems with $\delta\tilde{\Lambda}=0$ across $\tilde{\Lambda}$ at the $5^{\mathrm{th}}$ colocation point. The values of the dephasing at this colocation point are shown (black dots), as well as the final Pade fit (pink solid line).}
    \label{fig:1dfit}
\end{figure}

\subsection{Parameter Space Fits}
At each collocation point, we construct a fit for the value of the collocation point. This provides a map from $\{\eta, \tilde{\Lambda}, \delta\tilde{\Lambda}\}$ to the value of residual at the collocation node. 
We explored further reducing the parameter space to $\{\eta, \tilde{\Lambda}\}$, but this resulted in a significant degradation of the accuracy. As such, in this section we only discuss the full 3D fit. 
Due to the high dimensionality, we adopt a hierarchical approach in which the parameter space fits are constructed dimension-by-dimension~\cite{Jimenez-Forteza:2016oae,Pratten:2020fqn}. 
In practice, the starting point is a 1D fit across a well-defined subspace. From this, we can construct the 2D fit by expanding the 1D basis in terms of the second dimension. 
The complete 3D information is then incorporated by re-expanding the 2D basis in terms of the final dimension. 
All fits are computed using Mathematica's \texttt{NonlinearModelFit} package.

\subsubsection{1D Fit}
The 1D fit is taken to correspond to the equal mass, $\eta = 0.25$, and equal tidal deformability, i.e. $\delta\tilde{\Lambda} = 0$, subspace. 
In Fig.~\ref{fig:1dfit}, we show the dephasing at the fifth collocation point as a function of $\tilde{\Lambda}$. 
This collocation point corresponds to $M f_{\mathrm{MECO}}$. 
We note that there is a notable change in the slope of the fit %small discontinuity 
as $\tilde{\Lambda} \rightarrow 0$ corresponding to the transition from a binary neutron star to a binary black hole, which is related to the minimum tidal deformability enforced in \teob. 
In practice, we find that such features do not cause significant issues in the fitting procedure and we have verified that the model extrapolates sufficiently smoothly to the binary black hole limit. 

We fit polynomial ansatzes against the data up to various orders $I$
\begin{align}
    F(\eta = 0.25, \tilde{\Lambda}, \delta\tilde{\Lambda}=0) = \sum^{I}_{i=0}{a_i}\tilde{\Lambda}^i,
    \label{eq:1Dpoly}
\end{align}
where a best fit is defined for the lowest Bayesian Information Criterion (BIC), which attempts to avoid overfitting by adding a penalty for the number of parameters in a given model, e.g. see discussion in~\cite{Jimenez-Forteza:2016oae}. 
After selecting the best fit, we construct a Pad\'{e} approximant at a given order, which helps to effectively resum the information,
\begin{align}
\sum_{i=0}^{I} a_i \tilde{\Lambda}^i &\rightarrow \frac{\sum_{j=0}^{J} b_j \tilde{\Lambda}^j}{1 + \sum_{k=1}^{K} c_k \tilde{\Lambda}^k},
\label{eq:1Dpade}
\end{align}
where $J,K\leq I$. A benefit of using Pad\'{e} approximants is that they are more robust when extrapolating outside the calibration region than standard Taylor expanded results~\cite{Trefethen:2019apx}.

As before, we construct a range of Pad\'{e} approximants with differing orders and use \textsc{NonlinearModelFit} to infer the BIC. The residuals and BIC are used to determine the functional form of the final 1D fit. 

\subsubsection{2D Fit}
To obtain the $\{ \eta, \tilde{\Lambda} \}$ fit, we expand Eq.~\eqref{eq:1Dpade} about $\eta$ keeping $\delta\tilde{\Lambda} = 0$. 
The hierarchical process involves inserting a polynomial of order $M$ into the 1D fit via
\begin{align}
    a_l &\rightarrow a_l \sum^{M}_{m=0}{b_{lm}\eta^m}.
    \label{eq:2D hierarchical}
\end{align}

\begin{figure}[h]
    \centering
    \includegraphics[width = \linewidth]{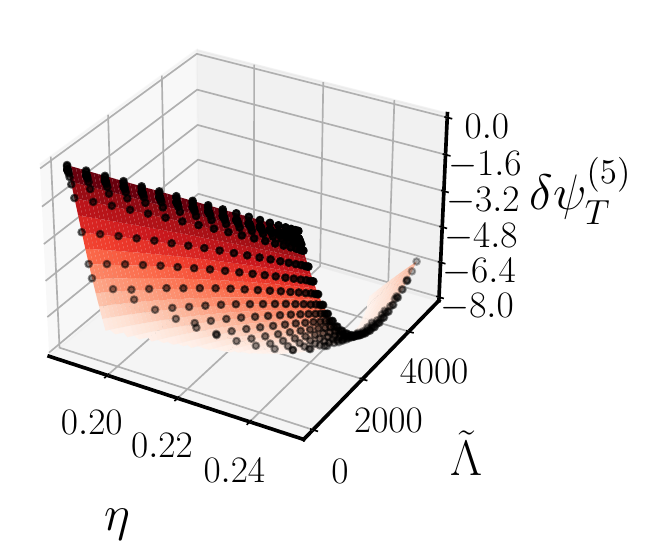}
    \caption{2D fit across $\{\eta, \tilde{\Lambda}\}$ for $\delta\tilde{\Lambda}=0$ at the $5^{\mathrm{th}}$ collocation point. The values of the dephasing at this collocation point are shown (black dots), as well as the final 2D fit (red surface).}
    \label{fig:2dfit}
\end{figure}

We only apply this insertion to the numerator of the 1D fit Eq. \eqref{eq:1Dpade} to avoid singularities. To asses the fit quality we found that in the 2D case, the penalty introduced in the BIC is not sufficient to avoid over-fitting. We introduce 3 sets of randomly generated data in the subparameter space $\lbrace \eta, \tilde{\Lambda}, \delta\tilde{\Lambda}=0 \rbrace$ and evaluate the goodness of fit between the independent data and the proposed ans\"{a}tze. To measure the goodness-of-fit, we use an $R^2$ criterion
\begin{equation}
R^2 = 1 - \frac{\sum_i (r_i - \hat{r}_i)^2}{\sum_i (r_i - \bar{r}_i)^2},
\end{equation}
where $r_i$ is the residual fit value, $\hat{r}_i$ is the true value, and $\bar{r}_i$ is the mean.
We impose a condition of $1-R^2 \leq 10^{-3}$, with fits exceeding this tolerance being discarded. Due to the increasing complexity of the higher-order fits, we evaluate the BIC for each fit with the highest-scoring, and therefore lowest contributing factor, being removed. 
This process is iterated until 1-$R^2$ decreases by at least an order of magnitude. 
This helps eliminate parameters in the fit that are not contributing and slowing the computation time down. An example of a 2D fit is shown in Fig.~\ref{fig:2dfit}.

%%%%%%%%%%%%%%%%%%%%%%%%%%%
\subsubsection{3D Fit}
%%%%%%%%%%%%%%%%%%%%%%%%%%%
This process is then repeated for the full 3D fit, with all input data, covering the full ${\eta, \tilde{\Lambda}, \delta\tilde{\Lambda}}$ space as detailed in Sec.~\ref{sec:inputs}. Once again a polynomial of order $N$ is inserted in the numerator
\begin{equation}
    a_lb_{lm} \rightarrow a_lb_{lm} \sum^{N}_{n=1}{c_{lmn}\delta\tilde{\Lambda}^n}.
    \label{eq:3D hierarchical}
\end{equation}
Note here that by definition $\delta\tilde{\Lambda} = 0$ recovers the 2D case. Once again the same $R^2$ criterion as in the 2D fit is implemented alongside the iterative cutting of $c_{lmn}$. At this point a full parameter space fit is achieved, full expressions for the 3D fits at each collocation point is given in Appendix~\ref{sec:FitsAppendix}.

\begin{figure*}[hbt!]
    \centering
    \includegraphics[width = \linewidth]{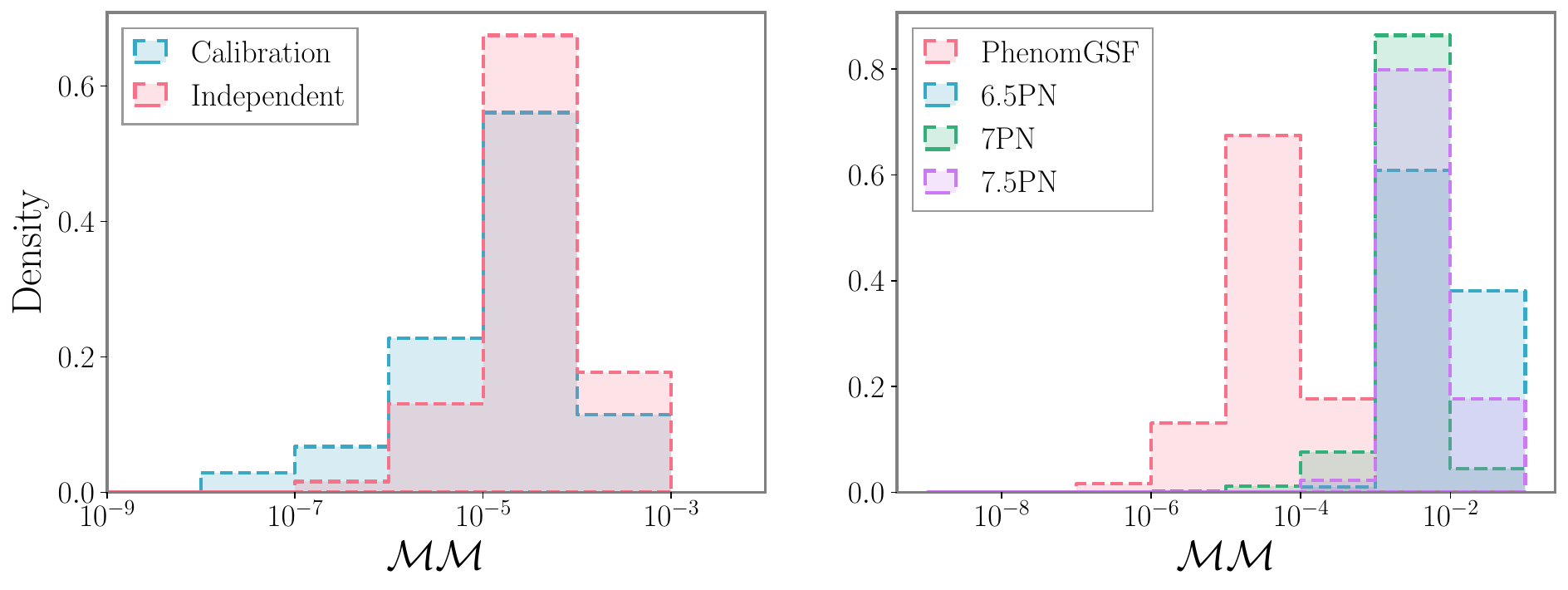}
    \caption{Mismatches between \teob and several other tidal models. \emph{Left panel}: Mismatches of \phenom against the calibration data (blue) as well as randomly selected binaries within the calibration region but not seen in the fitting (pink). \emph{Right panel}: Mismatches of \phenom data (pink), and \textsc{TaylorF2} 6.5PN (blue), 7PN (green) and 7.5PN (purple) and against \teob.}
    \label{fig:ModelMM}
\end{figure*}

%%%%%%%%%%%%%%%%%%%%%%%%%%%%%%%%
\subsection{Inclusion of Spin}
\label{sec:spin}
%%%%%%%%%%%%%%%%%%%%%%%%%%%%%%%%%
Whilst \phenom is only calibrated to nonspinning data, spin-effects, such as the spin-spin and self-spin tidal terms, can be incorporated into the phase using the PN expressions directly 
\begin{align}
    \psi_{\mathrm{T}} &= \psi_{\mathrm{BBH}} + \psi_{\mathrm{PhenomGSF}} + \psi_{\mathrm{SS}}.
\end{align}

Contributions associated to the mass-monopole and spin-dipole are already incorporated in the point-particle phase. We use the PN baseline from \cite{Pratten:2020fqn}, which includes all point-particle spin information up to 3.5PN \cite{Marsat:2014xea,Bohe:2015ana}. We subsequently incorporate the equation-of-state dependent quadrupole and octupole terms, which are related to the mass and spin moments via $M^{(A,B)}_{2} = - C^A_{\rm Q} M^3_{A,B} \chi^2_{A,B}$ and $S^{(A,B)}_3 = - C^{A,B}_{\rm Oct} M^4_{A,B} \chi^3_{A,B}$ respectively. Here $C_{\rm Q}$ and $C_{\rm Oct}$ parameterize the quadrupolar and octupolar deformations of the stars induced by their spin, e.g.~\cite{Poisson:1997ha,Damour:2009vw,Porto:2010zg,Levi:2014sba,Marsat:2014xea,Bohe:2015ana,Nagar2018,Nagar:2018plt,Dietrich:2019kaq}. For neutron stars or other exotic compact objects, $C_{X} \neq 1$ and need to be calculated assuming a given equation-of-state.  The self-spin terms can be schematically written as~\cite{Dietrich:2019kaq}
\begin{equation}
    \psi_{\mathrm{SS}} = \frac{3x^{5/2}}{128\nu}\left( \psi^{A}_{\mathrm{SS,2PN}}x^2 + \psi^{A}_{\mathrm{SS,3PN}}x^3 + \psi^{A}_{\mathrm{SS,3.5PN}} x^{7/2} \right),
    \label{eq:spinphase}
\end{equation}
where $x=(2\pi/f)^{2/3}$ and the individual coefficients can be found in~\cite{Dietrich:2019kaq}, ensuring that contributions from the black hole multipoles are explicitly removed to avoid double-counting their effect, which is already contained in the point-particle phase. 

This extends the model to include partial aligned-spin information, though we note that these contribution are not calibrated within the \phenom model and therefore not informed by \teob. This is similar to how spin information is incorporated in other models, e.g.~\cite{Dietrich:2019kaq, Abac:2023ujg}.
Should we wish to reduce the number of free parameters in the model, we can impose URs that relate $C^A_{\rm Q}$ and $C^A_{\rm Oct}$ to the tidal deformability $\Lambda_A$, e.g.~\cite{Yagi:2016bkt}. This is the strategy we employ by default, though the model also allows us to treat $C^A_{\rm Q}$ and $C^A_{\rm Oct}$ as free EOS-dependent parameters to be constrained by the data.
 
Model validation including spin is considered in Sec.~\ref{sec:validation} to check for consistency. 
We note that when including the 3.5PN spin-spin term, we do make use of universal relations, which are required to calculate the compactness of each neutron star~\cite{Yagi:2016qmr}. 
Extending this to generic compact objects, such as boson stars, is left to future work. Nonetheless, this is a comparatively small contribution to the phase and as we do not impose universal relations between the mass and tidal deformability, we do not expect this to have a significant impact. If desired, the spin-spin contributions in \phenom can be disabled, given the modularity and flexibility of the framework. 

%%%%%%%%%%%%%%%%%%%%%%%%%%%%
\section{Model Validation}
\label{sec:validation}
%%%%%%%%%%%%%%%%%%%%%%%%%%%%
To assess the accuracy of \phenom we perform several mismatch comparisons as well as full Bayesian parameter estimation. 

\begin{figure*}
    \centering
    \includegraphics[width = \linewidth]{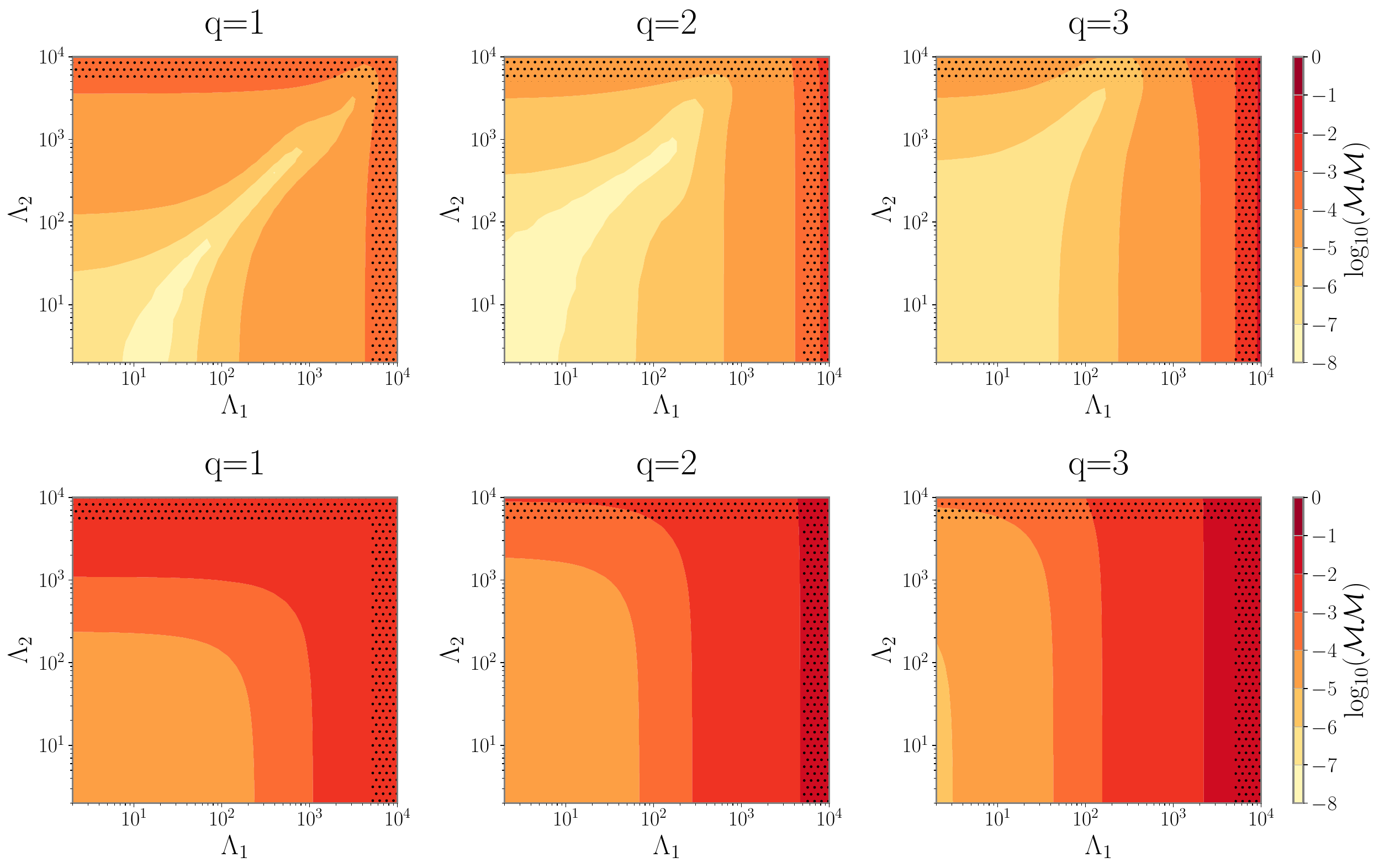}
    \caption{Mismatches across the parameter space over $M f= [0.0005, M f_{\mathrm{MECO}}]$ at $q=1$, $q=2$ and $q=3$. The dotted areas denote regions of parameter space outside the calibration range of the model. \emph{Top row}: Mismatches between \teob and \phenom. \emph{Bottom row}:  Mismatches between  \teob and \textsc{TaylorF2} at 7.5PN.}
    \label{fig:Extrapolation}
\end{figure*}

\begin{figure*}
    \centering
    \includegraphics[width = \linewidth]{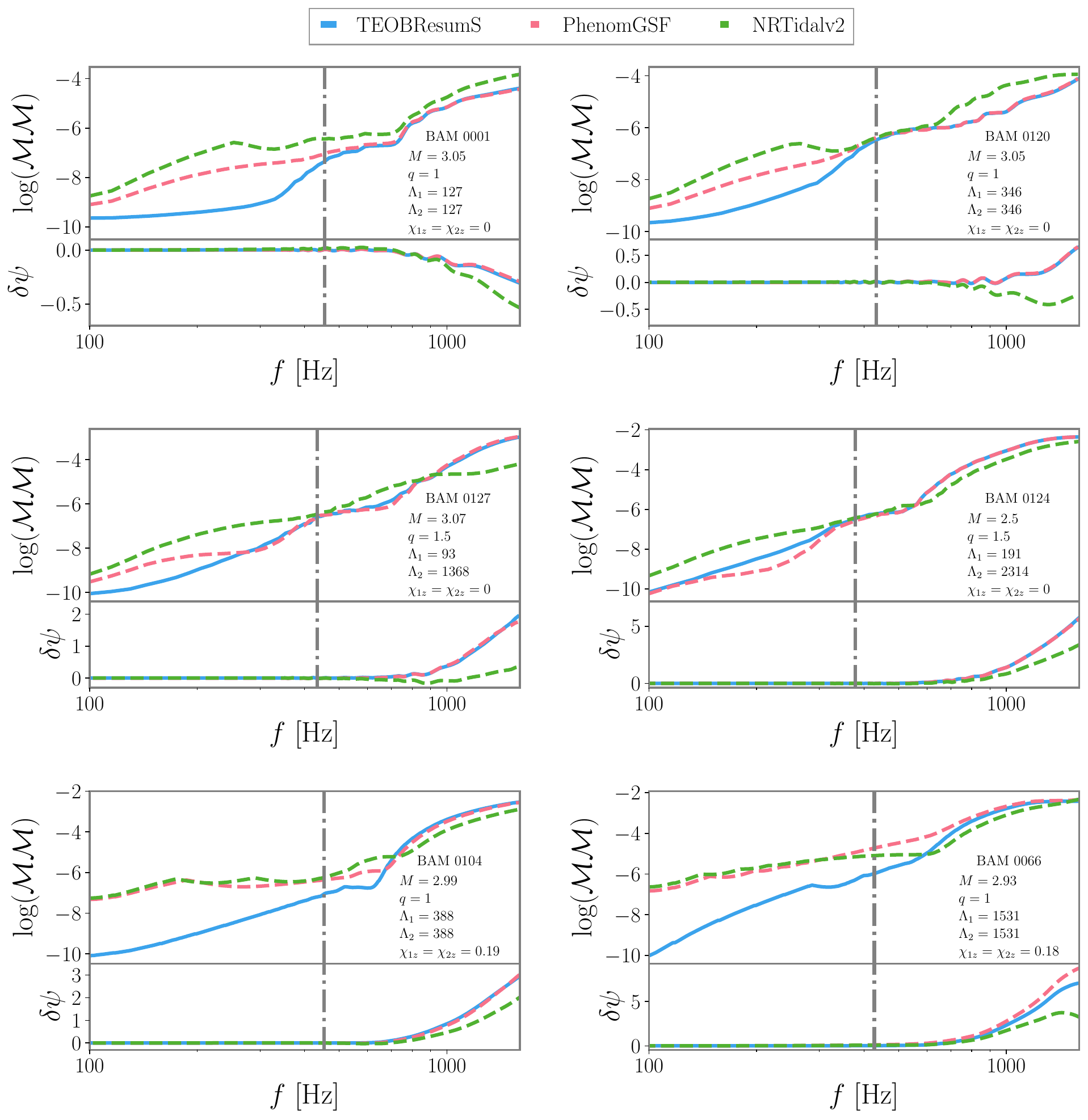}
    \caption{Comparisons for \teob-NR hybrid waveforms against \teob (blue solid) , \phenom (orange dashed) and \textsc{NRTidalv2} (green dashed). In each plot the upper panel shows log mismatches computed from $40$ Hz to a maximum frequency $f$; the lower panel shows the dephasing between the NR hybrid and one of the tidal approximants, i.e. $\delta\psi = \psi_{NR} - \psi_X$ where $X$ is one of the approximants. The NR starting frequency is marked by the grey dot-dashed line. \emph{Top row}: Equal-mass nonspinning systems. \emph{Middle row}: Unequal mass nonspinning systems. \emph{Bottom row}: Equal-mass spinning systems. The details of the NR simulations are given in Tab.~\ref{tab:hybrids}.} 
    \label{fig:DephasingNR}
\end{figure*}

%%%%%%%%%%%%%%%%%%%%%%%%%%%%%%%%%%%
\subsection{Mismatch Comparisons}
\label{sec:mismatch}
%%%%%%%%%%%%%%%%%%%%%%%%%%%%%%%%%%%
To assess how similar two waveforms are, we compute the match $\mathcal{M}$ defined as the time- and phase-optimised noise-weighted inner product between two waveforms given by
\begin{equation}
    \mathcal{M}(h_1, h_2) = \max_{t_{0}, \phi_{0}}\frac{\langle h_1, h_2\rangle}{\sqrt{\langle h_1, h_1\rangle \langle h_2, h_2\rangle}},
\end{equation}
where
\begin{equation}
    \langle h_1|h_2\rangle = 4 \Re \int_{f_{\rm min}}^{f_{\rm max}} \frac{\tilde{h}_1(f)\tilde{h}_2^*(f)}{S_n(f)} df,
    \label{eq:innerprod}
\end{equation}
where $S_n(f)$ denotes the one-sided power spectral density (PSD) of the detector noise, $\tilde{h}$ the Fourier transform of $h$ and ${}^*$ complex conjugation. For a detector agnostic approach, we choose white noise, i.e. $S_n(f)=1$.
The \emph{mismatch} $\mathcal{MM}$ is then defined as  
\begin{equation}
\label{eq:match}
    \mathcal{MM}(h_1,h_2) \equiv 1 - \mathcal{M}.
\end{equation}

To compute mismatches, we first need to construct complete waveforms $h_1, h_2$. We choose the \teob BBH phase, the \teob BNS amplitude and vary the choice of tidal phase. Therefore, in all mismatches shown the only difference is the tidal phase allowing for a direct comparison. 

The left panel of Fig.~\ref{fig:ModelMM} shows histograms of the mismatches between \teob and \phenom for the $8446$ waveforms used in the model calibration, and 1000 independent waveforms that lie within the model's calibration range but were not used in the fitting. 
The mismatches in both data sets are $\mathcal{MM} \leq 10^{-3}$, with a median mismatch of $\sim 2\times 10^{-5}$ for the calibration dataset and $\sim 3\times 10^{-5}$ for the independent dataset, which quantifies the average modelling error. We also see excellent agreement between the two distributions, with a tail towards lower mismatches, demonstrating that the \phenom model is robust and not overfitted. The improvement of \phenom over \textsc{TaylorF2} at different PN orders is shown in the right panel of Fig.~\ref{fig:ModelMM}. Here we find that the agreement between \phenom and \teob is on average $\sim 3$ orders of magnitude better than with any of the  considered PN approximants.

In Fig.~\ref{fig:Extrapolation} we show how the mismatches of \teob against \phenom (top row) and \textsc{TaylorF2} with 7.5PN tides (bottom row) vary as a function of $\{\Lambda_1, \Lambda_2 \}$ for $q=1,2,3$. Here, we also include waveforms outside the calibration region of \phenom to test for the robustness of the model under extrapolation to $\Lambda_{1,2} \leq 10000$ (dotted regions). We find that the \phenom mismatches degrade for more unequal-mass systems, although they remain $< 10^{-3}$ throughout the calibration range. 
We note that in the equal-mass limit the mismatches are not fully symmetric in the $\Lambda_1 - \Lambda_2$ plane, even though the input data is  symmetric under the interchange $1 \leftrightarrow 2$. 
This is due to the fact that in the fitting procedure, we did not enforce this symmetry in the equal-mass limit. 
Nonetheless, the mismatches in this regime are negligible and of the order $\sim \mathcal{O}(10^{-7})$. The asymmetry observed for unequal masses is as expected.
In contrast, tidal \textsc{TaylorF2} performs significantly worse even for moderate parameters, exceeding mismatches of $10^{-3}$ for $\Lambda_{1,2}\gtrsim 1000$ at  $q=1$, and $\Lambda_1\gtrsim 100$ for $q=3$. 
We observe better agreement within the regions nominally consistent with hadronic EOSs: For example, for equal masses $\Lambda_1 = \Lambda_2$ gives the lowest mismatches, and for unequal mass ratios the lowest mismatches are found for $\Lambda_2>\Lambda_1$.

Outside the calibration region, \phenom achieves mismatches $< 10^{-3}$ for equal-mass binaries; for $q=3$, mismatches remain below $10^{-3}$ for $\Lambda_{2} < 10000$ provided that $\Lambda_{1} < 5000$ but start to degrade significantly for larger values of $\Lambda_1$. We note, however, that this requires the radius of the primary neutron star to be $\sim 2.6$ times larger than that of the secondary.

Finally, we quantify the model's accuracy by also computing mismatches against a selection of \teob-NR hybrid waveforms as detailed in Sec.~\ref{sec:hybrids}. 
In Fig.~\ref{fig:DephasingNR} we show comparisons against \teob, \phenom and \textsc{NRTidalv2}. We note that the \textsc{NRTidalv2} model is the only tidal phase model in this comparison that is calibrated to NR simulations. Hence, we expect this model to have better agreement with the hybrids close to merger. At low frequencies, however, we expect \phenom to yield dephasing results similar to those of \teob by construction. 
The top two panels of Fig.~\ref{fig:DephasingNR} are for equal-mass nonspinning systems, the middle two for unequal mass nonspinning systems, and the bottom two are for equal-mass systems with aligned spins. For each panel, the top figure shows the mismatch where $f_{\rm min} = 40$ Hz to varying $f_{\rm max}$ up to a maximum of $f_{\rm max} = f_{\rm MECO}$, while the bottom shows the phase difference as a function of GW frequency. 

As anticipated, \phenom tracks the phase of \teob to a very high degree. The only notable phase difference between \phenom and \teob is seen for \textsc{BAM 0066}. 
In all cases, except for the equal-mass nonspinning hybrids, \textsc{NRTidalv2} gives the smallest dephasing for the NR portion of the hybrid waveforms.  
The observed oscillations in the dephasing are due to non-negligible residual eccentricities in the NR initial data (see Tab.~\ref{tab:hybrids}). 
For the mismatches we find that for all comparisons the \phenom results consistently approach those of \teob around or after the NR starting frequency indicated by the vertical line. The mismatches seemingly disagree at lower frequencies, however, we note that this difference is $\sim \mathcal{O}(10^{-10})$ for the nonspinning cases and $\sim \mathcal{O}(10^{-7})$ for the spinning cases. 
Surprisingly, for the equal mass non-spinning cases (top row) we find that \phenom performs better than \textsc{NRTidalv2} across the entire frequency range. 
For the unequal mass cases, \phenom obtains lower mismatches in the low-frequency region, however, in the high frequency regime \textsc{NRTidalv2} performs better. 
For the spinning comparisons we find that \phenom and \textsc{NRTidalv2} are comparable for low frequencies, while \teob is by far the most faithful. When including the NR region, all three models are comparable with \textsc{NRTidalv2} being slightly better than the other two models. 

%%%%%%%%%%%%%%%%%%%%%%%%%%%%%%%%
\subsection{Parameter Estimation}
\label{sec:PE}
%%%%%%%%%%%%%%%%%%%%%%%%%%%%%%%%
To further validate \phenom we carry out full Bayesian inference on simulated nonspinning and aligned-spin \textsc{PhenomXAS\_PhenomGSF} signals (injections), a \textsc{\teob}-NR hybrid waveform, and GW170817 data~\cite{LIGOScientific:2019lzm}. 

The posterior density distribution function (PDF) of a set of model parameters $\boldsymbol{\theta}$ given the data $d$ is 
\begin{align}
    p(\boldsymbol{\theta}|d) &= \frac{ \mathcal{L}(d|\boldsymbol{\theta}) \pi(\boldsymbol{\theta})}{\mathcal{Z}_d},
\end{align}
where $\mathcal{L}(d|\boldsymbol{\theta})$ denotes the likelihood, $\pi(\boldsymbol{\theta})$ the prior, and $\mathcal{Z}_d$ the evidence or marginalised likelihood which is defined as
\begin{equation}
    \mathcal{Z}_d = \int \mathcal{L}(d|\boldsymbol{\theta}) \pi (\boldsymbol{\theta}) d\boldsymbol{\theta}.
\end{equation}

For our analyses, we consider the following model parameters: $\boldsymbol{\theta} = \{\mathcal{M}_c, q, \chi_1,\chi_2,\tilde{\Lambda}, \delta{\tilde{\Lambda}}, \mathrm{RA}, \mathrm{DEC}, \theta_{jn}, \psi, \phi, t_c\}$, where $\mathcal{M}_c = (m_1 m_2)^{3/5}/M^{1/5}$ denotes the chirp mass, $q$ the mass ratio, $\chi_1$ and $\chi_2$ the two spin magnitudes for the aligned-spin runs, RA and DEC the right ascension and declination of the binary in the sky, $\psi$ the polarisation angle, $\phi$ and $t_c$ the phase and time of coalescence respectively, and $\theta_{jn}$ the inclination of the total angular momentum w.r.t. to the line-of-sight. 
The luminosity distance $D_L$ and phase $\phi$ are marginalised over in the likelihood evaluation and reconstructed in post-processing. 
All runs were performed with \textsc{Bilby}~\cite{Ashton:2018jfp} in conjunction with the nested sampler \textsc{Dynesty}~\cite{Speagle:2020aa} with slice sampling. For all analyses we used 2000 live points, 20 slices and 50 autocorrelation lengths with 32 CPU cores. 
Although we sample in $\mathcal{M}_c$ and $q$, we select priors that are uniform in component masses, whilst $\tilde{\Lambda}$ and $\delta{\tilde{\Lambda}}$ are sampled directly from uniform priors. The priors for each analysis are given in Tabs.~\ref{tab:injections} and~\ref{tab:PE}. Additionally, we constraint the individual tidal deformabilities $\Lambda_1$, $\Lambda_2$ to be between $0$ and $5000$.

\begin{table*}[]
\begin{tabularx}{\textwidth}{>{\hsize=.5\hsize}X >{\hsize=.5\hsize}X X X X}
\toprule[1pt]\midrule[0.3pt]
Parameter & Injected Value & Prior & Nonspinning  & Aligned-Spin \\ \hline 
$\mathcal{M}_c [ M_{\odot}]$ & 1.20 & U$_{(m_1,m_2)}[\mathcal{M}_{\rm{min}}, \mathcal{M}_{\rm{max}}]$ & 1.20 $\pm^{2.17\times 10^{-5}}_{2.17\times 10^{-5}} $ & 1.20 $\pm^{4.34\times 10^{-5}}_{3.20\times 10^{-5}} $\\
$q$ & 0.86 & U$_{(m_1,m_2)}$[0.125, 1.0] & 0.86$\pm^{0.03}_{0.02}$ & 0.89$\pm^{0.10}_{0.13}$\\ 
$\tilde{\Lambda}$ & 293.5 & U[0, 5000] & 296$\pm^{62}_{59}$ & 290$\pm^{61}_{60}$\\
$\delta{\tilde{\Lambda}}$ & 34.7 & U[-5000, 5000] & 19$\pm^{108}_{112}$ & 10$\pm^{112}_{103}$\\ 
$\chi_1$ & NA/0.01 & NA/U[0, 0.05] & NA & 0.01$\pm^{0.02}_{0.01}$\\  
$\chi_2$ & NA/0.01 & NA/U[0, 0.05] & NA & 0.01$\pm^{0.02}_{0.01}$\\  
RA [rad] & 2.55 & U[0, 2$\pi$] & 2.55$\pm^{0.01}_{0.01}$ & 2.55$\pm^{0.01}_{0.01}$\\ 
DEC [rad] & -0.40 & Sine & -0.41$\pm^{0.01}_{0.01}$ & -0.41$\pm^{0.01}_{0.01}$\\ 
$D_L$ [Mpc] & 40 & U[0, 2$\pi$] & 35$\pm^{5}_{8}$ & 35$\pm^{5}_{8}$\\
$\theta_{jn}$ [rad] & 0.1 & Sine & 0.52$\pm^{0.35}_{0.37}$ & 0.52$\pm^{0.35}_{0.38}$\\ 
$\phi$ [rad] & 0 & U[0, 2$\pi$] & 3.16$\pm^{2.79}_{2.86}$ & 3.16$\pm^{2.79}_{2.86}$\\ 
$\psi$ [rad] & 0 & U[0, $\pi$] & 1.63$\pm^{1.37}_{1.45}$ & 1.66$\pm^{1.34}_{1.47}$\\  
$t_c$ [s] & 0 & U[-0.005, 0.005]  & $t_c\pm^{1.8\times10^{-4}}_{1.7\times10^{-4}}$ & $t_c\pm^{1.8\times10^{-4}}_{1.8\times10^{-4}}$ \\ 
\midrule[0.3pt] \bottomrule[1pt]
\end{tabularx}
\caption{Injected parameters for a nonspinning and aligned spin GW170817-like system with the corresponding prior ranges and recovered values. Here U$_{(m_1,m_2)}$ denotes that the prior is conditional on the detector-frame component masses and we set $\mathcal{M}_{\rm{min}} = \mathcal{M}_{c, \mathrm{inj}} - 10^{-3}$ and $\mathcal{M}_{\rm{max}} = \mathcal{M}_{c, \mathrm{inj}} + 10^{-3}$. The recovered values denote the median and 90\% credible interval. NA indicates that a parameter was not applicable to the nonspinning analysis.}
\label{tab:injections}
\end{table*}

\begin{figure*}
    \centering
    \includegraphics[width=0.48\textwidth]{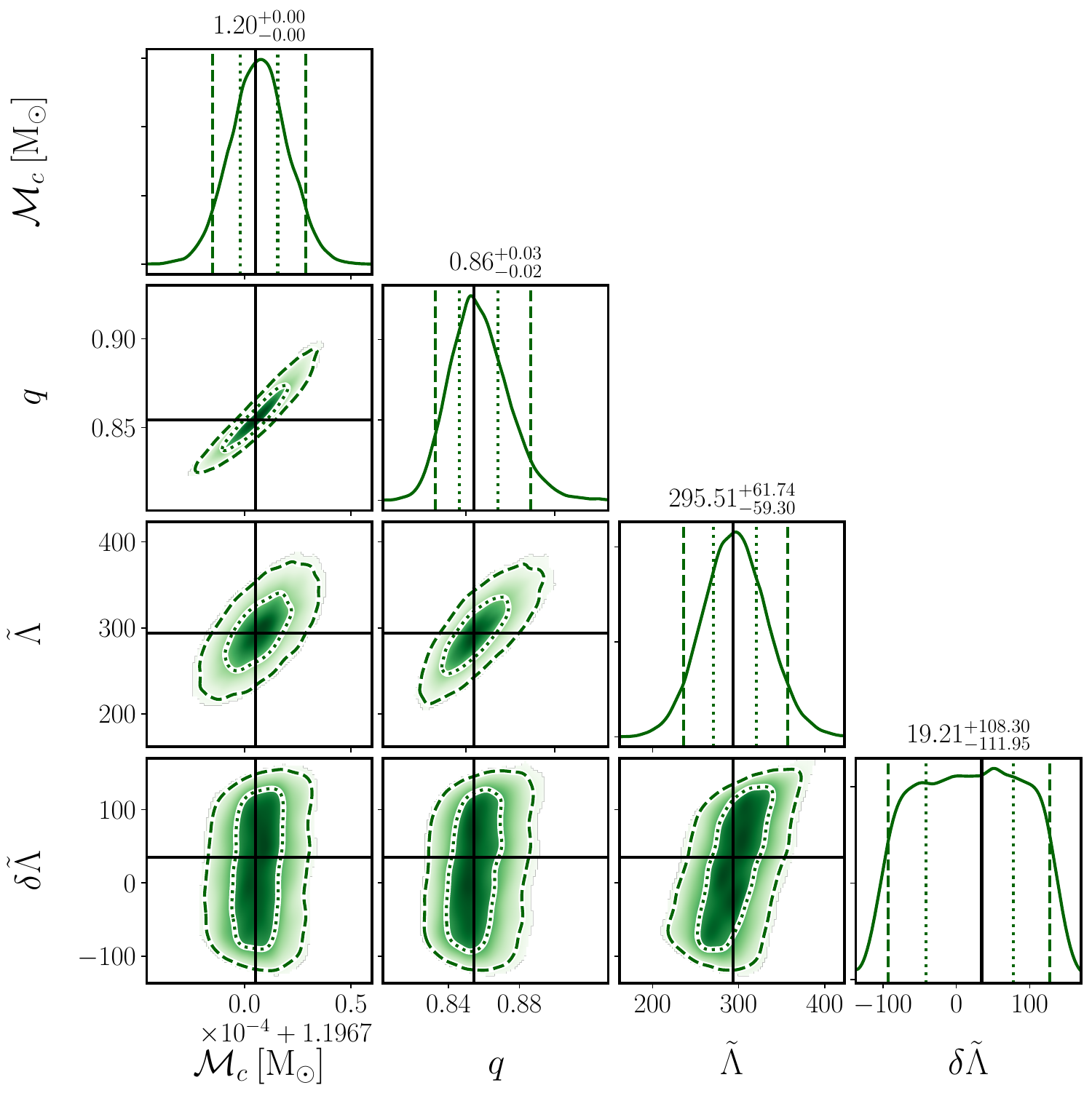} 
    \includegraphics[width=0.48\textwidth]{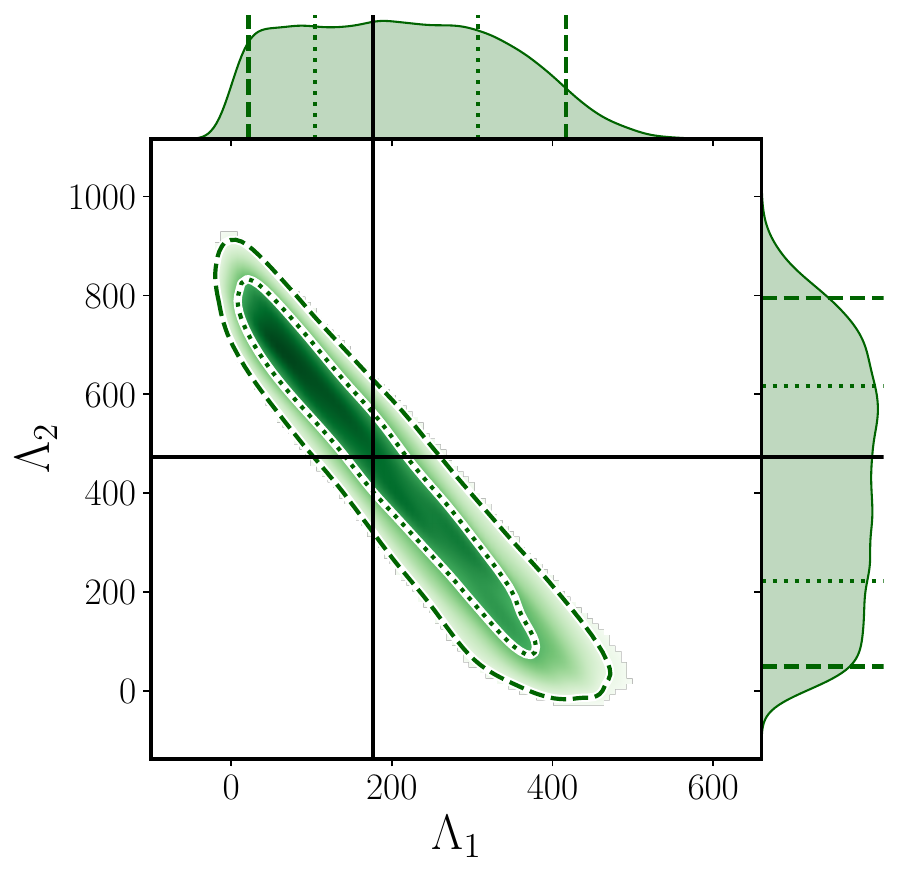} 
\caption{1D and 2D posteriors for a nonspinning GW170817-like \textsc{IMPPhenomXAS\_PhenomGSF} injection. \emph{Left panel}: Posteriors of the intrinsic parameters with their 50\% (dotted) and 90\% confidence intervals (dashed). The injected values are indicated by solid black lines. \emph{Right panel}: Posteriors for the individual tidal deformabilites with their 50\% (dotted) and 90\% confidence intervals (dashed).}  
\label{fig:injection_posteriors}
\end{figure*}

\begin{figure*}
    \centering
    \includegraphics[width=0.48\textwidth]{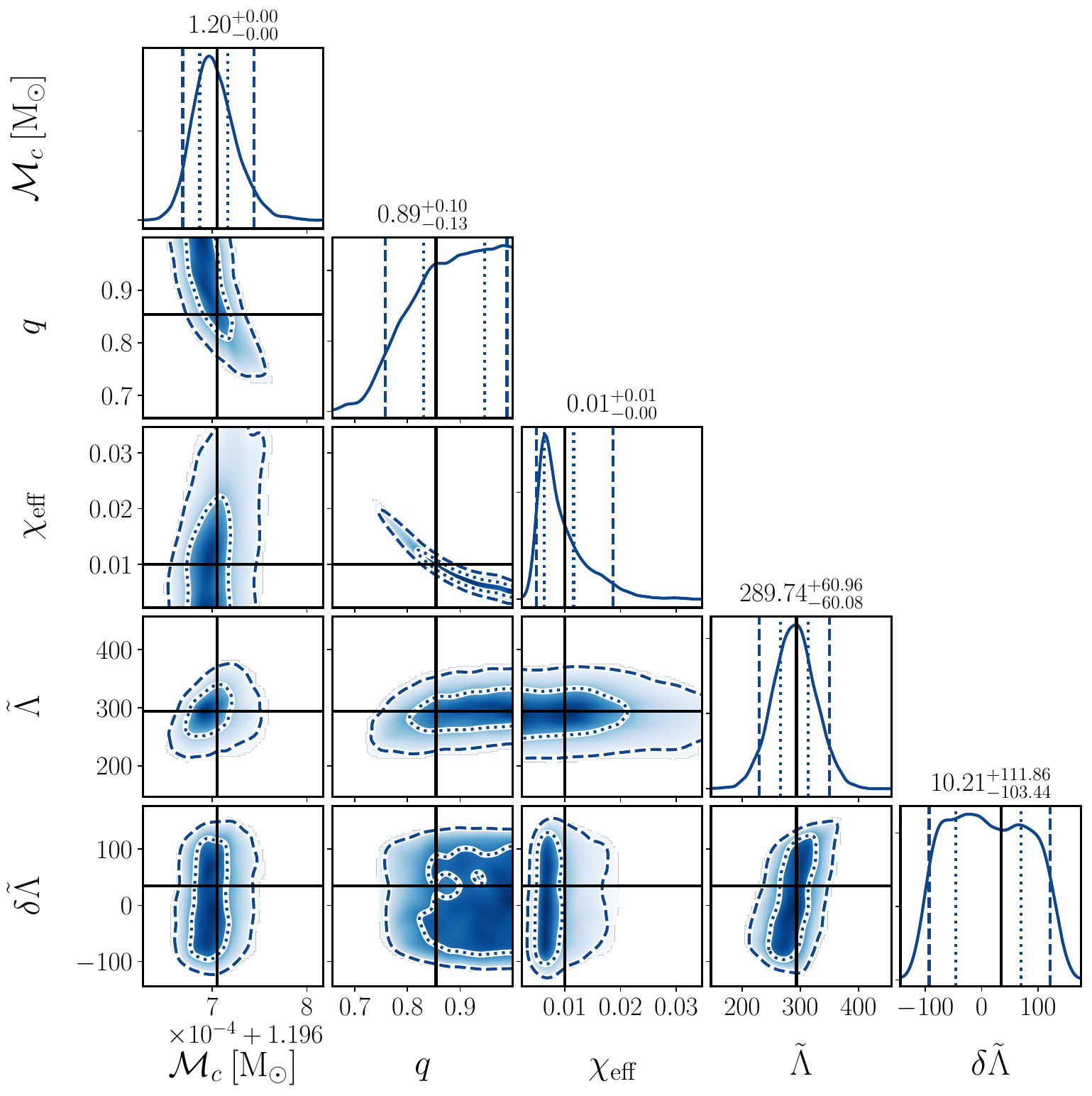} 
    \includegraphics[width=0.48\textwidth]{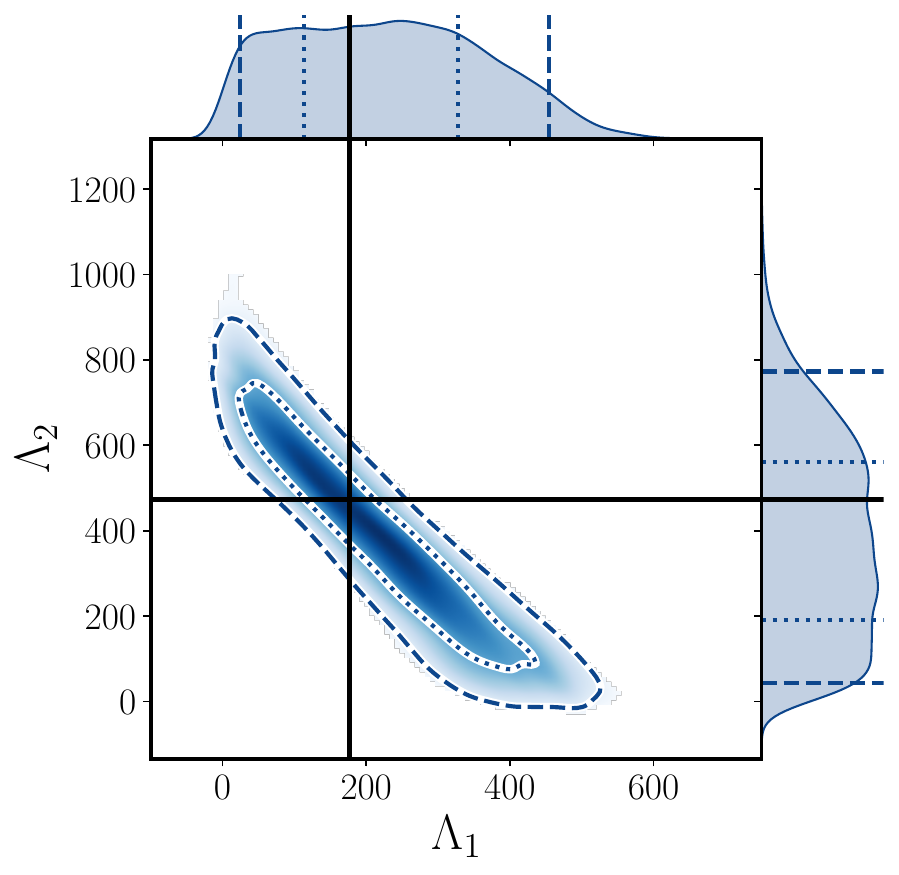}
\caption{1D and 2D posteriors for an aligned-spin GW170817-like \textsc{IMPPhenomXAS\_PhenomGSF} injection. \emph{Left panel}: Posteriors of the intrinsic parameters with their 50\% (dotted) and 90\% confidence intervals (dashed). The injected values are indicated by solid black lines. \emph{Right panel}: Posteriors for the individual tidal deformabilites with their 50\% (dotted) and 90\% confidence intervals (dashed).}   
\label{fig:injection_posteriors_spinning}%
\end{figure*}

%%%%%%%%%%%%%%%%%%%%%%%%%%%%%%%%%%%%%%%%%%%%%%%%%%%
\subsubsection{Zero-Noise Injections}
\label{sec:zeronoise}
%%%%%%%%%%%%%%%%%%%%%%%%%%%%%%%%%%%%%%%%%%%%%%%%%%%
We first perform two zero-noise injections with \textsc{IMRPhenomXAS\_PhenomGSF} into a three-detector network consisting of LIGO-Hanford, LIGO-Livingston~\cite{LIGOScientific:2014pky} and Virgo~\cite{VIRGO:2014yos} (HLV) with the sensitivity of the fourth observing run (O4)~\cite{O4PSDs} and recover with the same model. 
We note that zero-noise is representative of the results when averaging over many different noise realisations. The frequency-dependence of the sensitivity is taken into account through the PSD.
The injected parameters are similar to those of GW170817 (see Tab.~\ref{tab:injections} for details). 
For the EOS we choose the relatively soft APR4~\cite{Akmal:1998cf} one, which is compatible with the EOS-constraints obtained from GW170817~\cite{GW170817-EOS, LIGOScientific:2019eut}. 
We perform two analyses: one with zero NS spin and another one with aligned-spins of magnitudes $\chi_{1} = \chi_{2} = 0.01$. We integrate the likelihood from a  minimum frequency $f_{\rm min} = 23$ Hz to $f_{\rm max} = 2048$ Hz, which corresponds to the Nyquist frequency of our sampling rate. Although \phenom is only calibrated from a minimum frequency of $37$ Hz for such a GW170817-like system, we have confirmed that extrapolating to lower frequencies is robust due to the nature of our ansatz.  Both configurations have a network signal-to-noise ratio (SNR) of $\sim 97$. 

A selection of 1D and 2D posterior distributions for the nonspinning injection is shown in Fig.~\ref{fig:injection_posteriors}. We find that for all parameters the injected values lies confidently within the 90\% credible interval. 
As expected, $\tilde{\Lambda}$ is recovered with better accuracy than $\delta\tilde{\Lambda}$ as it is the leading-order tidal contribution to the GW phase, while $\delta\tilde{\Lambda}$ is subdominant. 
We also observe the expected linear correlation between the chirp mass and $\tilde{\Lambda}$. 
Mapping the binary tidal parameters to the individual tidal deformabilities, we observe the known correlation between the two and recover the injected values at the 50\% credible level as can be seen in the right panel of Fig.~\ref{fig:injection_posteriors}. The median values and 90\% credible intervals for all parameters are given in the fourth column of Tab.~\ref{tab:injections}. 

In Fig.~\ref{fig:injection_posteriors_spinning} we show a similar selection of 1D and 2D posteriors for the aligned-spin analysis, but now also including the effective spin $\chi_{\rm eff}$~\cite{Ajith:2009bn}. 
We find that the chirp mass and tidal parameters are recovered equally accurately as in the nonspinning case, but we now observe the known non-linear correlation between the mass ratio and the effective spin~\cite{Cutler:1994ys, Poisson:1995ef}, pushing the mass ratio posterior towards more equal masses for low spin magnitudes. The tidal parameters are, however, still recovered well. The median values and 90\% credible intervals for all parameters are given in the fifth column of Tab.~\ref{tab:injections}.

\setlength\doublerulesep{1.5pt}
\begin{table*}[]
\begin{tabularx}{\textwidth}{XXXXXX} \toprule[1pt]\midrule[0.3pt]
\multicolumn{1}{c}{} & \multicolumn{3}{c}{\bf \teob-NR hybrid} & \multicolumn{2}{c}{\bf GW170817}   \\ 
\cmidrule(r){2-4} \cmidrule(r){5-6}
\multicolumn{1}{l}{Parameter}  & \multicolumn{1}{l}{\hspace{0.2em}Injected Value} & \multicolumn{1}{l}{Prior} & \multicolumn{1}{l}{Recovered} & \multicolumn{1}{l}{\hspace{0.2em}Prior} & \multicolumn{1}{l}{\hspace{0.2em} Inferred} \\ \hline  
\multicolumn{1}{l}{$\mathcal{M}_c [ M_{\odot}]$}  & \multicolumn{1}{l}{\hspace{0.2em}1.18} & \multicolumn{1}{l}{U$_{(m_1,m_2)}$[1.1,1.3]} & \multicolumn{1}{l}{$1.18\pm ^{3.28\times10^{-5}}_{5.74\times10^{-5}}$}  & \hspace{0.2em}U$_{(m_1,m_2)}$[1.197,1.198] & \hspace{0.2em}$1.1976\pm ^{1.25\times10^{-4}}_{1.07\times10^{-4}}$ \\
\multicolumn{1}{l}{$q$} & \hspace{0.2em}1.0 & U$_{(m_1,m_2)}$[0.125, 1] & $0.94\pm^{0.05}_{0.08}$ & \hspace{0.2em}U$_{(m_1,m_2)}$[0.125, 1] & \hspace{0.2em}$0.86\pm^{0.13}_{0.16}$ \\ 
\multicolumn{1}{l}{$\tilde{\Lambda}$} & \hspace{0.2em}390 & U[0,5000] & $371\pm^{104}_{115}$ & \hspace{0.2em}U[0,5000] & \hspace{0.2em}$410\pm^{516}_{237}$ \\
\multicolumn{1}{l}{$\delta{\tilde{\Lambda}}$} & \hspace{0.2em}0 & U[-5000,5000] & $-1\pm^{152}_{133}$ & \hspace{0.2em}U[-5000,5000] & \hspace{0.2em}$18\pm^{207}_{205}$\\ 
\multicolumn{1}{l}{$\chi_1$} & \hspace{0.2em}0 & NA & NA & \hspace{0.2em}U[0,0.05] & \hspace{0.2em}$0.007\pm^{0.02}_{0.01}$ \\  
\multicolumn{1}{l}{$\chi_2$} & \hspace{0.2em}0 & NA & NA & \hspace{0.2em}U[0,0.05] & \hspace{0.2em}$0.007\pm^{0.02}_{0.01}$ \\  
\multicolumn{1}{l}{RA [rad]} & \hspace{0.2em}2.55 & U[0, $2\pi$] & $2.55 \pm^{0.01}_{0.01}$ & \hspace{0.2em}U[0, $2\pi$] & \hspace{0.2em}$3.42 \pm^{0.03}_{0.03}$ \\ 
\multicolumn{1}{l}{DEC [rad]} & \hspace{0.2em}-0.41 & Cosine & $-0.41\pm^{0.02}_{0.02}$  & \hspace{0.2em}Cosine & \hspace{0.2em}$-0.40\pm^{0.06}_{0.06}$ \\ 
\multicolumn{1}{l}{$D_L$ [Mpc]} & \hspace{0.2em}40 & U[10,100] & $36 \pm^{5}_{8}$ & \hspace{0.2em}U[10,100] & \hspace{0.2em}$37 \pm^{8}_{16}$\\
\multicolumn{1}{l}{$\theta_{jn}$ [rad]} & \hspace{0.2em}0.1 & Sine & $2.64\pm^{0.36}_{0.35}$ & \hspace{0.2em}Sine & \hspace{0.2em}$2.53\pm^{0.44}_{0.51}$\\ 
\multicolumn{1}{l}{$\phi$ [rad]}& \hspace{0.2em}0 & U[0, $2\pi$] & $2.96 \pm^{2.96}_{2.67}$ & \hspace{0.2em}U[0, $2\pi$] & \hspace{0.2em}$3.17 \pm^{2.81}_{2.93}$   \\ 
\multicolumn{1}{l}{$\psi$ [rad]}  & \hspace{0.2em}0 & U[0, $\pi$] & $1.63\pm^{1.37}_{1.48}$  & \hspace{0.2em}U[0, $\pi$] & \hspace{0.2em}$1.53\pm^{1.49}_{1.44}$  \\  
\multicolumn{1}{l}{$t_c$ [s]}   & \hspace{0.2em}$1187008882.4$ & \hspace{0.2em}U[$t_c$ - 0.1, $t_c$ + 0.1] & $t_c\pm^{3.9\times10^{-4}}_{4.3\times10^{-4}}$ & \hspace{0.2em}U[$t_c$ - 0.1, $t_c$ + 0.1] & \hspace{0.2em}$t_c\pm^{1.1\times10^{-3}}_{0.9\times10^{-3}}$\\ 
\midrule[0.3pt]\bottomrule[1pt]
\end{tabularx}
\caption{Injected parameters for the \teob-NR hybrid with the corresponding priors and 1D posterior medians with their 90\% credible intervals, and priors as well as the median inferred parameter values with their 90\% credible intervals for the GW event GW170817. Here, U$_{(m_1,m_2)}$ denotes uniform priors in component masses, and $t_c=1187008882.4$ s is the GPS time of GW170817. NA indicates that a parameter was not applicable to the nonspinning analysis.}
\label{tab:PE}
\end{table*}

\begin{figure*}[]
    \centering\includegraphics[width=0.48\textwidth]{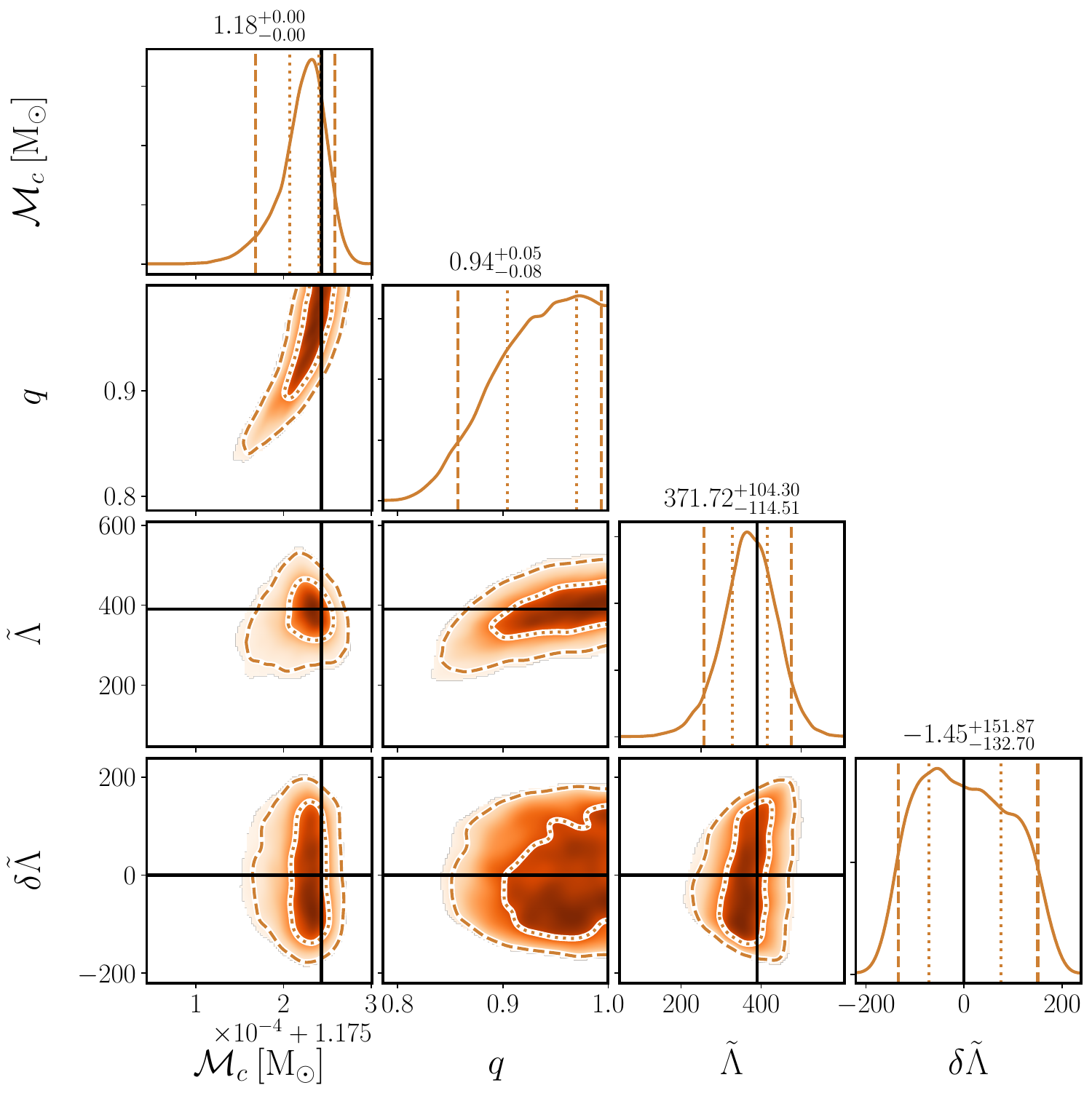} \includegraphics[width=0.48\textwidth]{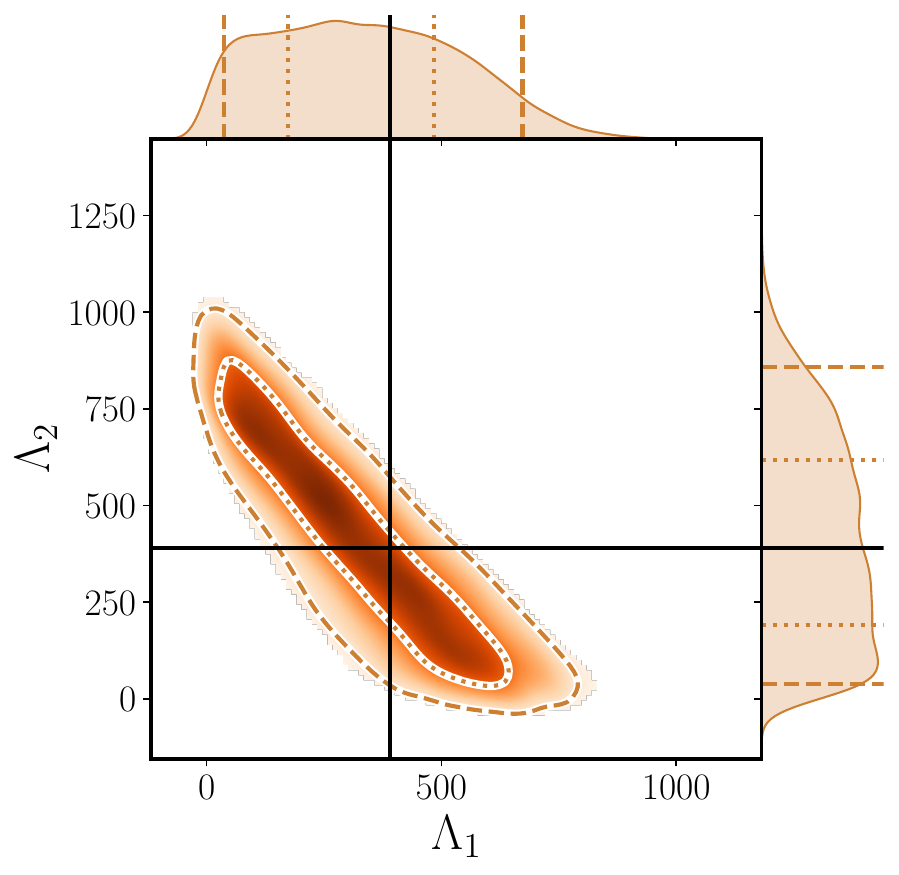}
\caption{1D and 2D posteriors for the \teob-NR hybrid injection with \textsc{BAM 0095} analysed with \textsc{IMPPhenomXAS\_PhenomGSF}. \emph{Left panel}: Posteriors of the intrinsic parameters with their 50\% (dotted) and 90\% confidence intervals (dashed). The injected values are indicated by solid black lines. We note that the injected mass ratio is $q=1$ and hence overlaps with the plot axis. \emph{Right panel}: Posteriors for the individual tidal deformabilites with their 50\% (dotted) and 90\% confidence intervals (dashed).} 
\label{fig:NRHybrid-posteriors}
\end{figure*}

%%%%%%%%%%%%%%%%%%%%%%%%%%%%%%%%%%%%%%%%%
\subsubsection{\teob-NR Hybrid Injection}
\label{sec:hybridinj}
%%%%%%%%%%%%%%%%%%%%%%%%%%%%%%%%%%%%%%%%%
Next, we perform a \teob-NR hybrid zero-noise injection into the same HLV network and recover again with \textsc{IMPPhenomXAS\_PhenomGSF}.
For this analysis we use the \textsc{BAM 0095} simulations. Details on the hybrid generation are given in Sec.~\ref{sec:hybrids}. The hybrid starting frequency is $f_{\mathrm{min}} = 30$ Hz, while the NR starting frequency is $f_{\mathrm{NR}} = 433$ Hz. To perform the analysis, we convert the hybrid into the LVK-NR format~\cite{Schmidt:2017btt} and use the tools provided by the LIGO Algorithms Library~\cite{lalsuite}. 
Starting at $f_{\mathrm{min}} = 33$ Hz, we obtain a network SNR of $\sim 51$ with the O4 design sensitivity, where the extrinsic parameters are the same as for the previous injections. 
We note that as this is a nonspinning injection, we set the spins to zero in the recovery. 
The injected values, prior choices and recovered values are given in Tab.~\ref{tab:PE}. 

In Fig.~\ref{fig:NRHybrid-posteriors} we show the obtained 1D and 2D posterior distributions for the chirp mass, mass ratio and tidal deformability parameters. We find that despite the differences in the BBH baseline and the lack of NR calibration of \phenom, all injected values are recovered accurately and contained within either the 90\% or even the 50\% credible interval, indicating that modelling systematics due to the BBH baseline and the lack of NR calibration are subdominant in this region of the parameter space. 

\begin{figure*}[t]
    \centering
    \includegraphics[width=0.4\textwidth]{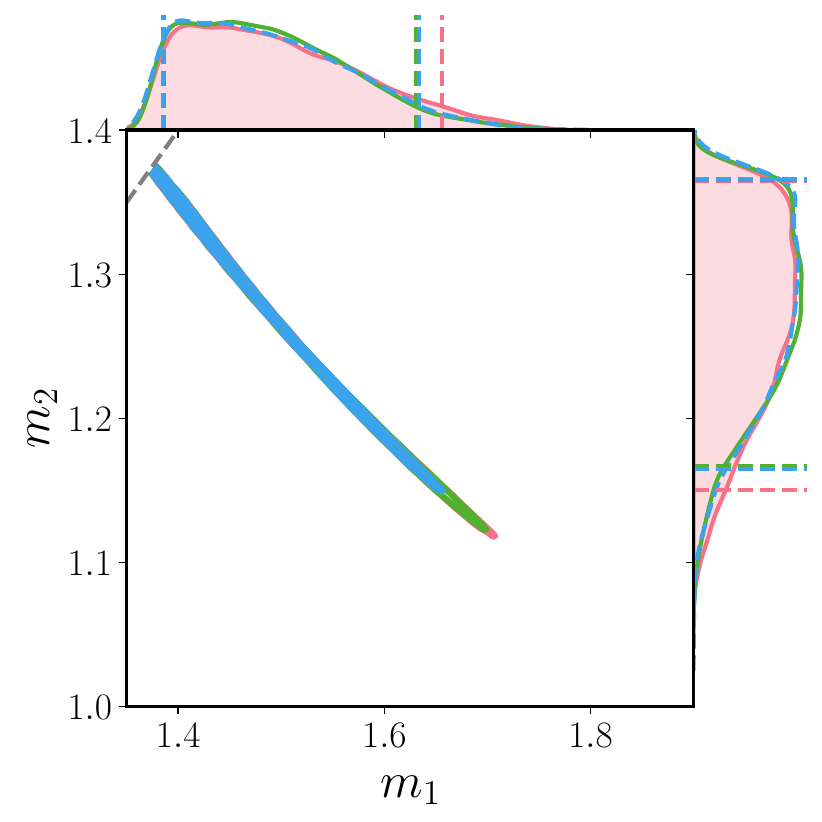} 
    \includegraphics[width=0.58\textwidth]{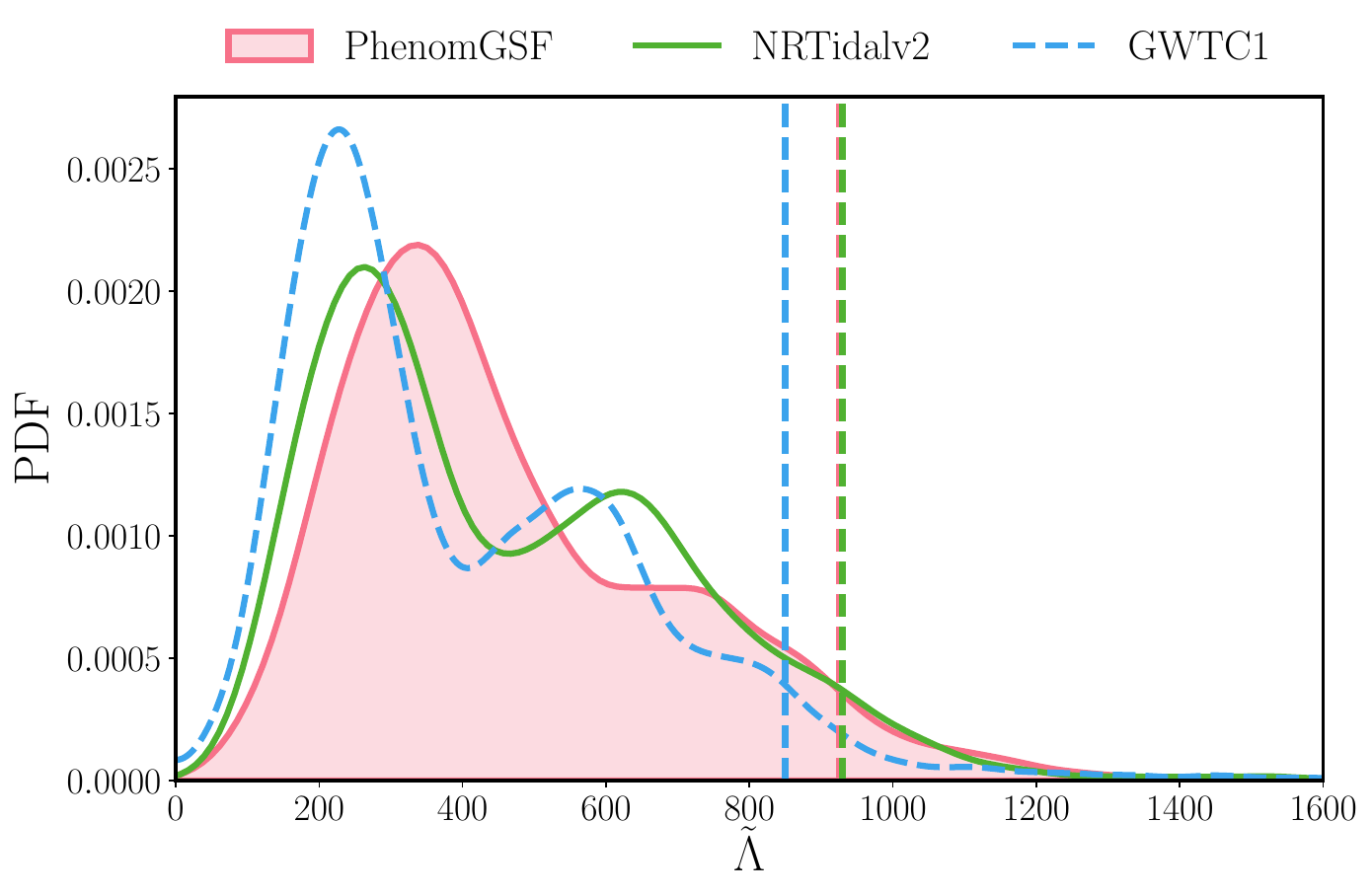} 
\caption{Selection of 1D and 2D posterior distributions for GW170817 for \textsc{IMRPhenomXAS\_PhenomGSF} (pink, solid), \textsc{IMRPhenomXAS\_NRTidalv2} (green, solid) and \textsc{IMRPhenomPv2\_NRTidalv2} (blue, dashed) from GWTC-1. 
\emph{Left panel}: Component mass posteriors with their 1D 90\% credible intervals (vertical dashed lines) and 90\% credible region, respectively. Joint 2D posteriors are. The grey dashed line indicates the equal-mass limit. 
\emph{Right panel}: 1D posteriors for the binary tidal deformability $\tilde{\Lambda}$ with their respective $95^{\rm th}$ percentiles (vertical dashed lines).}
\label{fig:GWTC1posteriors}
\end{figure*}

%%%%%%%%%%%%%%%%%%%%%%%%
\subsubsection{GW170817}
\label{sec:gw170817}
%%%%%%%%%%%%%%%%%%%%%%%%
%\todo{Add sampling times}
Finally, we analyse the data of GW170817 with \phenom again using \textsc{IMRPhenomXAS} as the BBH baseline model. We follow previous analyses~\cite{GW170817-PE, GWTC1} and employ a low-spin prior. Details for all priors are given in the fifth column of Tab.~\ref{tab:PE}. We analyse $128$ s of data around the trigger time with a sampling rate of $4096$ Hz. For the likelihood integration we choose a minimum frequency $f_{\mathrm{min}} = 23$ Hz consistent with the analyses presented in GWTC-1. 

The inferred posterior distributions for the masses and tidal deformability are shown in Fig.~\ref{fig:GWTC1posteriors}; the results for all other parameters are shown in Fig.~\ref{fig:GW170817full} in the Appendix. For comparison, we also show the posteriors obtained with the waveform model \textsc{IMRPhenomPv2\_NRTidal} from the GWTC-1 analysis~\cite{GWTC1, GWOSC} as well as posteriors from an analysis identical to ours but with the model \textsc{IMRPhenomXAS\_NRTidalv2}. 
The GWTC-1 analysis used flat priors in $\Lambda_1$ and $\Lambda_2$, hence our posteriors for $\tilde{\Lambda}$ has been reweighted to a flat $\tilde{\Lambda}$-prior as used in our setup. 

Overall, we observe a very high degree of consistency between the three results.
The component mass posteriors show excellent agreement between all three models, with a slightly higher 90\% upper bound for \phenom. Similarly for the tides, where we also find a slightly higher value of $\tilde{\Lambda}$ with \phenom than what was found in GWTC-1. This is also consistent with the analysis using \teob presented in Ref.~\cite{Gamba:2020wgg}.
Notably, however, it also appears that the secondary peak in the $\tilde{\Lambda}$ posterior is reduced in comparison to the other models, suggesting that this feature is likely inherent to this generation of \textsc{NRTidal} approximants or the previously used BBH baselines. 
We note that both of these observations are also in agreement with the recent analysis of GW170817 with \textsc{NRTidalv3} presented in Fig. 16 of Ref.~\cite{Abac:2023ujg}.  

%%%%%%%%%%%%%%%%%%%%%%%%%%%%%%%
\section{Discussion}
\label{sec:discussion}
%%%%%%%%%%%%%%%%%%%%%%%%%%%%%%%
We presented \phenom, a new phenomenological tidal phase model for inspiralling unequal mass BNS systems valid for mass ratios between $1$ and $3$ and dimensionless neutron star tidal deformabilities up to $5000$.
\phenom is a closed-form phenomenological fit to the tidal phase of the effective-one-body model \teob, which includes GSF-informed tides. The fit is constructed by employing a hierarchical procedure using the collocation point method to a large suite of \teob waveforms.

\phenom is fully modular and can readily be added to any frequency-domain BBH baseline model, such as the state-of-the-art \textsc{IMRPhenomX} waveform family. 

We note that we did not include EOB-calibrated spin effects in PhenomGSF, instead relying on standard PN spin-tidal expressions. Whilst we could have assumed an EOS or imposed URs, this is not explicitly required.

We assessed the accuracy and performance of \phenom extensively by comparing it against independent \teob waveforms, obtaining mismatches $\mathcal{MM}\leq 10^{-3}$ across the parameter space the model was calibrated against. For $q\sim 1$ we find that the model extrapolates well and retains its accuracy for tidal deformabilities up to $10000$. We also computed frequency-dependent mismatches against a suite of \teob-NR hybrid waveforms, finding that \phenom tracks the accuracy of \teob as expected. 

Finally, we performed full Bayesian inference on a suite of mock signals and the GW event GW170817. Parameter estimation on \textsc{IMRPhenomXAS\_PhenomGSF} injections into zero-noise show consistent recovery of the injected parameters. We then analyse a \teob-NR hybrid injection and find that neither the lack NR calibration of \phenom nor the difference in the underlying BBH baseline lead to any significant systematic differences. A more systematic Bayesian study across a wider parameter space is left for future work.
Lastly, we independently analyses the data of GW170817 with \textsc{IMRPhenomXAS\_PhenomGSF}. Our results are consistent with previous analysis, but we obtain a slightly higher upper limit for $\tilde{\Lambda}$ than for example with \textsc{NRTidal} and \textsc{NRTidalv2}. We also found that the secondary peak in $\tilde{\Lambda}$ is less pronounced for \phenom consistent with the \teob results of GWTC-1. The computational efficiency is comparable to that of the \textsc{NRTidal} approximants.

We have demonstrated that \phenom is a robust, highly accurate, modular and computationally efficient model of quadrupolar gravitoelectric tides. Moreover, it is flexible in that it allows for the addition of PN spin terms and the use of non-hadronic EOS, making it an ideal candidate for exploring exotic matter. Nevertheless, there are several avenues for further improving the model which we leave for future work, including:
\begin{enumerate}
    \item A full treatment of neutron star spin effects, extending the 3-dimensional phenomenological fit presented here to higher dimensions to incorporate aligned-spin and precessing binaries.
    \item The incorporation of dynamical tidal effects, such as adding the PN model of $f$-mode dynamical tides \textsc{fmtidal}~\cite{Schmidt:2019wrl}, as their neglect is known to introduce biases in the inferred EOS~\cite{Pratten:2019sed,Pratten:2021pro}.
    \item Extension to higher-order multipoles beyond the $(2,2)$-mode. This is of particular importance for unequal-mass and precessing systems, where more power is radiated in subdominant harmonics, and could play an important role in constraining the geometry of the merger and breaking the distance-inclination degeneracy.
    \item Modelling tidal corrections to the amplitude, e.g.~\cite{Dietrich:2019kaq}.
\end{enumerate}
There are also numerous other avenues for extending the model that are being pursued. For example, there are higher-order tidal moments such at the octopolar and hexadecapolar terms~\cite{Akcay:2018yyh}. The neutron star excites a spectrum of fluid modes beyond just the $f$-modes that need to be modelled, e.g. see~\cite{Flanagan:2006sb,Ma:2020oni,Poisson:2020eki,Ho:2023shr}. Including the effects of orbital eccentricity could also be important~\cite{Chirenti:2016xys,DuttaRoy:2024aew}, though this may be of less astrophysical significance~\cite{Lorimer:2008se}.

Finally, we note that although we did not include precession effects in the analysis presented here, it is straightforward to follow the framework outlined in~\cite{Schmidt:2014iyl} to construct an approximate precessing BNS model by augmenting the BBH baseline~\cite{Dietrich:2017aum,Dietrich:2019kaq,Abac:2023ujg,Colleoni:2023czp}. A notable benefit is that precession can help break the mass-spin degeneracy, leading to tighter constraints on the component masses~\cite{Chatziioannou:2014coa,Pratten:2020igi}.

%%%%%%%%%%%%%%%%%%%%%%%%%%%%
\section*{Acknowledgements}
%%%%%%%%%%%%%%%%%%%%%%%%%%%%
We thank Rossella Gamba for help with the C implementation of the \teob waveform model and useful comments on the manuscript. 
N.W. is supported by STFC, the School of Physics and Astronomy at the University of Birmingham and the Birmingham Institute for Gravitational Wave Astronomy. 
G.P. is very grateful for support from a Royal Society University Research Fellowship URF{\textbackslash}R1{\textbackslash}221500 and RF{\textbackslash}ERE{\textbackslash}221015, and gratefully acknowledges support from an NVIDIA Academic Hardware Grant.
G.P. and P.S. acknowledge support from STFC grant ST/V005677/1. 
P.S. also acknowledges support from a Royal Society Research Grant RG{\textbackslash}R1{\textbackslash}241327.
Computations were performed using the University of Birmingham's BlueBEAR HPC service, which provides a High Performance Computing service to the University's research community, the Bondi HPC cluster at the Birmingham Institute for Gravitational Wave Astronomy as well as on resources provided by Supercomputing Wales, funded by STFC grants ST/I006285/1 and ST/V001167/1 supporting the UK Involvement in the Operation of Advanced LIGO.
The fits were performed using Wolfram's \texttt{Mathematica}~\cite{Mathematica}. Figures were generated using \texttt{Mathematica} and \texttt{Matplotlib}~\cite{Hunter:2007}. The version of \teob used in this work corresponds to git commit 128a406~\cite{TEOBgit}.
This research has made use of data or software obtained from the Gravitational Wave Open Science Center (gwosc.org), a service of the LIGO Scientific Collaboration, the Virgo Collaboration, and KAGRA. This material is based upon work supported by NSF's LIGO Laboratory which is a major facility fully funded by the National Science Foundation, as well as the Science and Technology Facilities Council (STFC) of the United Kingdom, the Max-Planck-Society (MPS), and the State of Niedersachsen/Germany for support of the construction of Advanced LIGO and construction and operation of the GEO600 detector. Additional support for Advanced LIGO was provided by the Australian Research Council. Virgo is funded, through the European Gravitational Observatory (EGO), by the French Centre National de Recherche Scientifique (CNRS), the Italian Istituto Nazionale di Fisica Nucleare (INFN) and the Dutch Nikhef, with contributions by institutions from Belgium, Germany, Greece, Hungary, Ireland, Japan, Monaco, Poland, Portugal, Spain. KAGRA is supported by Ministry of Education, Culture, Sports, Science and Technology (MEXT), Japan Society for the Promotion of Science (JSPS) in Japan; National Research Foundation (NRF) and Ministry of Science and ICT (MSIT) in Korea; Academia Sinica (AS) and National Science and Technology Council (NSTC) in Taiwan.
% LIGO DCC
This manuscript has the LIGO document number P2400286.

%%%%%%%%%% APPENDICES %%%%%%%%%%
\appendix
%%%%%%%%%%%%%%%%%%%%%%%%%%%%%%%
\section{Parameter Space Fits}
\label{sec:FitsAppendix}
%%%%%%%%%%%%%%%%%%%%%%%%%%%%%%%
The full phenomenological fits of the five collocation points $\lambda^{(i)}$ needed to evaluate \phenom are given below. The phenomenological coefficients are listed in Tab.~\ref{tab:coeff}.

\begin{widetext}
\begin{align}
\begin{split}
\lambda^{(1)}(\eta,\tilde{\Lambda},\delta{\tilde{\Lambda}}) = \frac{ 1}{1 + a_{1}^{(1)} \tilde{\Lambda} + a_{2}^{(1)} \tilde{\Lambda} ^2+a_{3}^{(1)} \tilde{\Lambda}^3+a_{4}^{(1)} \tilde{\Lambda}^4}  \bigg[ b_{1}^{(1)}  (1 + d_{11}^{(1)} \delta\tilde{\Lambda}
   +d_{12}^{(1)} \delta\Lambda^2)\eta \tilde{\Lambda} - b_{2}^{(1)} (c_{11}^{(1)}  (1 + d_{21}^{(1)} \delta\tilde{\Lambda} - d_{22}^{(1)} \delta\tilde{\Lambda}^2)  - \\ 
   c_{12}^{(1)} (1 - d_{31}^{(1)}\delta\tilde{\Lambda} + d_{32}^{(1)} \delta\tilde{\Lambda}^2) \eta)\tilde{\Lambda}^4 
   + b_{3}^{(1)} (c_{21}^{(1)}
   (1 + d_{41}^{(1)} \delta\tilde{\Lambda} +  d_{42}^{(1)}  \delta\tilde{\Lambda} ^2) + \\ c_{22}^{(1)} (1 +  d_{51}^{(1)}  \delta\tilde{\Lambda} + d_{52}^{(1)}  \delta\tilde{\Lambda}^2)\eta)\tilde{\Lambda}^5\bigg]
\end{split}
\\
\begin{split}
\lambda^{(2)}(\eta,\tilde{\Lambda},\delta{\tilde{\Lambda}}) = \frac{ 1}{1 + a_{1}^{(2)} \tilde{\Lambda} + a_{2}^{(2)} \tilde{\Lambda} ^2+a_{3}^{(2)} \tilde{\Lambda}^3} \bigg[ b_{1}^{(2)}  (c_{11}^{(2)}(1 + d_{11}^{(2)}\delta\tilde{\Lambda}) + c_{12}^{(2)}(1 + d_{21}^{(2)}\delta\tilde{\Lambda})\eta) \tilde{\Lambda} + b_{2}^{(2)} (c_{21}^{(2)}(1+d_{31}^{(2)}\delta\tilde{\Lambda}) + \\
c_{22}^{(2)}(1+d_{41}^{(2)}\delta\tilde{\Lambda})\eta)\tilde{\Lambda}^2 +  b_{3}^{(2)} (c_{31}^{(2)}(1+d_{51}^{(2)}\delta\tilde{\Lambda}) + c_{32}^{(2)}(1+d_{61}^{(2)}\delta\tilde{\Lambda})\eta)\tilde{\Lambda}^3 + \\
b_{4}^{(2)} (1+d_{71}^{(2)}\delta\tilde{\Lambda})\tilde{\Lambda}^4
\bigg]
\end{split}
\\
\begin{split}
\lambda^{(3)}(\eta,\tilde{\Lambda},\delta{\tilde{\Lambda}}) = \frac{ 1}{1 + a_{1}^{(3)} \tilde{\Lambda} + a_{2}^{(3)} \tilde{\Lambda} ^2+a_{3}^{(3)} \tilde{\Lambda}^3} \bigg[ b_{1}^{(3)}  (c_{11}^{(3)}(1 + d_{11}^{(3)}\delta\tilde{\Lambda})\eta + c_{12}^{(3)}(1 + d_{21}^{(3)}\delta\tilde{\Lambda})\eta^2) \tilde{\Lambda} + b_{2}^{(3)} (c_{21}^{(3)}(1+d_{31}^{(3)}\delta\tilde{\Lambda}) + \\
c_{22}^{(3)}(1+d_{41}^{(3)}\delta\tilde{\Lambda})\eta + c_{23}^{(3)}\eta^3)\tilde{\Lambda}^2 +  b_{3}^{(3)} (c_{31}^{(3)} - c_{32}^{(3)}\eta)\tilde{\Lambda}^3
\bigg]
\end{split}
\\
\begin{split}
\lambda^{(4)}(\eta,\tilde{\Lambda},\delta{\tilde{\Lambda}}) = \frac{ 1}{1 + a_{1}^{(4)} \tilde{\Lambda} + a_{2}^{(4)} \tilde{\Lambda} ^2+a_{3}^{(4)} \tilde{\Lambda}^3+a_{4}^{(4)} \tilde{\Lambda}^4} \bigg[
 b_{1}^{(4)}  (1 + d_{11}^{(4)}\delta\tilde{\Lambda})\eta^2\tilde{\Lambda} + b_{2}^{(4)} (c_{21}^{(4)}(1+d_{21}^{(4)}\delta\tilde{\Lambda}) + c_{22}^{(4)}(1+d_{31}^{(4)}\delta\tilde{\Lambda})\eta)\tilde{\Lambda}^2 
 + \\
 b_{3}^{(4)} (c_{31}^{(4)}(1+d_{41}^{(4)}\delta\tilde{\Lambda}) + c_{32}^{(4)}(1+d_{51}^{(4)}\delta\tilde{\Lambda})\eta + c_{33}^{(4)}\eta^2 )\tilde{\Lambda}^3 + b_{4}^{(4)}\eta\tilde{\Lambda}^4 
\bigg]
\end{split}
\\
\begin{split}
\lambda^{(5)}(\eta,\tilde{\Lambda},\delta{\tilde{\Lambda}}) = \frac{ 1}{1 + a_{1}^{(5)} \tilde{\Lambda} + a_{2}^{(5)} \tilde{\Lambda} ^2 +a_{3}^{(5)} \tilde{\Lambda}^3} \bigg[
 b_{1}^{(5)}\eta^2\tilde{\Lambda} + b_{2}^{(5)} (c_{11}^{(5)}(1+d_{11}^{(5)}\delta\tilde{\Lambda}) + c_{12}^{(5)}(1+d_{12}^{(5)}\delta\tilde{\Lambda})\eta)\tilde{\Lambda}^2 
 + b_{3}^{(5)} (c_{21}^{(5)} + \\ c_{22}^{(5)}(1+d_{21}^{(5)}\delta\tilde{\Lambda})\eta + c_{23}^{(5)}\eta^2 )\tilde{\Lambda}^3 + b_{4}^{(5)}\eta^2\tilde{\Lambda}^4 + b_{5}^{(5)}\tilde{\Lambda}^5 
\bigg]
\end{split}
\end{align}
\label{eq:finalfits}
\end{widetext}

\begin{table*}[!ht]
\begin{minipage}[t]{.2\linewidth}
\begin{tabular}[t]{ll}
\toprule[1pt]\midrule[0.3pt]
$x^{(1)}$ & Value  \\ \hline
$a_{1}^{(1)}$ & $704.35868$   \\
$a_{2}^{(1)}$ & $0.61592$\\
$a_{3}^{(1)}$ & $0.00037$\\
$a_{4}^{(1)}$ & $9.36183\mathrm{e}{-8}$\\
$b_{1}^{(1)}$ & $-0.06132$\\
$b_{2}^{(1)}$ & $-3.15315\mathrm{e}{-10}$
\\
$b_{3}^{(1)}$ & $5.46894\mathrm{e}{-14}$ \\
$c_{11}^{(1)}$ & $4.10522$ \\
$c_{12}^{(1)}$ & $-12.80678$ \\
$c_{21}^{(1)}$ & $4.38039$ \\
$c_{22}^{(1)}$ & $-13.84189$ \\
$d_{11}^{(1)}$ & $-0.01033$ \\
$d_{12}^{(1)}$ & $6.81023\mathrm{e}{-6}$ \\
$d_{21}^{(1)}$ & $-0.00122$ \\
$d_{22}^{(1)}$ & $-8.60463\mathrm{e}{-7}$ \\
$d_{31}^{(1)}$ & $-0.00154$ \\
$d_{32}^{(1)}$ & $-1.09916\mathrm{e}{-6}$ \\
$d_{41}^{(1)}$ & $-0.00072$ \\
$d_{42}^{(1)}$ & $-5.84154\mathrm{e}{-7}$ \\
$d_{51}^{(1)}$ & $-0.00086$ \\
$d_{52}^{(1)}$ & $-7.22994\mathrm{e}{-7}$ \\
\midrule[0.3pt]\bottomrule[1pt]
\end{tabular}
\end{minipage}%
\begin{minipage}[t]{.2\linewidth}
\begin{tabular}[t]{ll}
\toprule[1pt]\midrule[0.3pt]
$x^{(2)}$ & Value  \\ \hline
$a_{1}^{(2)}$ & $29.78722$   \\
$a_{2}^{(2)}$ & $0.00029$\\
$a_{3}^{(2)}$ & $4.42273\mathrm{e}{-6}$\\
$b_{1}^{(2)}$ & $-0.20692$\\
$b_{2}^{(2)}$ & $-0.00013$ \\
$b_{3}^{(2)}$ & $-6.05658\mathrm{e}{-7}$ \\
$b_{4}^{(2)}$ & $7.15217\mathrm{e}{-11}$ \\
$c_{11}^{(2)}$ & $-1.78228$ \\
$c_{12}^{(2)}$ & $11.01742$ \\
$c_{21}^{(2)}$ & $16.30702$ \\
$c_{22}^{(2)}$ & $-61.74570$ \\
$c_{31}^{(2)}$ & $1.92755$ \\
$c_{32}^{(2)}$ & $-3.65176$ \\
$d_{11}^{(2)}$ & $0.02351$ \\
$d_{21}^{(2)}$ & $0.01609$ \\
$d_{31}^{(2)}$ & $-0.00046$ \\
$d_{41}^{(2)}$ & $-0.00008$ \\
$d_{51}^{(2)}$ & $-0.00024$ \\
$d_{61}^{(2)}$ & $-0.00095$ \\
$d_{71}^{(2)}$ & $0.00012$ \\
\midrule[0.3pt]\bottomrule[1pt]
\end{tabular}
\end{minipage}%
\begin{minipage}[t]{.2\linewidth}
\begin{tabular}[t]{ll}
\toprule[1pt]\midrule[0.3pt]
$x^{(3)}$ & Value  \\ \hline
$a_{1}^{(3)}$ & $3.93783\mathrm{e}{13}$   \\
$a_{2}^{(3)}$ & $1.57134\mathrm{e}{10}$\\
$a_{3}^{(3)}$ & $4.06194\mathrm{e}{5}$\\
$b_{1}^{(3)}$ & $-3.07059\mathrm{e}{12}$\\
$b_{2}^{(3)}$ & $-1.37322$ \\
$b_{3}^{(3)}$ & $-1.68496\mathrm{e}{7}$ \\
$c_{11}^{(3)}$ & $-3.94830$ \\
$c_{12}^{(3)}$ & $31.60021$ \\
$c_{21}^{(3)}$ & $10.06460$ \\
$c_{22}^{(3)}$ & $-63.87040$ \\
$c_{23}^{(3)}$ & $110.41652$ \\
$c_{31}^{(3)}$ & $3.08006$ \\
$c_{32}^{(3)}$ & $-8.31909$ \\
$d_{11}^{(3)}$ & $0.04560$ \\
$d_{21}^{(3)}$ & $0.02267$ \\
$d_{31}^{(3)}$ & $-0.00144$ \\
$d_{41}^{(3)}$ & $-0.00088$ \\
\midrule[0.3pt]\bottomrule[1pt]
\end{tabular}
\end{minipage}%
\begin{minipage}[t]{.2\linewidth}
\begin{tabular}[t]{ll}
\toprule[1pt]\midrule[0.3pt]
$x^{(4)}$ & Value  \\ \hline
$a_{1}^{(4)}$ & $6.11502\mathrm{e}{13}$   \\
$a_{2}^{(4)}$ & $1.59883\mathrm{e}{11}$\\
$a_{3}^{(4)}$ & $1.50039\mathrm{e}{7}$\\
$a_{4}^{(4)}$ & $-407.21516$\\
$b_{1}^{(4)}$ & $-2.01764\mathrm{e}{14}$\\
$b_{2}^{(4)}$ & $-1.09596\mathrm{e}{11}$ \\
$b_{3}^{(4)}$ & $-6.39775\mathrm{e}{8}$ \\
$b_{4}^{(4)}$ & $2.10833\mathrm{e}{5}$ \\
$c_{21}^{(4)}$ & $3.93019$ \\
$c_{22}^{(4)}$ & $-11.63559$ \\
$c_{31}^{(4)}$ & $7.32205$ \\
$c_{32}^{(4)}$ & $-42.95971$ \\
$c_{33}^{(4)}$ & $70.66872$ \\
$d_{11}^{(4)}$ & $-0.02398$ \\
$d_{21}^{(4)}$ & $-0.01840$ \\
$d_{31}^{(4)}$ & $-0.02534$ \\
$d_{41}^{(4)}$ & $0.00017$ \\
$d_{51}^{(4)}$ & $0.00013$ \\
\midrule[0.3pt]\bottomrule[1pt]
\end{tabular}
\end{minipage}%
\begin{minipage}[t]{.2\linewidth}
\begin{tabular}[t]{ll}
\toprule[1pt]\midrule[0.3pt]
$x^{(5)}$ & Value  \\ \hline
$a_{1}^{(5)}$ & $1.65802\mathrm{e}{12}$   \\
$a_{2}^{(5)}$ & $1.03757\mathrm{e}{10}$\\
$a_{3}^{(5)}$ & $2.06815\mathrm{e}{6}$\\
$b_{1}^{(5)}$ & $-7.50478\mathrm{e}{12}$\\
$b_{2}^{(5)}$ & $- 5.45113\mathrm{e}{9}$ \\
$b_{3}^{(5)}$ & $-5.61410\mathrm{e}{7}$ \\
$b_{4}^{(5)}$ & $7.04369\mathrm{e}{4}$ \\
$b_{5}^{(5)}$ & $0.36838$ \\
$c_{11}^{(5)}$ & $2.96697$ \\
$c_{12}^{(5)}$ & $-7.84667$ \\
$c_{21}^{(5)}$ & $7.56107$ \\
$c_{22}^{(5)}$ & $-43.79735$ \\
$c_{23}^{(5)}$ & $70.23612$ \\
$d_{11}^{(5)}$ & $-0.03780$ \\
$d_{12}^{(5)}$ & $-0.05580$ \\
$d_{21}^{(5)}$ & $4.36021$ \\
\midrule[0.3pt]\bottomrule[1pt]
\end{tabular}
\end{minipage}%
\caption{Values of the phenomenological coefficients of the parameter space fits, where $x$ is a placeholder variable.}
\label{tab:coeff}
\end{table*}

%%%%%%%%%%%%%%%%%%%%%%%%%%%%%%%%
\section{GW170817 Posteriors}
\label{sec:GW170817appx}
\begin{figure*}
\centering
\includegraphics[width = \linewidth]{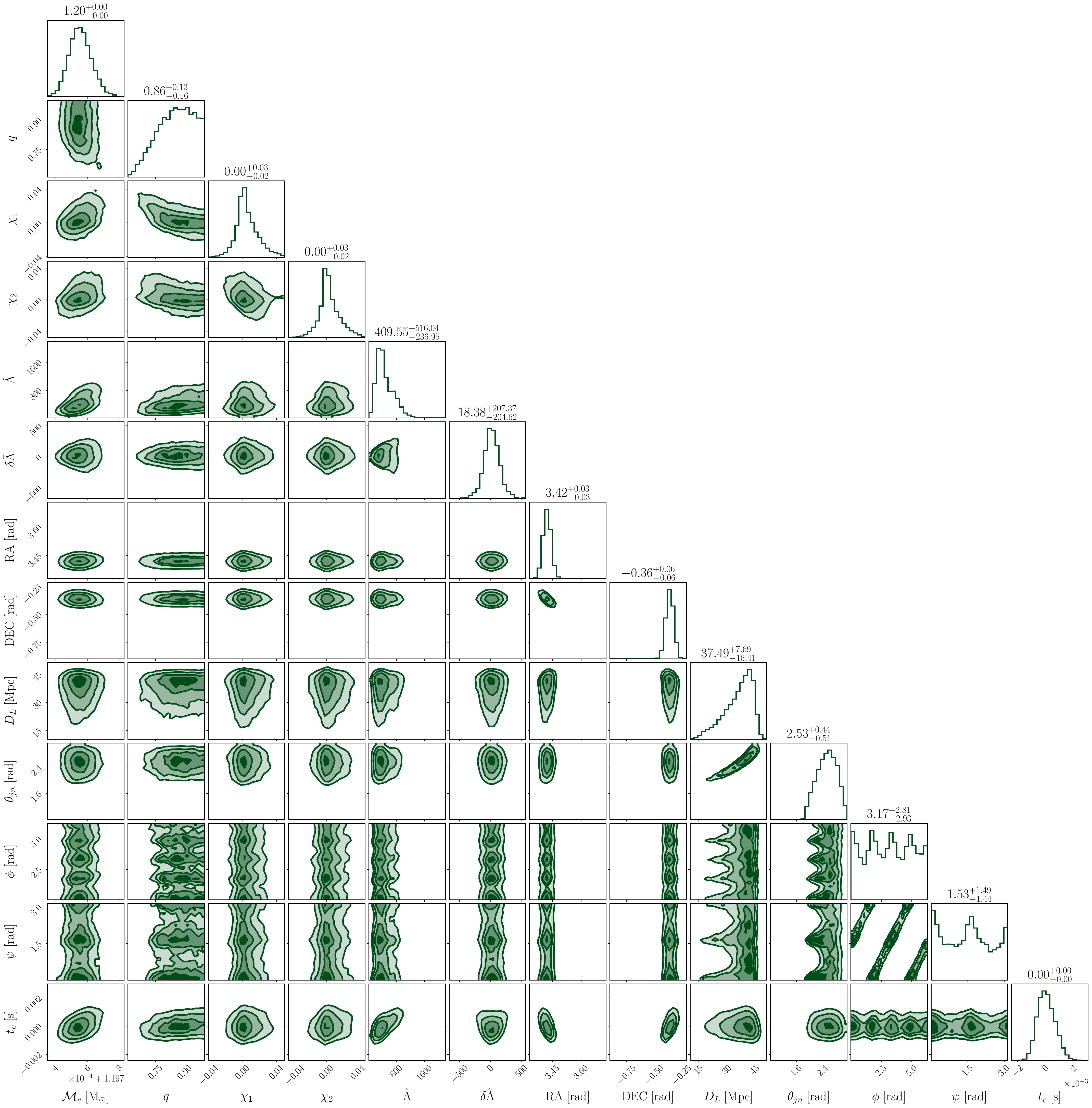}
\caption{1D and 2D posteriors of all parameters obtained for GW170817 with \textsc{IMRPhenomXAS\_PhenomGSF}. The geocentric time is shown with an offset of the true injected $t_c$ = $1187008882.4$s for better visualisation.}
\label{fig:GW170817full}
\end{figure*}

%%%%%%%%%%%%%%%%%%%%%%%%%%%%%%%%
We present the full set of posteriors for the low-spin analysis of GW170817 in Fig.~\ref{fig:GW170817full}. Information on priors and recovered values can be found in Tab.~\ref{tab:PE}.
\clearpage

%%%%%%%%%%%%%%%%%%%%%%%%%%%%%
\bibliography{references.bib}

\begin{thebibliography}{130}%
\makeatletter
\providecommand \@ifxundefined [1]{%
 \@ifx{#1\undefined}
}%
\providecommand \@ifnum [1]{%
 \ifnum #1\expandafter \@firstoftwo
 \else \expandafter \@secondoftwo
 \fi
}%
\providecommand \@ifx [1]{%
 \ifx #1\expandafter \@firstoftwo
 \else \expandafter \@secondoftwo
 \fi
}%
\providecommand \natexlab [1]{#1}%
\providecommand \enquote  [1]{``#1''}%
\providecommand \bibnamefont  [1]{#1}%
\providecommand \bibfnamefont [1]{#1}%
\providecommand \citenamefont [1]{#1}%
\providecommand \href@noop [0]{\@secondoftwo}%
\providecommand \href [0]{\begingroup \@sanitize@url \@href}%
\providecommand \@href[1]{\@@startlink{#1}\@@href}%
\providecommand \@@href[1]{\endgroup#1\@@endlink}%
\providecommand \@sanitize@url [0]{\catcode `\\12\catcode `\$12\catcode
  `\&12\catcode `\#12\catcode `\^12\catcode `\_12\catcode `\%12\relax}%
\providecommand \@@startlink[1]{}%
\providecommand \@@endlink[0]{}%
\providecommand \url  [0]{\begingroup\@sanitize@url \@url }%
\providecommand \@url [1]{\endgroup\@href {#1}{\urlprefix }}%
\providecommand \urlprefix  [0]{URL }%
\providecommand \Eprint [0]{\href }%
\providecommand \doibase [0]{https://doi.org/}%
\providecommand \selectlanguage [0]{\@gobble}%
\providecommand \bibinfo  [0]{\@secondoftwo}%
\providecommand \bibfield  [0]{\@secondoftwo}%
\providecommand \translation [1]{[#1]}%
\providecommand \BibitemOpen [0]{}%
\providecommand \bibitemStop [0]{}%
\providecommand \bibitemNoStop [0]{.\EOS\space}%
\providecommand \EOS [0]{\spacefactor3000\relax}%
\providecommand \BibitemShut  [1]{\csname bibitem#1\endcsname}%
\let\auto@bib@innerbib\@empty
%</preamble>
\bibitem [{\citenamefont {Flanagan}\ and\ \citenamefont
  {Hinderer}(2008)}]{Flanagan:2007ix}%
  \BibitemOpen
  \bibfield  {author} {\bibinfo {author} {\bibfnamefont {E.~E.}\ \bibnamefont
  {Flanagan}}\ and\ \bibinfo {author} {\bibfnamefont {T.}~\bibnamefont
  {Hinderer}},\ }\bibfield  {title} {\bibinfo {title} {{Constraining neutron
  star tidal Love numbers with gravitational wave detectors}},\ }\href
  {https://doi.org/10.1103/PhysRevD.77.021502} {\bibfield  {journal} {\bibinfo
  {journal} {Phys. Rev. D}\ }\textbf {\bibinfo {volume} {77}},\ \bibinfo
  {pages} {021502} (\bibinfo {year} {2008})},\ \Eprint
  {https://arxiv.org/abs/0709.1915} {arXiv:0709.1915 [astro-ph]} \BibitemShut
  {NoStop}%
\bibitem [{\citenamefont {Wade}\ \emph {et~al.}(2014)\citenamefont {Wade},
  \citenamefont {Creighton}, \citenamefont {Ochsner}, \citenamefont {Lackey},
  \citenamefont {Farr}, \citenamefont {Littenberg},\ and\ \citenamefont
  {Raymond}}]{Wade:2014vqa}%
  \BibitemOpen
  \bibfield  {author} {\bibinfo {author} {\bibfnamefont {L.}~\bibnamefont
  {Wade}}, \bibinfo {author} {\bibfnamefont {J.~D.~E.}\ \bibnamefont
  {Creighton}}, \bibinfo {author} {\bibfnamefont {E.}~\bibnamefont {Ochsner}},
  \bibinfo {author} {\bibfnamefont {B.~D.}\ \bibnamefont {Lackey}}, \bibinfo
  {author} {\bibfnamefont {B.~F.}\ \bibnamefont {Farr}}, \bibinfo {author}
  {\bibfnamefont {T.~B.}\ \bibnamefont {Littenberg}},\ and\ \bibinfo {author}
  {\bibfnamefont {V.}~\bibnamefont {Raymond}},\ }\bibfield  {title} {\bibinfo
  {title} {{Systematic and statistical errors in a bayesian approach to the
  estimation of the neutron-star equation of state using advanced gravitational
  wave detectors}},\ }\href {https://doi.org/10.1103/PhysRevD.89.103012}
  {\bibfield  {journal} {\bibinfo  {journal} {Phys. Rev. D}\ }\textbf {\bibinfo
  {volume} {89}},\ \bibinfo {pages} {103012} (\bibinfo {year} {2014})},\
  \Eprint {https://arxiv.org/abs/1402.5156} {arXiv:1402.5156 [gr-qc]}
  \BibitemShut {NoStop}%
\bibitem [{\citenamefont {Abbott}\ \emph
  {et~al.}(2017{\natexlab{a}})\citenamefont {Abbott} \emph
  {et~al.}}]{TheLIGOScientific:2017qsa}%
  \BibitemOpen
  \bibfield  {author} {\bibinfo {author} {\bibfnamefont {B.~P.}\ \bibnamefont
  {Abbott}} \emph {et~al.} (\bibinfo {collaboration} {LIGO Scientific,
  Virgo}),\ }\bibfield  {title} {\bibinfo {title} {{GW170817: Observation of
  Gravitational Waves from a Binary Neutron Star Inspiral}},\ }\href
  {https://doi.org/10.1103/PhysRevLett.119.161101} {\bibfield  {journal}
  {\bibinfo  {journal} {Phys. Rev. Lett.}\ }\textbf {\bibinfo {volume} {119}},\
  \bibinfo {pages} {161101} (\bibinfo {year} {2017}{\natexlab{a}})},\ \Eprint
  {https://arxiv.org/abs/1710.05832} {arXiv:1710.05832 [gr-qc]} \BibitemShut
  {NoStop}%
\bibitem [{\citenamefont {Abbott}\ \emph
  {et~al.}(2019{\natexlab{a}})\citenamefont {Abbott} \emph
  {et~al.}}]{GW170817-PE}%
  \BibitemOpen
  \bibfield  {author} {\bibinfo {author} {\bibfnamefont {B.~P.}\ \bibnamefont
  {Abbott}} \emph {et~al.} (\bibinfo {collaboration} {LIGO Scientific,
  Virgo}),\ }\bibfield  {title} {\bibinfo {title} {{Properties of the binary
  neutron star merger GW170817}},\ }\href
  {https://doi.org/10.1103/PhysRevX.9.011001} {\bibfield  {journal} {\bibinfo
  {journal} {Phys. Rev. X}\ }\textbf {\bibinfo {volume} {9}},\ \bibinfo {pages}
  {011001} (\bibinfo {year} {2019}{\natexlab{a}})},\ \Eprint
  {https://arxiv.org/abs/1805.11579} {arXiv:1805.11579 [gr-qc]} \BibitemShut
  {NoStop}%
\bibitem [{\citenamefont {Raaijmakers}\ \emph {et~al.}(2021)\citenamefont
  {Raaijmakers}, \citenamefont {Greif}, \citenamefont {Hebeler}, \citenamefont
  {Hinderer}, \citenamefont {Nissanke}, \citenamefont {Schwenk}, \citenamefont
  {Riley}, \citenamefont {Watts}, \citenamefont {Lattimer},\ and\ \citenamefont
  {Ho}}]{Raaijmakers:2021uju}%
  \BibitemOpen
  \bibfield  {author} {\bibinfo {author} {\bibfnamefont {G.}~\bibnamefont
  {Raaijmakers}}, \bibinfo {author} {\bibfnamefont {S.~K.}\ \bibnamefont
  {Greif}}, \bibinfo {author} {\bibfnamefont {K.}~\bibnamefont {Hebeler}},
  \bibinfo {author} {\bibfnamefont {T.}~\bibnamefont {Hinderer}}, \bibinfo
  {author} {\bibfnamefont {S.}~\bibnamefont {Nissanke}}, \bibinfo {author}
  {\bibfnamefont {A.}~\bibnamefont {Schwenk}}, \bibinfo {author} {\bibfnamefont
  {T.~E.}\ \bibnamefont {Riley}}, \bibinfo {author} {\bibfnamefont {A.~L.}\
  \bibnamefont {Watts}}, \bibinfo {author} {\bibfnamefont {J.~M.}\ \bibnamefont
  {Lattimer}},\ and\ \bibinfo {author} {\bibfnamefont {W.~C.~G.}\ \bibnamefont
  {Ho}},\ }\bibfield  {title} {\bibinfo {title} {{Constraints on the Dense
  Matter Equation of State and Neutron Star Properties from
  NICER\textquoteright{}s Mass\textendash{}Radius Estimate of PSR J0740+6620
  and Multimessenger Observations}},\ }\href
  {https://doi.org/10.3847/2041-8213/ac089a} {\bibfield  {journal} {\bibinfo
  {journal} {Astrophys. J. Lett.}\ }\textbf {\bibinfo {volume} {918}},\
  \bibinfo {pages} {L29} (\bibinfo {year} {2021})},\ \Eprint
  {https://arxiv.org/abs/2105.06981} {arXiv:2105.06981 [astro-ph.HE]}
  \BibitemShut {NoStop}%
\bibitem [{\citenamefont {Miller}\ \emph {et~al.}(2021)\citenamefont {Miller}
  \emph {et~al.}}]{Miller:2021qha}%
  \BibitemOpen
  \bibfield  {author} {\bibinfo {author} {\bibfnamefont {M.~C.}\ \bibnamefont
  {Miller}} \emph {et~al.},\ }\bibfield  {title} {\bibinfo {title} {{The Radius
  of PSR J0740+6620 from NICER and XMM-Newton Data}},\ }\href
  {https://doi.org/10.3847/2041-8213/ac089b} {\bibfield  {journal} {\bibinfo
  {journal} {Astrophys. J. Lett.}\ }\textbf {\bibinfo {volume} {918}},\
  \bibinfo {pages} {L28} (\bibinfo {year} {2021})},\ \Eprint
  {https://arxiv.org/abs/2105.06979} {arXiv:2105.06979 [astro-ph.HE]}
  \BibitemShut {NoStop}%
\bibitem [{\citenamefont {Adhikari}\ \emph {et~al.}(2021)\citenamefont
  {Adhikari} \emph {et~al.}}]{PREX:2021umo}%
  \BibitemOpen
  \bibfield  {author} {\bibinfo {author} {\bibfnamefont {D.}~\bibnamefont
  {Adhikari}} \emph {et~al.} (\bibinfo {collaboration} {PREX}),\ }\bibfield
  {title} {\bibinfo {title} {{Accurate Determination of the Neutron Skin
  Thickness of $^{208}$Pb through Parity-Violation in Electron Scattering}},\
  }\href {https://doi.org/10.1103/PhysRevLett.126.172502} {\bibfield  {journal}
  {\bibinfo  {journal} {Phys. Rev. Lett.}\ }\textbf {\bibinfo {volume} {126}},\
  \bibinfo {pages} {172502} (\bibinfo {year} {2021})},\ \Eprint
  {https://arxiv.org/abs/2102.10767} {arXiv:2102.10767 [nucl-ex]} \BibitemShut
  {NoStop}%
\bibitem [{\citenamefont {Chatziioannou}\ \emph {et~al.}(2024)\citenamefont
  {Chatziioannou}, \citenamefont {Cromartie}, \citenamefont {Gandolfi},
  \citenamefont {Tews}, \citenamefont {Radice}, \citenamefont {Steiner},\ and\
  \citenamefont {Watts}}]{Chatziioannou:2024tjq}%
  \BibitemOpen
  \bibfield  {author} {\bibinfo {author} {\bibfnamefont {K.}~\bibnamefont
  {Chatziioannou}}, \bibinfo {author} {\bibfnamefont {H.~T.}\ \bibnamefont
  {Cromartie}}, \bibinfo {author} {\bibfnamefont {S.}~\bibnamefont {Gandolfi}},
  \bibinfo {author} {\bibfnamefont {I.}~\bibnamefont {Tews}}, \bibinfo {author}
  {\bibfnamefont {D.}~\bibnamefont {Radice}}, \bibinfo {author} {\bibfnamefont
  {A.~W.}\ \bibnamefont {Steiner}},\ and\ \bibinfo {author} {\bibfnamefont
  {A.~L.}\ \bibnamefont {Watts}},\ }\bibfield  {title} {\bibinfo {title}
  {{Neutron stars and the dense matter equation of state: from microscopic
  theory to macroscopic observations}},\ }\href@noop {} {\  (\bibinfo {year}
  {2024})},\ \Eprint {https://arxiv.org/abs/2407.11153} {arXiv:2407.11153
  [nucl-th]} \BibitemShut {NoStop}%
\bibitem [{\citenamefont {Reed}\ \emph {et~al.}(2021)\citenamefont {Reed},
  \citenamefont {Fattoyev}, \citenamefont {Horowitz},\ and\ \citenamefont
  {Piekarewicz}}]{Reed:2021nqk}%
  \BibitemOpen
  \bibfield  {author} {\bibinfo {author} {\bibfnamefont {B.~T.}\ \bibnamefont
  {Reed}}, \bibinfo {author} {\bibfnamefont {F.~J.}\ \bibnamefont {Fattoyev}},
  \bibinfo {author} {\bibfnamefont {C.~J.}\ \bibnamefont {Horowitz}},\ and\
  \bibinfo {author} {\bibfnamefont {J.}~\bibnamefont {Piekarewicz}},\
  }\bibfield  {title} {\bibinfo {title} {{Implications of PREX-2 on the
  Equation of State of Neutron-Rich Matter}},\ }\href
  {https://doi.org/10.1103/PhysRevLett.126.172503} {\bibfield  {journal}
  {\bibinfo  {journal} {Phys. Rev. Lett.}\ }\textbf {\bibinfo {volume} {126}},\
  \bibinfo {pages} {172503} (\bibinfo {year} {2021})},\ \Eprint
  {https://arxiv.org/abs/2101.03193} {arXiv:2101.03193 [nucl-th]} \BibitemShut
  {NoStop}%
\bibitem [{\citenamefont {Abbott}\ \emph {et~al.}(2018)\citenamefont {Abbott}
  \emph {et~al.}}]{GW170817-EOS}%
  \BibitemOpen
  \bibfield  {author} {\bibinfo {author} {\bibfnamefont {B.~P.}\ \bibnamefont
  {Abbott}} \emph {et~al.} (\bibinfo {collaboration} {LIGO Scientific,
  Virgo}),\ }\bibfield  {title} {\bibinfo {title} {{GW170817: Measurements of
  neutron star radii and equation of state}},\ }\href
  {https://doi.org/10.1103/PhysRevLett.121.161101} {\bibfield  {journal}
  {\bibinfo  {journal} {Phys. Rev. Lett.}\ }\textbf {\bibinfo {volume} {121}},\
  \bibinfo {pages} {161101} (\bibinfo {year} {2018})},\ \Eprint
  {https://arxiv.org/abs/1805.11581} {arXiv:1805.11581 [gr-qc]} \BibitemShut
  {NoStop}%
\bibitem [{\citenamefont {Vines}\ \emph {et~al.}(2011)\citenamefont {Vines},
  \citenamefont {Flanagan},\ and\ \citenamefont {Hinderer}}]{Vines:2011ud}%
  \BibitemOpen
  \bibfield  {author} {\bibinfo {author} {\bibfnamefont {J.}~\bibnamefont
  {Vines}}, \bibinfo {author} {\bibfnamefont {E.~E.}\ \bibnamefont
  {Flanagan}},\ and\ \bibinfo {author} {\bibfnamefont {T.}~\bibnamefont
  {Hinderer}},\ }\bibfield  {title} {\bibinfo {title} {{Post-1-Newtonian tidal
  effects in the gravitational waveform from binary inspirals}},\ }\href
  {https://doi.org/10.1103/PhysRevD.83.084051} {\bibfield  {journal} {\bibinfo
  {journal} {Phys. Rev.}\ }\textbf {\bibinfo {volume} {D83}},\ \bibinfo {pages}
  {084051} (\bibinfo {year} {2011})},\ \Eprint
  {https://arxiv.org/abs/1101.1673} {arXiv:1101.1673 [gr-qc]} \BibitemShut
  {NoStop}%
%%CITATION = ARXIV:1101.1673;%%
\bibitem [{\citenamefont {Damour}\ \emph {et~al.}(2012)\citenamefont {Damour},
  \citenamefont {Nagar},\ and\ \citenamefont {Villain}}]{Damour:2012yf}%
  \BibitemOpen
  \bibfield  {author} {\bibinfo {author} {\bibfnamefont {T.}~\bibnamefont
  {Damour}}, \bibinfo {author} {\bibfnamefont {A.}~\bibnamefont {Nagar}},\ and\
  \bibinfo {author} {\bibfnamefont {L.}~\bibnamefont {Villain}},\ }\bibfield
  {title} {\bibinfo {title} {{Measurability of the tidal polarizability of
  neutron stars in late-inspiral gravitational-wave signals}},\ }\href
  {https://doi.org/10.1103/PhysRevD.85.123007} {\bibfield  {journal} {\bibinfo
  {journal} {Phys. Rev.}\ }\textbf {\bibinfo {volume} {D85}},\ \bibinfo {pages}
  {123007} (\bibinfo {year} {2012})},\ \Eprint
  {https://arxiv.org/abs/1203.4352} {arXiv:1203.4352 [gr-qc]} \BibitemShut
  {NoStop}%
%%CITATION = ARXIV:1203.4352;%%
\bibitem [{\citenamefont {Abdelsalhin}\ \emph {et~al.}(2018)\citenamefont
  {Abdelsalhin}, \citenamefont {Gualtieri},\ and\ \citenamefont
  {Pani}}]{Abdelsalhin:2018reg}%
  \BibitemOpen
  \bibfield  {author} {\bibinfo {author} {\bibfnamefont {T.}~\bibnamefont
  {Abdelsalhin}}, \bibinfo {author} {\bibfnamefont {L.}~\bibnamefont
  {Gualtieri}},\ and\ \bibinfo {author} {\bibfnamefont {P.}~\bibnamefont
  {Pani}},\ }\bibfield  {title} {\bibinfo {title} {{Post-Newtonian spin-tidal
  couplings for compact binaries}},\ }\href
  {https://doi.org/10.1103/PhysRevD.98.104046} {\bibfield  {journal} {\bibinfo
  {journal} {Phys. Rev. D}\ }\textbf {\bibinfo {volume} {98}},\ \bibinfo
  {pages} {104046} (\bibinfo {year} {2018})},\ \Eprint
  {https://arxiv.org/abs/1805.01487} {arXiv:1805.01487 [gr-qc]} \BibitemShut
  {NoStop}%
\bibitem [{\citenamefont {Banihashemi}\ and\ \citenamefont
  {Vines}(2020)}]{Banihashemi:2018xfb}%
  \BibitemOpen
  \bibfield  {author} {\bibinfo {author} {\bibfnamefont {B.}~\bibnamefont
  {Banihashemi}}\ and\ \bibinfo {author} {\bibfnamefont {J.}~\bibnamefont
  {Vines}},\ }\bibfield  {title} {\bibinfo {title} {{Gravitomagnetic tidal
  effects in gravitational waves from neutron star binaries}},\ }\href
  {https://doi.org/10.1103/PhysRevD.101.064003} {\bibfield  {journal} {\bibinfo
   {journal} {Phys. Rev. D}\ }\textbf {\bibinfo {volume} {101}},\ \bibinfo
  {pages} {064003} (\bibinfo {year} {2020})},\ \Eprint
  {https://arxiv.org/abs/1805.07266} {arXiv:1805.07266 [gr-qc]} \BibitemShut
  {NoStop}%
\bibitem [{\citenamefont {Landry}(2018)}]{Landry:2018bil}%
  \BibitemOpen
  \bibfield  {author} {\bibinfo {author} {\bibfnamefont {P.}~\bibnamefont
  {Landry}},\ }\bibfield  {title} {\bibinfo {title} {{Rotational-tidal phasing
  of the binary neutron star waveform}},\ }\href@noop {} {\  (\bibinfo {year}
  {2018})},\ \Eprint {https://arxiv.org/abs/1805.01882} {arXiv:1805.01882
  [gr-qc]} \BibitemShut {NoStop}%
\bibitem [{\citenamefont {Schmidt}\ and\ \citenamefont
  {Hinderer}(2019)}]{Schmidt:2019wrl}%
  \BibitemOpen
  \bibfield  {author} {\bibinfo {author} {\bibfnamefont {P.}~\bibnamefont
  {Schmidt}}\ and\ \bibinfo {author} {\bibfnamefont {T.}~\bibnamefont
  {Hinderer}},\ }\bibfield  {title} {\bibinfo {title} {{Frequency domain model
  of $f$-mode dynamic tides in gravitational waveforms from compact binary
  inspirals}},\ }\href {https://doi.org/10.1103/PhysRevD.100.021501} {\bibfield
   {journal} {\bibinfo  {journal} {Phys. Rev. D}\ }\textbf {\bibinfo {volume}
  {100}},\ \bibinfo {pages} {021501} (\bibinfo {year} {2019})},\ \Eprint
  {https://arxiv.org/abs/1905.00818} {arXiv:1905.00818 [gr-qc]} \BibitemShut
  {NoStop}%
\bibitem [{\citenamefont {Henry}\ \emph
  {et~al.}(2020{\natexlab{a}})\citenamefont {Henry}, \citenamefont {Faye},\
  and\ \citenamefont {Blanchet}}]{Henry:2019xhg}%
  \BibitemOpen
  \bibfield  {author} {\bibinfo {author} {\bibfnamefont {Q.}~\bibnamefont
  {Henry}}, \bibinfo {author} {\bibfnamefont {G.}~\bibnamefont {Faye}},\ and\
  \bibinfo {author} {\bibfnamefont {L.}~\bibnamefont {Blanchet}},\ }\bibfield
  {title} {\bibinfo {title} {{Tidal effects in the equations of motion of
  compact binary systems to next-to-next-to-leading post-Newtonian order}},\
  }\href {https://doi.org/10.1103/PhysRevD.101.064047} {\bibfield  {journal}
  {\bibinfo  {journal} {Phys. Rev. D}\ }\textbf {\bibinfo {volume} {101}},\
  \bibinfo {pages} {064047} (\bibinfo {year} {2020}{\natexlab{a}})},\ \Eprint
  {https://arxiv.org/abs/1912.01920} {arXiv:1912.01920 [gr-qc]} \BibitemShut
  {NoStop}%
\bibitem [{\citenamefont {Henry}\ \emph
  {et~al.}(2020{\natexlab{b}})\citenamefont {Henry}, \citenamefont {Faye},\
  and\ \citenamefont {Blanchet}}]{Henry:2020ski}%
  \BibitemOpen
  \bibfield  {author} {\bibinfo {author} {\bibfnamefont {Q.}~\bibnamefont
  {Henry}}, \bibinfo {author} {\bibfnamefont {G.}~\bibnamefont {Faye}},\ and\
  \bibinfo {author} {\bibfnamefont {L.}~\bibnamefont {Blanchet}},\ }\bibfield
  {title} {\bibinfo {title} {{Tidal effects in the gravitational-wave phase
  evolution of compact binary systems to next-to-next-to-leading post-Newtonian
  order}},\ }\href {https://doi.org/10.1103/PhysRevD.102.044033} {\bibfield
  {journal} {\bibinfo  {journal} {Phys. Rev. D}\ }\textbf {\bibinfo {volume}
  {102}},\ \bibinfo {pages} {044033} (\bibinfo {year} {2020}{\natexlab{b}})},\
  \bibinfo {note} {[Erratum: Phys.Rev.D 108, 089901 (2023)]},\ \Eprint
  {https://arxiv.org/abs/2005.13367} {arXiv:2005.13367 [gr-qc]} \BibitemShut
  {NoStop}%
\bibitem [{\citenamefont {Damour}\ and\ \citenamefont
  {Nagar}(2010)}]{Damour:2009wj}%
  \BibitemOpen
  \bibfield  {author} {\bibinfo {author} {\bibfnamefont {T.}~\bibnamefont
  {Damour}}\ and\ \bibinfo {author} {\bibfnamefont {A.}~\bibnamefont {Nagar}},\
  }\bibfield  {title} {\bibinfo {title} {{Effective One Body description of
  tidal effects in inspiralling compact binaries}},\ }\href
  {https://doi.org/10.1103/PhysRevD.81.084016} {\bibfield  {journal} {\bibinfo
  {journal} {Phys. Rev. D}\ }\textbf {\bibinfo {volume} {81}},\ \bibinfo
  {pages} {084016} (\bibinfo {year} {2010})},\ \Eprint
  {https://arxiv.org/abs/0911.5041} {arXiv:0911.5041 [gr-qc]} \BibitemShut
  {NoStop}%
\bibitem [{\citenamefont {Bini}\ \emph {et~al.}(2012)\citenamefont {Bini},
  \citenamefont {Damour},\ and\ \citenamefont {Faye}}]{Bini:2012gu}%
  \BibitemOpen
  \bibfield  {author} {\bibinfo {author} {\bibfnamefont {D.}~\bibnamefont
  {Bini}}, \bibinfo {author} {\bibfnamefont {T.}~\bibnamefont {Damour}},\ and\
  \bibinfo {author} {\bibfnamefont {G.}~\bibnamefont {Faye}},\ }\bibfield
  {title} {\bibinfo {title} {{Effective action approach to higher-order
  relativistic tidal interactions in binary systems and their effective one
  body description}},\ }\href {https://doi.org/10.1103/PhysRevD.85.124034}
  {\bibfield  {journal} {\bibinfo  {journal} {Phys. Rev. D}\ }\textbf {\bibinfo
  {volume} {85}},\ \bibinfo {pages} {124034} (\bibinfo {year} {2012})},\
  \Eprint {https://arxiv.org/abs/1202.3565} {arXiv:1202.3565 [gr-qc]}
  \BibitemShut {NoStop}%
\bibitem [{\citenamefont {Hinderer}\ \emph {et~al.}(2016)\citenamefont
  {Hinderer} \emph {et~al.}}]{Hinderer:2016eia}%
  \BibitemOpen
  \bibfield  {author} {\bibinfo {author} {\bibfnamefont {T.}~\bibnamefont
  {Hinderer}} \emph {et~al.},\ }\bibfield  {title} {\bibinfo {title} {{Effects
  of neutron-star dynamic tides on gravitational waveforms within the
  effective-one-body approach}},\ }\href
  {https://doi.org/10.1103/PhysRevLett.116.181101} {\bibfield  {journal}
  {\bibinfo  {journal} {Phys. Rev. Lett.}\ }\textbf {\bibinfo {volume} {116}},\
  \bibinfo {pages} {181101} (\bibinfo {year} {2016})},\ \Eprint
  {https://arxiv.org/abs/1602.00599} {arXiv:1602.00599 [gr-qc]} \BibitemShut
  {NoStop}%
\bibitem [{\citenamefont {Steinhoff}\ \emph {et~al.}(2016)\citenamefont
  {Steinhoff}, \citenamefont {Hinderer}, \citenamefont {Buonanno},\ and\
  \citenamefont {Taracchini}}]{Steinhoff:2016rfi}%
  \BibitemOpen
  \bibfield  {author} {\bibinfo {author} {\bibfnamefont {J.}~\bibnamefont
  {Steinhoff}}, \bibinfo {author} {\bibfnamefont {T.}~\bibnamefont {Hinderer}},
  \bibinfo {author} {\bibfnamefont {A.}~\bibnamefont {Buonanno}},\ and\
  \bibinfo {author} {\bibfnamefont {A.}~\bibnamefont {Taracchini}},\ }\bibfield
   {title} {\bibinfo {title} {{Dynamical Tides in General Relativity: Effective
  Action and Effective-One-Body Hamiltonian}},\ }\href
  {https://doi.org/10.1103/PhysRevD.94.104028} {\bibfield  {journal} {\bibinfo
  {journal} {Phys. Rev. D}\ }\textbf {\bibinfo {volume} {94}},\ \bibinfo
  {pages} {104028} (\bibinfo {year} {2016})},\ \Eprint
  {https://arxiv.org/abs/1608.01907} {arXiv:1608.01907 [gr-qc]} \BibitemShut
  {NoStop}%
\bibitem [{\citenamefont {Boh\'e}\ \emph {et~al.}(2017)\citenamefont {Boh\'e}
  \emph {et~al.}}]{Bohe:2016gbl}%
  \BibitemOpen
  \bibfield  {author} {\bibinfo {author} {\bibfnamefont {A.}~\bibnamefont
  {Boh\'e}} \emph {et~al.},\ }\bibfield  {title} {\bibinfo {title} {{Improved
  effective-one-body model of spinning, nonprecessing binary black holes for
  the era of gravitational-wave astrophysics with advanced detectors}},\ }\href
  {https://doi.org/10.1103/PhysRevD.95.044028} {\bibfield  {journal} {\bibinfo
  {journal} {Phys. Rev. D}\ }\textbf {\bibinfo {volume} {95}},\ \bibinfo
  {pages} {044028} (\bibinfo {year} {2017})},\ \Eprint
  {https://arxiv.org/abs/1611.03703} {arXiv:1611.03703 [gr-qc]} \BibitemShut
  {NoStop}%
\bibitem [{\citenamefont {Nagar}\ \emph {et~al.}(2018)\citenamefont {Nagar},
  \citenamefont {Bernuzzi}, \citenamefont {Del~Pozzo}, \citenamefont
  {Riemenschneider}, \citenamefont {Akcay}, \citenamefont {Carullo},
  \citenamefont {Fleig}, \citenamefont {Babak}, \citenamefont {Tsang},
  \citenamefont {Colleoni}, \citenamefont {Messina}, \citenamefont {Pratten},
  \citenamefont {Radice}, \citenamefont {Rettegno}, \citenamefont {Agathos},
  \citenamefont {Fauchon-Jones}, \citenamefont {Hannam}, \citenamefont {Husa},
  \citenamefont {Dietrich}, \citenamefont {Cerd\'a-Duran}, \citenamefont
  {Font}, \citenamefont {Pannarale}, \citenamefont {Schmidt},\ and\
  \citenamefont {Damour}}]{Nagar2018}%
  \BibitemOpen
  \bibfield  {author} {\bibinfo {author} {\bibfnamefont {A.}~\bibnamefont
  {Nagar}}, \bibinfo {author} {\bibfnamefont {S.}~\bibnamefont {Bernuzzi}},
  \bibinfo {author} {\bibfnamefont {W.}~\bibnamefont {Del~Pozzo}}, \bibinfo
  {author} {\bibfnamefont {G.}~\bibnamefont {Riemenschneider}}, \bibinfo
  {author} {\bibfnamefont {S.}~\bibnamefont {Akcay}}, \bibinfo {author}
  {\bibfnamefont {G.}~\bibnamefont {Carullo}}, \bibinfo {author} {\bibfnamefont
  {P.}~\bibnamefont {Fleig}}, \bibinfo {author} {\bibfnamefont
  {S.}~\bibnamefont {Babak}}, \bibinfo {author} {\bibfnamefont {K.~W.}\
  \bibnamefont {Tsang}}, \bibinfo {author} {\bibfnamefont {M.}~\bibnamefont
  {Colleoni}}, \bibinfo {author} {\bibfnamefont {F.}~\bibnamefont {Messina}},
  \bibinfo {author} {\bibfnamefont {G.}~\bibnamefont {Pratten}}, \bibinfo
  {author} {\bibfnamefont {D.}~\bibnamefont {Radice}}, \bibinfo {author}
  {\bibfnamefont {P.}~\bibnamefont {Rettegno}}, \bibinfo {author}
  {\bibfnamefont {M.}~\bibnamefont {Agathos}}, \bibinfo {author} {\bibfnamefont
  {E.}~\bibnamefont {Fauchon-Jones}}, \bibinfo {author} {\bibfnamefont
  {M.}~\bibnamefont {Hannam}}, \bibinfo {author} {\bibfnamefont
  {S.}~\bibnamefont {Husa}}, \bibinfo {author} {\bibfnamefont {T.}~\bibnamefont
  {Dietrich}}, \bibinfo {author} {\bibfnamefont {P.}~\bibnamefont
  {Cerd\'a-Duran}}, \bibinfo {author} {\bibfnamefont {J.~A.}\ \bibnamefont
  {Font}}, \bibinfo {author} {\bibfnamefont {F.}~\bibnamefont {Pannarale}},
  \bibinfo {author} {\bibfnamefont {P.}~\bibnamefont {Schmidt}},\ and\ \bibinfo
  {author} {\bibfnamefont {T.}~\bibnamefont {Damour}},\ }\bibfield  {title}
  {\bibinfo {title} {Time-domain effective-one-body gravitational waveforms for
  coalescing compact binaries with nonprecessing spins, tides, and self-spin
  effects},\ }\href {https://doi.org/10.1103/PhysRevD.98.104052} {\bibfield
  {journal} {\bibinfo  {journal} {Phys. Rev. D}\ }\textbf {\bibinfo {volume}
  {98}},\ \bibinfo {pages} {104052} (\bibinfo {year} {2018})}\BibitemShut
  {NoStop}%
\bibitem [{\citenamefont {Akcay}\ \emph {et~al.}(2019)\citenamefont {Akcay},
  \citenamefont {Bernuzzi}, \citenamefont {Messina}, \citenamefont {Nagar},
  \citenamefont {Ortiz},\ and\ \citenamefont {Rettegno}}]{Akcay:2018yyh}%
  \BibitemOpen
  \bibfield  {author} {\bibinfo {author} {\bibfnamefont {S.}~\bibnamefont
  {Akcay}}, \bibinfo {author} {\bibfnamefont {S.}~\bibnamefont {Bernuzzi}},
  \bibinfo {author} {\bibfnamefont {F.}~\bibnamefont {Messina}}, \bibinfo
  {author} {\bibfnamefont {A.}~\bibnamefont {Nagar}}, \bibinfo {author}
  {\bibfnamefont {N.}~\bibnamefont {Ortiz}},\ and\ \bibinfo {author}
  {\bibfnamefont {P.}~\bibnamefont {Rettegno}},\ }\bibfield  {title} {\bibinfo
  {title} {{Effective-one-body multipolar waveform for tidally interacting
  binary neutron stars up to merger}},\ }\href
  {https://doi.org/10.1103/PhysRevD.99.044051} {\bibfield  {journal} {\bibinfo
  {journal} {Phys. Rev. D}\ }\textbf {\bibinfo {volume} {99}},\ \bibinfo
  {pages} {044051} (\bibinfo {year} {2019})},\ \Eprint
  {https://arxiv.org/abs/1812.02744} {arXiv:1812.02744 [gr-qc]} \BibitemShut
  {NoStop}%
\bibitem [{\citenamefont {Steinhoff}\ \emph {et~al.}(2021)\citenamefont
  {Steinhoff}, \citenamefont {Hinderer}, \citenamefont {Dietrich},\ and\
  \citenamefont {Foucart}}]{Steinhoff:2021dsn}%
  \BibitemOpen
  \bibfield  {author} {\bibinfo {author} {\bibfnamefont {J.}~\bibnamefont
  {Steinhoff}}, \bibinfo {author} {\bibfnamefont {T.}~\bibnamefont {Hinderer}},
  \bibinfo {author} {\bibfnamefont {T.}~\bibnamefont {Dietrich}},\ and\
  \bibinfo {author} {\bibfnamefont {F.}~\bibnamefont {Foucart}},\ }\bibfield
  {title} {\bibinfo {title} {{Spin effects on neutron star fundamental-mode
  dynamical tides: Phenomenology and comparison to numerical simulations}},\
  }\href {https://doi.org/10.1103/PhysRevResearch.3.033129} {\bibfield
  {journal} {\bibinfo  {journal} {Phys. Rev. Res.}\ }\textbf {\bibinfo {volume}
  {3}},\ \bibinfo {pages} {033129} (\bibinfo {year} {2021})},\ \Eprint
  {https://arxiv.org/abs/2103.06100} {arXiv:2103.06100 [gr-qc]} \BibitemShut
  {NoStop}%
\bibitem [{\citenamefont {Gamba}\ \emph {et~al.}(2023)\citenamefont {Gamba}
  \emph {et~al.}}]{Gamba:2023mww}%
  \BibitemOpen
  \bibfield  {author} {\bibinfo {author} {\bibfnamefont {R.}~\bibnamefont
  {Gamba}} \emph {et~al.},\ }\bibfield  {title} {\bibinfo {title}
  {{Analytically improved and numerical-relativity informed effective-one-body
  model for coalescing binary neutron stars}},\ }\href@noop {} {\  (\bibinfo
  {year} {2023})},\ \Eprint {https://arxiv.org/abs/2307.15125}
  {arXiv:2307.15125 [gr-qc]} \BibitemShut {NoStop}%
\bibitem [{\citenamefont {Dietrich}\ \emph {et~al.}(2017)\citenamefont
  {Dietrich}, \citenamefont {Bernuzzi},\ and\ \citenamefont
  {Tichy}}]{Dietrich:2017aum}%
  \BibitemOpen
  \bibfield  {author} {\bibinfo {author} {\bibfnamefont {T.}~\bibnamefont
  {Dietrich}}, \bibinfo {author} {\bibfnamefont {S.}~\bibnamefont {Bernuzzi}},\
  and\ \bibinfo {author} {\bibfnamefont {W.}~\bibnamefont {Tichy}},\ }\bibfield
   {title} {\bibinfo {title} {{Closed-form tidal approximants for binary
  neutron star gravitational waveforms constructed from high-resolution
  numerical relativity simulations}},\ }\href
  {https://doi.org/10.1103/PhysRevD.96.121501} {\bibfield  {journal} {\bibinfo
  {journal} {Phys. Rev. D}\ }\textbf {\bibinfo {volume} {96}},\ \bibinfo
  {pages} {121501} (\bibinfo {year} {2017})},\ \Eprint
  {https://arxiv.org/abs/1706.02969} {arXiv:1706.02969 [gr-qc]} \BibitemShut
  {NoStop}%
\bibitem [{\citenamefont {Dietrich}\ \emph
  {et~al.}(2019{\natexlab{a}})\citenamefont {Dietrich}, \citenamefont
  {Samajdar}, \citenamefont {Khan}, \citenamefont {Johnson-McDaniel},
  \citenamefont {Dudi},\ and\ \citenamefont {Tichy}}]{Dietrich:2019kaq}%
  \BibitemOpen
  \bibfield  {author} {\bibinfo {author} {\bibfnamefont {T.}~\bibnamefont
  {Dietrich}}, \bibinfo {author} {\bibfnamefont {A.}~\bibnamefont {Samajdar}},
  \bibinfo {author} {\bibfnamefont {S.}~\bibnamefont {Khan}}, \bibinfo {author}
  {\bibfnamefont {N.~K.}\ \bibnamefont {Johnson-McDaniel}}, \bibinfo {author}
  {\bibfnamefont {R.}~\bibnamefont {Dudi}},\ and\ \bibinfo {author}
  {\bibfnamefont {W.}~\bibnamefont {Tichy}},\ }\bibfield  {title} {\bibinfo
  {title} {{Improving the NRTidal model for binary neutron star systems}},\
  }\href {https://doi.org/10.1103/PhysRevD.100.044003} {\bibfield  {journal}
  {\bibinfo  {journal} {Phys. Rev. D}\ }\textbf {\bibinfo {volume} {100}},\
  \bibinfo {pages} {044003} (\bibinfo {year} {2019}{\natexlab{a}})},\ \Eprint
  {https://arxiv.org/abs/1905.06011} {arXiv:1905.06011 [gr-qc]} \BibitemShut
  {NoStop}%
\bibitem [{\citenamefont {Abac}\ \emph
  {et~al.}(2024{\natexlab{a}})\citenamefont {Abac}, \citenamefont {Dietrich},
  \citenamefont {Buonanno}, \citenamefont {Steinhoff},\ and\ \citenamefont
  {Ujevic}}]{Abac:2023ujg}%
  \BibitemOpen
  \bibfield  {author} {\bibinfo {author} {\bibfnamefont {A.}~\bibnamefont
  {Abac}}, \bibinfo {author} {\bibfnamefont {T.}~\bibnamefont {Dietrich}},
  \bibinfo {author} {\bibfnamefont {A.}~\bibnamefont {Buonanno}}, \bibinfo
  {author} {\bibfnamefont {J.}~\bibnamefont {Steinhoff}},\ and\ \bibinfo
  {author} {\bibfnamefont {M.}~\bibnamefont {Ujevic}},\ }\bibfield  {title}
  {\bibinfo {title} {{New and robust gravitational-waveform model for
  high-mass-ratio binary neutron star systems with dynamical tidal effects}},\
  }\href {https://doi.org/10.1103/PhysRevD.109.024062} {\bibfield  {journal}
  {\bibinfo  {journal} {Phys. Rev. D}\ }\textbf {\bibinfo {volume} {109}},\
  \bibinfo {pages} {024062} (\bibinfo {year} {2024}{\natexlab{a}})},\ \Eprint
  {https://arxiv.org/abs/2311.07456} {arXiv:2311.07456 [gr-qc]} \BibitemShut
  {NoStop}%
\bibitem [{\citenamefont {Bernuzzi}\ \emph {et~al.}(2015)\citenamefont
  {Bernuzzi}, \citenamefont {Nagar}, \citenamefont {Dietrich},\ and\
  \citenamefont {Damour}}]{Bernuzzi:2014owa}%
  \BibitemOpen
  \bibfield  {author} {\bibinfo {author} {\bibfnamefont {S.}~\bibnamefont
  {Bernuzzi}}, \bibinfo {author} {\bibfnamefont {A.}~\bibnamefont {Nagar}},
  \bibinfo {author} {\bibfnamefont {T.}~\bibnamefont {Dietrich}},\ and\
  \bibinfo {author} {\bibfnamefont {T.}~\bibnamefont {Damour}},\ }\bibfield
  {title} {\bibinfo {title} {{Modeling the Dynamics of Tidally Interacting
  Binary Neutron Stars up to the Merger}},\ }\href
  {https://doi.org/10.1103/PhysRevLett.114.161103} {\bibfield  {journal}
  {\bibinfo  {journal} {Phys. Rev. Lett.}\ }\textbf {\bibinfo {volume} {114}},\
  \bibinfo {pages} {161103} (\bibinfo {year} {2015})},\ \Eprint
  {https://arxiv.org/abs/1412.4553} {arXiv:1412.4553 [gr-qc]} \BibitemShut
  {NoStop}%
\bibitem [{\citenamefont {Kawaguchi}\ \emph {et~al.}(2018)\citenamefont
  {Kawaguchi}, \citenamefont {Kiuchi}, \citenamefont {Kyutoku}, \citenamefont
  {Sekiguchi}, \citenamefont {Shibata},\ and\ \citenamefont
  {Taniguchi}}]{Kawaguchi:2018gvj}%
  \BibitemOpen
  \bibfield  {author} {\bibinfo {author} {\bibfnamefont {K.}~\bibnamefont
  {Kawaguchi}}, \bibinfo {author} {\bibfnamefont {K.}~\bibnamefont {Kiuchi}},
  \bibinfo {author} {\bibfnamefont {K.}~\bibnamefont {Kyutoku}}, \bibinfo
  {author} {\bibfnamefont {Y.}~\bibnamefont {Sekiguchi}}, \bibinfo {author}
  {\bibfnamefont {M.}~\bibnamefont {Shibata}},\ and\ \bibinfo {author}
  {\bibfnamefont {K.}~\bibnamefont {Taniguchi}},\ }\bibfield  {title} {\bibinfo
  {title} {{Frequency-domain gravitational waveform models for inspiraling
  binary neutron stars}},\ }\href {https://doi.org/10.1103/PhysRevD.97.044044}
  {\bibfield  {journal} {\bibinfo  {journal} {Phys. Rev. D}\ }\textbf {\bibinfo
  {volume} {97}},\ \bibinfo {pages} {044044} (\bibinfo {year} {2018})},\
  \Eprint {https://arxiv.org/abs/1802.06518} {arXiv:1802.06518 [gr-qc]}
  \BibitemShut {NoStop}%
\bibitem [{\citenamefont {Ajith}\ \emph {et~al.}(2011)\citenamefont {Ajith}
  \emph {et~al.}}]{Ajith:2009bn}%
  \BibitemOpen
  \bibfield  {author} {\bibinfo {author} {\bibfnamefont {P.}~\bibnamefont
  {Ajith}} \emph {et~al.},\ }\bibfield  {title} {\bibinfo {title}
  {{Inspiral-merger-ringdown waveforms for black-hole binaries with
  non-precessing spins}},\ }\href
  {https://doi.org/10.1103/PhysRevLett.106.241101} {\bibfield  {journal}
  {\bibinfo  {journal} {Phys. Rev. Lett.}\ }\textbf {\bibinfo {volume} {106}},\
  \bibinfo {pages} {241101} (\bibinfo {year} {2011})},\ \Eprint
  {https://arxiv.org/abs/0909.2867} {arXiv:0909.2867 [gr-qc]} \BibitemShut
  {NoStop}%
\bibitem [{\citenamefont {Santamaria}\ \emph {et~al.}(2010)\citenamefont
  {Santamaria} \emph {et~al.}}]{Santamaria:2010yb}%
  \BibitemOpen
  \bibfield  {author} {\bibinfo {author} {\bibfnamefont {L.}~\bibnamefont
  {Santamaria}} \emph {et~al.},\ }\bibfield  {title} {\bibinfo {title}
  {{Matching post-Newtonian and numerical relativity waveforms: systematic
  errors and a new phenomenological model for non-precessing black hole
  binaries}},\ }\href {https://doi.org/10.1103/PhysRevD.82.064016} {\bibfield
  {journal} {\bibinfo  {journal} {Phys. Rev. D}\ }\textbf {\bibinfo {volume}
  {82}},\ \bibinfo {pages} {064016} (\bibinfo {year} {2010})},\ \Eprint
  {https://arxiv.org/abs/1005.3306} {arXiv:1005.3306 [gr-qc]} \BibitemShut
  {NoStop}%
\bibitem [{\citenamefont {Hannam}\ \emph {et~al.}(2014)\citenamefont {Hannam},
  \citenamefont {Schmidt}, \citenamefont {Boh\'e}, \citenamefont {Haegel},
  \citenamefont {Husa}, \citenamefont {Ohme}, \citenamefont {Pratten},\ and\
  \citenamefont {P\"urrer}}]{Hannam:2013oca}%
  \BibitemOpen
  \bibfield  {author} {\bibinfo {author} {\bibfnamefont {M.}~\bibnamefont
  {Hannam}}, \bibinfo {author} {\bibfnamefont {P.}~\bibnamefont {Schmidt}},
  \bibinfo {author} {\bibfnamefont {A.}~\bibnamefont {Boh\'e}}, \bibinfo
  {author} {\bibfnamefont {L.}~\bibnamefont {Haegel}}, \bibinfo {author}
  {\bibfnamefont {S.}~\bibnamefont {Husa}}, \bibinfo {author} {\bibfnamefont
  {F.}~\bibnamefont {Ohme}}, \bibinfo {author} {\bibfnamefont {G.}~\bibnamefont
  {Pratten}},\ and\ \bibinfo {author} {\bibfnamefont {M.}~\bibnamefont
  {P\"urrer}},\ }\bibfield  {title} {\bibinfo {title} {{Simple Model of
  Complete Precessing Black-Hole-Binary Gravitational Waveforms}},\ }\href
  {https://doi.org/10.1103/PhysRevLett.113.151101} {\bibfield  {journal}
  {\bibinfo  {journal} {Phys. Rev. Lett.}\ }\textbf {\bibinfo {volume} {113}},\
  \bibinfo {pages} {151101} (\bibinfo {year} {2014})},\ \Eprint
  {https://arxiv.org/abs/1308.3271} {arXiv:1308.3271 [gr-qc]} \BibitemShut
  {NoStop}%
\bibitem [{\citenamefont {Pratten}\ \emph
  {et~al.}(2020{\natexlab{a}})\citenamefont {Pratten}, \citenamefont {Husa},
  \citenamefont {Garcia-Quiros}, \citenamefont {Colleoni}, \citenamefont
  {Ramos-Buades}, \citenamefont {Estelles},\ and\ \citenamefont
  {Jaume}}]{Pratten:2020fqn}%
  \BibitemOpen
  \bibfield  {author} {\bibinfo {author} {\bibfnamefont {G.}~\bibnamefont
  {Pratten}}, \bibinfo {author} {\bibfnamefont {S.}~\bibnamefont {Husa}},
  \bibinfo {author} {\bibfnamefont {C.}~\bibnamefont {Garcia-Quiros}}, \bibinfo
  {author} {\bibfnamefont {M.}~\bibnamefont {Colleoni}}, \bibinfo {author}
  {\bibfnamefont {A.}~\bibnamefont {Ramos-Buades}}, \bibinfo {author}
  {\bibfnamefont {H.}~\bibnamefont {Estelles}},\ and\ \bibinfo {author}
  {\bibfnamefont {R.}~\bibnamefont {Jaume}},\ }\bibfield  {title} {\bibinfo
  {title} {{Setting the cornerstone for a family of models for gravitational
  waves from compact binaries: The dominant harmonic for nonprecessing
  quasicircular black holes}},\ }\href
  {https://doi.org/10.1103/PhysRevD.102.064001} {\bibfield  {journal} {\bibinfo
   {journal} {Phys. Rev. D}\ }\textbf {\bibinfo {volume} {102}},\ \bibinfo
  {pages} {064001} (\bibinfo {year} {2020}{\natexlab{a}})},\ \Eprint
  {https://arxiv.org/abs/2001.11412} {arXiv:2001.11412 [gr-qc]} \BibitemShut
  {NoStop}%
\bibitem [{\citenamefont {Pratten}\ \emph {et~al.}(2021)\citenamefont {Pratten}
  \emph {et~al.}}]{Pratten:2020ceb}%
  \BibitemOpen
  \bibfield  {author} {\bibinfo {author} {\bibfnamefont {G.}~\bibnamefont
  {Pratten}} \emph {et~al.},\ }\bibfield  {title} {\bibinfo {title}
  {{Computationally efficient models for the dominant and subdominant harmonic
  modes of precessing binary black holes}},\ }\href
  {https://doi.org/10.1103/PhysRevD.103.104056} {\bibfield  {journal} {\bibinfo
   {journal} {Phys. Rev. D}\ }\textbf {\bibinfo {volume} {103}},\ \bibinfo
  {pages} {104056} (\bibinfo {year} {2021})},\ \Eprint
  {https://arxiv.org/abs/2004.06503} {arXiv:2004.06503 [gr-qc]} \BibitemShut
  {NoStop}%
\bibitem [{\citenamefont {Garc\'\i{}a-Quir\'os}\ \emph
  {et~al.}(2020)\citenamefont {Garc\'\i{}a-Quir\'os}, \citenamefont {Colleoni},
  \citenamefont {Husa}, \citenamefont {Estell\'es}, \citenamefont {Pratten},
  \citenamefont {Ramos-Buades}, \citenamefont {Mateu-Lucena},\ and\
  \citenamefont {Jaume}}]{Garcia-Quiros:2020qpx}%
  \BibitemOpen
  \bibfield  {author} {\bibinfo {author} {\bibfnamefont {C.}~\bibnamefont
  {Garc\'\i{}a-Quir\'os}}, \bibinfo {author} {\bibfnamefont {M.}~\bibnamefont
  {Colleoni}}, \bibinfo {author} {\bibfnamefont {S.}~\bibnamefont {Husa}},
  \bibinfo {author} {\bibfnamefont {H.}~\bibnamefont {Estell\'es}}, \bibinfo
  {author} {\bibfnamefont {G.}~\bibnamefont {Pratten}}, \bibinfo {author}
  {\bibfnamefont {A.}~\bibnamefont {Ramos-Buades}}, \bibinfo {author}
  {\bibfnamefont {M.}~\bibnamefont {Mateu-Lucena}},\ and\ \bibinfo {author}
  {\bibfnamefont {R.}~\bibnamefont {Jaume}},\ }\bibfield  {title} {\bibinfo
  {title} {{Multimode frequency-domain model for the gravitational wave signal
  from nonprecessing black-hole binaries}},\ }\href
  {https://doi.org/10.1103/PhysRevD.102.064002} {\bibfield  {journal} {\bibinfo
   {journal} {Phys. Rev. D}\ }\textbf {\bibinfo {volume} {102}},\ \bibinfo
  {pages} {064002} (\bibinfo {year} {2020})},\ \Eprint
  {https://arxiv.org/abs/2001.10914} {arXiv:2001.10914 [gr-qc]} \BibitemShut
  {NoStop}%
\bibitem [{\citenamefont {Thompson}\ \emph {et~al.}(2024)\citenamefont
  {Thompson}, \citenamefont {Hamilton}, \citenamefont {London}, \citenamefont
  {Ghosh}, \citenamefont {Kolitsidou}, \citenamefont {Hoy},\ and\ \citenamefont
  {Hannam}}]{Thompson:2023ase}%
  \BibitemOpen
  \bibfield  {author} {\bibinfo {author} {\bibfnamefont {J.~E.}\ \bibnamefont
  {Thompson}}, \bibinfo {author} {\bibfnamefont {E.}~\bibnamefont {Hamilton}},
  \bibinfo {author} {\bibfnamefont {L.}~\bibnamefont {London}}, \bibinfo
  {author} {\bibfnamefont {S.}~\bibnamefont {Ghosh}}, \bibinfo {author}
  {\bibfnamefont {P.}~\bibnamefont {Kolitsidou}}, \bibinfo {author}
  {\bibfnamefont {C.}~\bibnamefont {Hoy}},\ and\ \bibinfo {author}
  {\bibfnamefont {M.}~\bibnamefont {Hannam}},\ }\bibfield  {title} {\bibinfo
  {title} {{PhenomXO4a: a phenomenological gravitational-wave model for
  precessing black-hole binaries with higher multipoles and asymmetries}},\
  }\href {https://doi.org/10.1103/PhysRevD.109.063012} {\bibfield  {journal}
  {\bibinfo  {journal} {Phys. Rev. D}\ }\textbf {\bibinfo {volume} {109}},\
  \bibinfo {pages} {063012} (\bibinfo {year} {2024})},\ \Eprint
  {https://arxiv.org/abs/2312.10025} {arXiv:2312.10025 [gr-qc]} \BibitemShut
  {NoStop}%
\bibitem [{\citenamefont {Thompson}\ \emph {et~al.}(2020)\citenamefont
  {Thompson}, \citenamefont {Fauchon-Jones}, \citenamefont {Khan},
  \citenamefont {Nitoglia}, \citenamefont {Pannarale}, \citenamefont
  {Dietrich},\ and\ \citenamefont {Hannam}}]{Thompson:2020nei}%
  \BibitemOpen
  \bibfield  {author} {\bibinfo {author} {\bibfnamefont {J.~E.}\ \bibnamefont
  {Thompson}}, \bibinfo {author} {\bibfnamefont {E.}~\bibnamefont
  {Fauchon-Jones}}, \bibinfo {author} {\bibfnamefont {S.}~\bibnamefont {Khan}},
  \bibinfo {author} {\bibfnamefont {E.}~\bibnamefont {Nitoglia}}, \bibinfo
  {author} {\bibfnamefont {F.}~\bibnamefont {Pannarale}}, \bibinfo {author}
  {\bibfnamefont {T.}~\bibnamefont {Dietrich}},\ and\ \bibinfo {author}
  {\bibfnamefont {M.}~\bibnamefont {Hannam}},\ }\bibfield  {title} {\bibinfo
  {title} {{Modeling the gravitational wave signature of neutron star black
  hole coalescences}},\ }\href {https://doi.org/10.1103/PhysRevD.101.124059}
  {\bibfield  {journal} {\bibinfo  {journal} {Phys. Rev. D}\ }\textbf {\bibinfo
  {volume} {101}},\ \bibinfo {pages} {124059} (\bibinfo {year} {2020})},\
  \Eprint {https://arxiv.org/abs/2002.08383} {arXiv:2002.08383 [gr-qc]}
  \BibitemShut {NoStop}%
\bibitem [{\citenamefont {Matas}\ \emph {et~al.}(2020)\citenamefont {Matas}
  \emph {et~al.}}]{Matas:2020wab}%
  \BibitemOpen
  \bibfield  {author} {\bibinfo {author} {\bibfnamefont {A.}~\bibnamefont
  {Matas}} \emph {et~al.},\ }\bibfield  {title} {\bibinfo {title}
  {{Aligned-spin neutron-star\textendash{}black-hole waveform model based on
  the effective-one-body approach and numerical-relativity simulations}},\
  }\href {https://doi.org/10.1103/PhysRevD.102.043023} {\bibfield  {journal}
  {\bibinfo  {journal} {Phys. Rev. D}\ }\textbf {\bibinfo {volume} {102}},\
  \bibinfo {pages} {043023} (\bibinfo {year} {2020})},\ \Eprint
  {https://arxiv.org/abs/2004.10001} {arXiv:2004.10001 [gr-qc]} \BibitemShut
  {NoStop}%
\bibitem [{\citenamefont {Finn}\ and\ \citenamefont
  {Chernoff}(1993)}]{Finn:1992xs}%
  \BibitemOpen
  \bibfield  {author} {\bibinfo {author} {\bibfnamefont {L.~S.}\ \bibnamefont
  {Finn}}\ and\ \bibinfo {author} {\bibfnamefont {D.~F.}\ \bibnamefont
  {Chernoff}},\ }\bibfield  {title} {\bibinfo {title} {{Observing binary
  inspiral in gravitational radiation: One interferometer}},\ }\href
  {https://doi.org/10.1103/PhysRevD.47.2198} {\bibfield  {journal} {\bibinfo
  {journal} {Phys. Rev. D}\ }\textbf {\bibinfo {volume} {47}},\ \bibinfo
  {pages} {2198} (\bibinfo {year} {1993})},\ \Eprint
  {https://arxiv.org/abs/gr-qc/9301003} {arXiv:gr-qc/9301003} \BibitemShut
  {NoStop}%
\bibitem [{\citenamefont {Field}\ \emph {et~al.}(2014)\citenamefont {Field},
  \citenamefont {Galley}, \citenamefont {Hesthaven}, \citenamefont {Kaye},\
  and\ \citenamefont {Tiglio}}]{Field:2013cfa}%
  \BibitemOpen
  \bibfield  {author} {\bibinfo {author} {\bibfnamefont {S.~E.}\ \bibnamefont
  {Field}}, \bibinfo {author} {\bibfnamefont {C.~R.}\ \bibnamefont {Galley}},
  \bibinfo {author} {\bibfnamefont {J.~S.}\ \bibnamefont {Hesthaven}}, \bibinfo
  {author} {\bibfnamefont {J.}~\bibnamefont {Kaye}},\ and\ \bibinfo {author}
  {\bibfnamefont {M.}~\bibnamefont {Tiglio}},\ }\bibfield  {title} {\bibinfo
  {title} {{Fast prediction and evaluation of gravitational waveforms using
  surrogate models}},\ }\href {https://doi.org/10.1103/PhysRevX.4.031006}
  {\bibfield  {journal} {\bibinfo  {journal} {Phys. Rev. X}\ }\textbf {\bibinfo
  {volume} {4}},\ \bibinfo {pages} {031006} (\bibinfo {year} {2014})},\ \Eprint
  {https://arxiv.org/abs/1308.3565} {arXiv:1308.3565 [gr-qc]} \BibitemShut
  {NoStop}%
\bibitem [{\citenamefont {Cornish}(2010)}]{Cornish:2010kf}%
  \BibitemOpen
  \bibfield  {author} {\bibinfo {author} {\bibfnamefont {N.~J.}\ \bibnamefont
  {Cornish}},\ }\bibfield  {title} {\bibinfo {title} {{Fast Fisher Matrices and
  Lazy Likelihoods}},\ }\href@noop {} {\  (\bibinfo {year} {2010})},\ \Eprint
  {https://arxiv.org/abs/1007.4820} {arXiv:1007.4820 [gr-qc]} \BibitemShut
  {NoStop}%
\bibitem [{\citenamefont {Zackay}\ \emph {et~al.}(2018)\citenamefont {Zackay},
  \citenamefont {Dai},\ and\ \citenamefont {Venumadhav}}]{Zackay:2018qdy}%
  \BibitemOpen
  \bibfield  {author} {\bibinfo {author} {\bibfnamefont {B.}~\bibnamefont
  {Zackay}}, \bibinfo {author} {\bibfnamefont {L.}~\bibnamefont {Dai}},\ and\
  \bibinfo {author} {\bibfnamefont {T.}~\bibnamefont {Venumadhav}},\ }\bibfield
   {title} {\bibinfo {title} {{Relative Binning and Fast Likelihood Evaluation
  for Gravitational Wave Parameter Estimation}},\ }\href@noop {} {\  (\bibinfo
  {year} {2018})},\ \Eprint {https://arxiv.org/abs/1806.08792}
  {arXiv:1806.08792 [astro-ph.IM]} \BibitemShut {NoStop}%
\bibitem [{\citenamefont {Finstad}\ and\ \citenamefont
  {Brown}(2020)}]{Finstad:2020sok}%
  \BibitemOpen
  \bibfield  {author} {\bibinfo {author} {\bibfnamefont {D.}~\bibnamefont
  {Finstad}}\ and\ \bibinfo {author} {\bibfnamefont {D.~A.}\ \bibnamefont
  {Brown}},\ }\bibfield  {title} {\bibinfo {title} {{Fast Parameter Estimation
  of Binary Mergers for Multimessenger Follow-up}},\ }\href
  {https://doi.org/10.3847/2041-8213/abca9e} {\bibfield  {journal} {\bibinfo
  {journal} {Astrophys. J. Lett.}\ }\textbf {\bibinfo {volume} {905}},\
  \bibinfo {pages} {L9} (\bibinfo {year} {2020})},\ \Eprint
  {https://arxiv.org/abs/2009.13759} {arXiv:2009.13759 [astro-ph.IM]}
  \BibitemShut {NoStop}%
\bibitem [{\citenamefont {Cornish}(2021)}]{Cornish:2021lje}%
  \BibitemOpen
  \bibfield  {author} {\bibinfo {author} {\bibfnamefont {N.~J.}\ \bibnamefont
  {Cornish}},\ }\bibfield  {title} {\bibinfo {title} {{Heterodyned likelihood
  for rapid gravitational wave parameter inference}},\ }\href
  {https://doi.org/10.1103/PhysRevD.104.104054} {\bibfield  {journal} {\bibinfo
   {journal} {Phys. Rev. D}\ }\textbf {\bibinfo {volume} {104}},\ \bibinfo
  {pages} {104054} (\bibinfo {year} {2021})},\ \Eprint
  {https://arxiv.org/abs/2109.02728} {arXiv:2109.02728 [gr-qc]} \BibitemShut
  {NoStop}%
\bibitem [{\citenamefont {Leslie}\ \emph {et~al.}(2021)\citenamefont {Leslie},
  \citenamefont {Dai},\ and\ \citenamefont {Pratten}}]{Leslie:2021ssu}%
  \BibitemOpen
  \bibfield  {author} {\bibinfo {author} {\bibfnamefont {N.}~\bibnamefont
  {Leslie}}, \bibinfo {author} {\bibfnamefont {L.}~\bibnamefont {Dai}},\ and\
  \bibinfo {author} {\bibfnamefont {G.}~\bibnamefont {Pratten}},\ }\bibfield
  {title} {\bibinfo {title} {{Mode-by-mode relative binning: Fast likelihood
  estimation for gravitational waveforms with spin-orbit precession and
  multiple harmonics}},\ }\href {https://doi.org/10.1103/PhysRevD.104.123030}
  {\bibfield  {journal} {\bibinfo  {journal} {Phys. Rev. D}\ }\textbf {\bibinfo
  {volume} {104}},\ \bibinfo {pages} {123030} (\bibinfo {year} {2021})},\
  \Eprint {https://arxiv.org/abs/2109.09872} {arXiv:2109.09872 [astro-ph.IM]}
  \BibitemShut {NoStop}%
\bibitem [{\citenamefont {Vinciguerra}\ \emph {et~al.}(2017)\citenamefont
  {Vinciguerra}, \citenamefont {Veitch},\ and\ \citenamefont
  {Mandel}}]{Vinciguerra:2017ngf}%
  \BibitemOpen
  \bibfield  {author} {\bibinfo {author} {\bibfnamefont {S.}~\bibnamefont
  {Vinciguerra}}, \bibinfo {author} {\bibfnamefont {J.}~\bibnamefont
  {Veitch}},\ and\ \bibinfo {author} {\bibfnamefont {I.}~\bibnamefont
  {Mandel}},\ }\bibfield  {title} {\bibinfo {title} {{Accelerating
  gravitational wave parameter estimation with multi-band template
  interpolation}},\ }\href {https://doi.org/10.1088/1361-6382/aa6d44}
  {\bibfield  {journal} {\bibinfo  {journal} {Class. Quant. Grav.}\ }\textbf
  {\bibinfo {volume} {34}},\ \bibinfo {pages} {115006} (\bibinfo {year}
  {2017})},\ \Eprint {https://arxiv.org/abs/1703.02062} {arXiv:1703.02062
  [gr-qc]} \BibitemShut {NoStop}%
\bibitem [{\citenamefont {Morisaki}(2021)}]{Morisaki:2021ngj}%
  \BibitemOpen
  \bibfield  {author} {\bibinfo {author} {\bibfnamefont {S.}~\bibnamefont
  {Morisaki}},\ }\bibfield  {title} {\bibinfo {title} {{Accelerating parameter
  estimation of gravitational waves from compact binary coalescence using
  adaptive frequency resolutions}},\ }\href
  {https://doi.org/10.1103/PhysRevD.104.044062} {\bibfield  {journal} {\bibinfo
   {journal} {Phys. Rev. D}\ }\textbf {\bibinfo {volume} {104}},\ \bibinfo
  {pages} {044062} (\bibinfo {year} {2021})},\ \Eprint
  {https://arxiv.org/abs/2104.07813} {arXiv:2104.07813 [gr-qc]} \BibitemShut
  {NoStop}%
\bibitem [{\citenamefont {Antil}\ \emph {et~al.}(2013)\citenamefont {Antil},
  \citenamefont {Field}, \citenamefont {Herrmann}, \citenamefont {Nochetto},\
  and\ \citenamefont {Tiglio}}]{Antil:2012wf}%
  \BibitemOpen
  \bibfield  {author} {\bibinfo {author} {\bibfnamefont {H.}~\bibnamefont
  {Antil}}, \bibinfo {author} {\bibfnamefont {S.~E.}\ \bibnamefont {Field}},
  \bibinfo {author} {\bibfnamefont {F.}~\bibnamefont {Herrmann}}, \bibinfo
  {author} {\bibfnamefont {R.~H.}\ \bibnamefont {Nochetto}},\ and\ \bibinfo
  {author} {\bibfnamefont {M.}~\bibnamefont {Tiglio}},\ }\bibfield  {title}
  {\bibinfo {title} {{Two-step greedy algorithm for reduced order
  quadratures}},\ }\href {https://doi.org/10.1007/s10915-013-9722-z} {\bibfield
   {journal} {\bibinfo  {journal} {J. Sci. Comput.}\ }\textbf {\bibinfo
  {volume} {57}},\ \bibinfo {pages} {604} (\bibinfo {year} {2013})},\ \Eprint
  {https://arxiv.org/abs/1210.0577} {arXiv:1210.0577 [cs.NA]} \BibitemShut
  {NoStop}%
\bibitem [{\citenamefont {Canizares}\ \emph {et~al.}(2013)\citenamefont
  {Canizares}, \citenamefont {Field}, \citenamefont {Gair},\ and\ \citenamefont
  {Tiglio}}]{Canizares:2013ywa}%
  \BibitemOpen
  \bibfield  {author} {\bibinfo {author} {\bibfnamefont {P.}~\bibnamefont
  {Canizares}}, \bibinfo {author} {\bibfnamefont {S.~E.}\ \bibnamefont
  {Field}}, \bibinfo {author} {\bibfnamefont {J.~R.}\ \bibnamefont {Gair}},\
  and\ \bibinfo {author} {\bibfnamefont {M.}~\bibnamefont {Tiglio}},\
  }\bibfield  {title} {\bibinfo {title} {{Gravitational wave parameter
  estimation with compressed likelihood evaluations}},\ }\href
  {https://doi.org/10.1103/PhysRevD.87.124005} {\bibfield  {journal} {\bibinfo
  {journal} {Phys. Rev. D}\ }\textbf {\bibinfo {volume} {87}},\ \bibinfo
  {pages} {124005} (\bibinfo {year} {2013})},\ \Eprint
  {https://arxiv.org/abs/1304.0462} {arXiv:1304.0462 [gr-qc]} \BibitemShut
  {NoStop}%
\bibitem [{\citenamefont {Canizares}\ \emph {et~al.}(2015)\citenamefont
  {Canizares}, \citenamefont {Field}, \citenamefont {Gair}, \citenamefont
  {Raymond}, \citenamefont {Smith},\ and\ \citenamefont
  {Tiglio}}]{Canizares:2014fya}%
  \BibitemOpen
  \bibfield  {author} {\bibinfo {author} {\bibfnamefont {P.}~\bibnamefont
  {Canizares}}, \bibinfo {author} {\bibfnamefont {S.~E.}\ \bibnamefont
  {Field}}, \bibinfo {author} {\bibfnamefont {J.}~\bibnamefont {Gair}},
  \bibinfo {author} {\bibfnamefont {V.}~\bibnamefont {Raymond}}, \bibinfo
  {author} {\bibfnamefont {R.}~\bibnamefont {Smith}},\ and\ \bibinfo {author}
  {\bibfnamefont {M.}~\bibnamefont {Tiglio}},\ }\bibfield  {title} {\bibinfo
  {title} {{Accelerated gravitational-wave parameter estimation with reduced
  order modeling}},\ }\href {https://doi.org/10.1103/PhysRevLett.114.071104}
  {\bibfield  {journal} {\bibinfo  {journal} {Phys. Rev. Lett.}\ }\textbf
  {\bibinfo {volume} {114}},\ \bibinfo {pages} {071104} (\bibinfo {year}
  {2015})},\ \Eprint {https://arxiv.org/abs/1404.6284} {arXiv:1404.6284
  [gr-qc]} \BibitemShut {NoStop}%
\bibitem [{\citenamefont {Smith}\ \emph {et~al.}(2016)\citenamefont {Smith},
  \citenamefont {Field}, \citenamefont {Blackburn}, \citenamefont {Haster},
  \citenamefont {P\"urrer}, \citenamefont {Raymond},\ and\ \citenamefont
  {Schmidt}}]{Smith:2016qas}%
  \BibitemOpen
  \bibfield  {author} {\bibinfo {author} {\bibfnamefont {R.}~\bibnamefont
  {Smith}}, \bibinfo {author} {\bibfnamefont {S.~E.}\ \bibnamefont {Field}},
  \bibinfo {author} {\bibfnamefont {K.}~\bibnamefont {Blackburn}}, \bibinfo
  {author} {\bibfnamefont {C.-J.}\ \bibnamefont {Haster}}, \bibinfo {author}
  {\bibfnamefont {M.}~\bibnamefont {P\"urrer}}, \bibinfo {author}
  {\bibfnamefont {V.}~\bibnamefont {Raymond}},\ and\ \bibinfo {author}
  {\bibfnamefont {P.}~\bibnamefont {Schmidt}},\ }\bibfield  {title} {\bibinfo
  {title} {{Fast and accurate inference on gravitational waves from precessing
  compact binaries}},\ }\href {https://doi.org/10.1103/PhysRevD.94.044031}
  {\bibfield  {journal} {\bibinfo  {journal} {Phys. Rev. D}\ }\textbf {\bibinfo
  {volume} {94}},\ \bibinfo {pages} {044031} (\bibinfo {year} {2016})},\
  \Eprint {https://arxiv.org/abs/1604.08253} {arXiv:1604.08253 [gr-qc]}
  \BibitemShut {NoStop}%
\bibitem [{\citenamefont {Yagi}\ and\ \citenamefont
  {Yunes}(2017{\natexlab{a}})}]{Yagi:2016qmr}%
  \BibitemOpen
  \bibfield  {author} {\bibinfo {author} {\bibfnamefont {K.}~\bibnamefont
  {Yagi}}\ and\ \bibinfo {author} {\bibfnamefont {N.}~\bibnamefont {Yunes}},\
  }\bibfield  {title} {\bibinfo {title} {{Approximate Universal Relations among
  Tidal Parameters for Neutron Star Binaries}},\ }\href
  {https://doi.org/10.1088/1361-6382/34/1/015006} {\bibfield  {journal}
  {\bibinfo  {journal} {Class. Quant. Grav.}\ }\textbf {\bibinfo {volume}
  {34}},\ \bibinfo {pages} {015006} (\bibinfo {year} {2017}{\natexlab{a}})},\
  \Eprint {https://arxiv.org/abs/1608.06187} {arXiv:1608.06187 [gr-qc]}
  \BibitemShut {NoStop}%
\bibitem [{\citenamefont {Raithel}\ and\ \citenamefont
  {Most}(2022)}]{Raithel:2022orm}%
  \BibitemOpen
  \bibfield  {author} {\bibinfo {author} {\bibfnamefont {C.~A.}\ \bibnamefont
  {Raithel}}\ and\ \bibinfo {author} {\bibfnamefont {E.~R.}\ \bibnamefont
  {Most}},\ }\bibfield  {title} {\bibinfo {title} {{Characterizing the
  Breakdown of Quasi-universality in Postmerger Gravitational Waves from Binary
  Neutron Star Mergers}},\ }\href {https://doi.org/10.3847/2041-8213/ac7c75}
  {\bibfield  {journal} {\bibinfo  {journal} {Astrophys. J. Lett.}\ }\textbf
  {\bibinfo {volume} {933}},\ \bibinfo {pages} {L39} (\bibinfo {year}
  {2022})},\ \Eprint {https://arxiv.org/abs/2201.03594} {arXiv:2201.03594
  [astro-ph.HE]} \BibitemShut {NoStop}%
\bibitem [{\citenamefont {Favata}(2014)}]{Favata:2013rwa}%
  \BibitemOpen
  \bibfield  {author} {\bibinfo {author} {\bibfnamefont {M.}~\bibnamefont
  {Favata}},\ }\bibfield  {title} {\bibinfo {title} {{Systematic parameter
  errors in inspiraling neutron star binaries}},\ }\href
  {https://doi.org/10.1103/PhysRevLett.112.101101} {\bibfield  {journal}
  {\bibinfo  {journal} {Phys. Rev. Lett.}\ }\textbf {\bibinfo {volume} {112}},\
  \bibinfo {pages} {101101} (\bibinfo {year} {2014})},\ \Eprint
  {https://arxiv.org/abs/1310.8288} {arXiv:1310.8288 [gr-qc]} \BibitemShut
  {NoStop}%
\bibitem [{\citenamefont {Hotokezaka}\ \emph {et~al.}(2016)\citenamefont
  {Hotokezaka}, \citenamefont {Kyutoku}, \citenamefont {Sekiguchi},\ and\
  \citenamefont {Shibata}}]{Hotokezaka:2016bzh}%
  \BibitemOpen
  \bibfield  {author} {\bibinfo {author} {\bibfnamefont {K.}~\bibnamefont
  {Hotokezaka}}, \bibinfo {author} {\bibfnamefont {K.}~\bibnamefont {Kyutoku}},
  \bibinfo {author} {\bibfnamefont {Y.-i.}\ \bibnamefont {Sekiguchi}},\ and\
  \bibinfo {author} {\bibfnamefont {M.}~\bibnamefont {Shibata}},\ }\bibfield
  {title} {\bibinfo {title} {{Measurability of the tidal deformability by
  gravitational waves from coalescing binary neutron stars}},\ }\href
  {https://doi.org/10.1103/PhysRevD.93.064082} {\bibfield  {journal} {\bibinfo
  {journal} {Phys. Rev. D}\ }\textbf {\bibinfo {volume} {93}},\ \bibinfo
  {pages} {064082} (\bibinfo {year} {2016})},\ \Eprint
  {https://arxiv.org/abs/1603.01286} {arXiv:1603.01286 [gr-qc]} \BibitemShut
  {NoStop}%
\bibitem [{\citenamefont {Dietrich}\ \emph
  {et~al.}(2019{\natexlab{b}})\citenamefont {Dietrich} \emph
  {et~al.}}]{Dietrich:2018uni}%
  \BibitemOpen
  \bibfield  {author} {\bibinfo {author} {\bibfnamefont {T.}~\bibnamefont
  {Dietrich}} \emph {et~al.},\ }\bibfield  {title} {\bibinfo {title} {{Matter
  imprints in waveform models for neutron star binaries: Tidal and self-spin
  effects}},\ }\href {https://doi.org/10.1103/PhysRevD.99.024029} {\bibfield
  {journal} {\bibinfo  {journal} {Phys. Rev. D}\ }\textbf {\bibinfo {volume}
  {99}},\ \bibinfo {pages} {024029} (\bibinfo {year} {2019}{\natexlab{b}})},\
  \Eprint {https://arxiv.org/abs/1804.02235} {arXiv:1804.02235 [gr-qc]}
  \BibitemShut {NoStop}%
\bibitem [{\citenamefont {Nagar}\ \emph {et~al.}(2019)\citenamefont {Nagar},
  \citenamefont {Messina}, \citenamefont {Rettegno}, \citenamefont {Bini},
  \citenamefont {Damour}, \citenamefont {Geralico}, \citenamefont {Akcay},\
  and\ \citenamefont {Bernuzzi}}]{Nagar:2018plt}%
  \BibitemOpen
  \bibfield  {author} {\bibinfo {author} {\bibfnamefont {A.}~\bibnamefont
  {Nagar}}, \bibinfo {author} {\bibfnamefont {F.}~\bibnamefont {Messina}},
  \bibinfo {author} {\bibfnamefont {P.}~\bibnamefont {Rettegno}}, \bibinfo
  {author} {\bibfnamefont {D.}~\bibnamefont {Bini}}, \bibinfo {author}
  {\bibfnamefont {T.}~\bibnamefont {Damour}}, \bibinfo {author} {\bibfnamefont
  {A.}~\bibnamefont {Geralico}}, \bibinfo {author} {\bibfnamefont
  {S.}~\bibnamefont {Akcay}},\ and\ \bibinfo {author} {\bibfnamefont
  {S.}~\bibnamefont {Bernuzzi}},\ }\bibfield  {title} {\bibinfo {title}
  {{Nonlinear-in-spin effects in effective-one-body waveform models of
  spin-aligned, inspiralling, neutron star binaries}},\ }\href
  {https://doi.org/10.1103/PhysRevD.99.044007} {\bibfield  {journal} {\bibinfo
  {journal} {Phys. Rev. D}\ }\textbf {\bibinfo {volume} {99}},\ \bibinfo
  {pages} {044007} (\bibinfo {year} {2019})},\ \Eprint
  {https://arxiv.org/abs/1812.07923} {arXiv:1812.07923 [gr-qc]} \BibitemShut
  {NoStop}%
\bibitem [{\citenamefont {Bini}\ and\ \citenamefont
  {Damour}(2014)}]{Bini:2014zxa}%
  \BibitemOpen
  \bibfield  {author} {\bibinfo {author} {\bibfnamefont {D.}~\bibnamefont
  {Bini}}\ and\ \bibinfo {author} {\bibfnamefont {T.}~\bibnamefont {Damour}},\
  }\bibfield  {title} {\bibinfo {title} {{Gravitational self-force corrections
  to two-body tidal interactions and the effective one-body formalism}},\
  }\href {https://doi.org/10.1103/PhysRevD.90.124037} {\bibfield  {journal}
  {\bibinfo  {journal} {Phys. Rev. D}\ }\textbf {\bibinfo {volume} {90}},\
  \bibinfo {pages} {124037} (\bibinfo {year} {2014})},\ \Eprint
  {https://arxiv.org/abs/1409.6933} {arXiv:1409.6933 [gr-qc]} \BibitemShut
  {NoStop}%
\bibitem [{\citenamefont {Damour}\ and\ \citenamefont
  {Nagar}(2009)}]{Damour:2009vw}%
  \BibitemOpen
  \bibfield  {author} {\bibinfo {author} {\bibfnamefont {T.}~\bibnamefont
  {Damour}}\ and\ \bibinfo {author} {\bibfnamefont {A.}~\bibnamefont {Nagar}},\
  }\bibfield  {title} {\bibinfo {title} {{Relativistic tidal properties of
  neutron stars}},\ }\href {https://doi.org/10.1103/PhysRevD.80.084035}
  {\bibfield  {journal} {\bibinfo  {journal} {\prd}\ }\textbf {\bibinfo
  {volume} {80}},\ \bibinfo {pages} {084035} (\bibinfo {year} {2009})},\
  \Eprint {https://arxiv.org/abs/0906.0096} {arXiv:0906.0096 [gr-qc]}
  \BibitemShut {NoStop}%
%%CITATION = ARXIV:0906.0096;%%
\bibitem [{\citenamefont {{Provost}}\ \emph {et~al.}(1981)\citenamefont
  {{Provost}}, \citenamefont {{Berthomieu}},\ and\ \citenamefont
  {{Rocca}}}]{Provost:1981lfo}%
  \BibitemOpen
  \bibfield  {author} {\bibinfo {author} {\bibfnamefont {J.}~\bibnamefont
  {{Provost}}}, \bibinfo {author} {\bibfnamefont {G.}~\bibnamefont
  {{Berthomieu}}},\ and\ \bibinfo {author} {\bibfnamefont {A.}~\bibnamefont
  {{Rocca}}},\ }\bibfield  {title} {\bibinfo {title} {{Low Frequency
  Oscillations of a Slowly Rotating Star - Quasi Toroidal Modes}},\ }\href@noop
  {} {\bibfield  {journal} {\bibinfo  {journal} {A\&A}\ }\textbf {\bibinfo
  {volume} {94}},\ \bibinfo {pages} {126} (\bibinfo {year} {1981})}\BibitemShut
  {NoStop}%
\bibitem [{\citenamefont {Ho}\ and\ \citenamefont {Lai}(1999)}]{Ho:1998hq}%
  \BibitemOpen
  \bibfield  {author} {\bibinfo {author} {\bibfnamefont {W.~C.~G.}\
  \bibnamefont {Ho}}\ and\ \bibinfo {author} {\bibfnamefont {D.}~\bibnamefont
  {Lai}},\ }\bibfield  {title} {\bibinfo {title} {{Resonant tidal excitations
  of rotating neutron stars in coalescing binaries}},\ }\href
  {https://doi.org/10.1046/j.1365-8711.1999.02703.x} {\bibfield  {journal}
  {\bibinfo  {journal} {Mon. Not. Roy. Astron. Soc.}\ }\textbf {\bibinfo
  {volume} {308}},\ \bibinfo {pages} {153} (\bibinfo {year} {1999})},\ \Eprint
  {https://arxiv.org/abs/astro-ph/9812116} {arXiv:astro-ph/9812116}
  \BibitemShut {NoStop}%
\bibitem [{\citenamefont {Flanagan}\ and\ \citenamefont
  {Racine}(2007)}]{Flanagan:2006sb}%
  \BibitemOpen
  \bibfield  {author} {\bibinfo {author} {\bibfnamefont {E.~E.}\ \bibnamefont
  {Flanagan}}\ and\ \bibinfo {author} {\bibfnamefont {E.}~\bibnamefont
  {Racine}},\ }\bibfield  {title} {\bibinfo {title} {{Gravitomagnetic resonant
  excitation of Rossby modes in coalescing neutron star binaries}},\ }\href
  {https://doi.org/10.1103/PhysRevD.75.044001} {\bibfield  {journal} {\bibinfo
  {journal} {Phys. Rev. D}\ }\textbf {\bibinfo {volume} {75}},\ \bibinfo
  {pages} {044001} (\bibinfo {year} {2007})},\ \Eprint
  {https://arxiv.org/abs/gr-qc/0601029} {arXiv:gr-qc/0601029} \BibitemShut
  {NoStop}%
\bibitem [{\citenamefont {Gamba}\ \emph
  {et~al.}(2021{\natexlab{a}})\citenamefont {Gamba}, \citenamefont {Bernuzzi},\
  and\ \citenamefont {Nagar}}]{Gamba:2020ljo}%
  \BibitemOpen
  \bibfield  {author} {\bibinfo {author} {\bibfnamefont {R.}~\bibnamefont
  {Gamba}}, \bibinfo {author} {\bibfnamefont {S.}~\bibnamefont {Bernuzzi}},\
  and\ \bibinfo {author} {\bibfnamefont {A.}~\bibnamefont {Nagar}},\ }\bibfield
   {title} {\bibinfo {title} {{Fast, faithful, frequency-domain
  effective-one-body waveforms for compact binary coalescences}},\ }\href
  {https://doi.org/10.1103/PhysRevD.104.084058} {\bibfield  {journal} {\bibinfo
   {journal} {Phys. Rev. D}\ }\textbf {\bibinfo {volume} {104}},\ \bibinfo
  {pages} {084058} (\bibinfo {year} {2021}{\natexlab{a}})},\ \Eprint
  {https://arxiv.org/abs/2012.00027} {arXiv:2012.00027 [gr-qc]} \BibitemShut
  {NoStop}%
\bibitem [{\citenamefont {Dietrich}\ \emph {et~al.}(2018)\citenamefont
  {Dietrich}, \citenamefont {Radice}, \citenamefont {Bernuzzi}, \citenamefont
  {Zappa}, \citenamefont {Perego}, \citenamefont {Br\"ugmann}, \citenamefont
  {Chaurasia}, \citenamefont {Dudi}, \citenamefont {Tichy},\ and\ \citenamefont
  {Ujevic}}]{Dietrich:2018phi}%
  \BibitemOpen
  \bibfield  {author} {\bibinfo {author} {\bibfnamefont {T.}~\bibnamefont
  {Dietrich}}, \bibinfo {author} {\bibfnamefont {D.}~\bibnamefont {Radice}},
  \bibinfo {author} {\bibfnamefont {S.}~\bibnamefont {Bernuzzi}}, \bibinfo
  {author} {\bibfnamefont {F.}~\bibnamefont {Zappa}}, \bibinfo {author}
  {\bibfnamefont {A.}~\bibnamefont {Perego}}, \bibinfo {author} {\bibfnamefont
  {B.}~\bibnamefont {Br\"ugmann}}, \bibinfo {author} {\bibfnamefont {S.~V.}\
  \bibnamefont {Chaurasia}}, \bibinfo {author} {\bibfnamefont {R.}~\bibnamefont
  {Dudi}}, \bibinfo {author} {\bibfnamefont {W.}~\bibnamefont {Tichy}},\ and\
  \bibinfo {author} {\bibfnamefont {M.}~\bibnamefont {Ujevic}},\ }\bibfield
  {title} {\bibinfo {title} {{CoRe database of binary neutron star merger
  waveforms}},\ }\href {https://doi.org/10.1088/1361-6382/aaebc0} {\bibfield
  {journal} {\bibinfo  {journal} {Class. Quant. Grav.}\ }\textbf {\bibinfo
  {volume} {35}},\ \bibinfo {pages} {24LT01} (\bibinfo {year} {2018})},\
  \Eprint {https://arxiv.org/abs/1806.01625} {arXiv:1806.01625 [gr-qc]}
  \BibitemShut {NoStop}%
\bibitem [{\citenamefont {Gonzalez}\ \emph {et~al.}(2023)\citenamefont
  {Gonzalez} \emph {et~al.}}]{Gonzalez:2022mgo}%
  \BibitemOpen
  \bibfield  {author} {\bibinfo {author} {\bibfnamefont {A.}~\bibnamefont
  {Gonzalez}} \emph {et~al.},\ }\bibfield  {title} {\bibinfo {title} {{Second
  release of the CoRe database of binary neutron star merger waveforms}},\
  }\href {https://doi.org/10.1088/1361-6382/acc231} {\bibfield  {journal}
  {\bibinfo  {journal} {Class. Quant. Grav.}\ }\textbf {\bibinfo {volume}
  {40}},\ \bibinfo {pages} {085011} (\bibinfo {year} {2023})},\ \Eprint
  {https://arxiv.org/abs/2210.16366} {arXiv:2210.16366 [gr-qc]} \BibitemShut
  {NoStop}%
\bibitem [{\citenamefont {Bernuzzi}\ \emph
  {et~al.}(2024{\natexlab{a}})\citenamefont {Bernuzzi} \emph {et~al.}}]{Core}%
  \BibitemOpen
  \bibfield  {author} {\bibinfo {author} {\bibfnamefont {S.}~\bibnamefont
  {Bernuzzi}} \emph {et~al.},\ }\href {http://www.computational-relativity.org}
  {\bibinfo {title} {Core computational relativity}} (\bibinfo {year}
  {2024}{\natexlab{a}})\BibitemShut {NoStop}%
\bibitem [{\citenamefont {Dietrich}\ and\ \citenamefont
  {Hinderer}(2017)}]{Dietrich:2017feu}%
  \BibitemOpen
  \bibfield  {author} {\bibinfo {author} {\bibfnamefont {T.}~\bibnamefont
  {Dietrich}}\ and\ \bibinfo {author} {\bibfnamefont {T.}~\bibnamefont
  {Hinderer}},\ }\bibfield  {title} {\bibinfo {title} {{Comprehensive
  comparison of numerical relativity and effective-one-body results to inform
  improvements in waveform models for binary neutron star systems}},\ }\href
  {https://doi.org/10.1103/PhysRevD.95.124006} {\bibfield  {journal} {\bibinfo
  {journal} {Phys. Rev. D}\ }\textbf {\bibinfo {volume} {95}},\ \bibinfo
  {pages} {124006} (\bibinfo {year} {2017})},\ \Eprint
  {https://arxiv.org/abs/1702.02053} {arXiv:1702.02053 [gr-qc]} \BibitemShut
  {NoStop}%
\bibitem [{\citenamefont {Read}\ \emph {et~al.}(2009)\citenamefont {Read},
  \citenamefont {Lackey}, \citenamefont {Owen},\ and\ \citenamefont
  {Friedman}}]{Read:2008iy}%
  \BibitemOpen
  \bibfield  {author} {\bibinfo {author} {\bibfnamefont {J.~S.}\ \bibnamefont
  {Read}}, \bibinfo {author} {\bibfnamefont {B.~D.}\ \bibnamefont {Lackey}},
  \bibinfo {author} {\bibfnamefont {B.~J.}\ \bibnamefont {Owen}},\ and\
  \bibinfo {author} {\bibfnamefont {J.~L.}\ \bibnamefont {Friedman}},\
  }\bibfield  {title} {\bibinfo {title} {{Constraints on a phenomenologically
  parameterized neutron-star equation of state}},\ }\href
  {https://doi.org/10.1103/PhysRevD.79.124032} {\bibfield  {journal} {\bibinfo
  {journal} {Phys. Rev. D}\ }\textbf {\bibinfo {volume} {79}},\ \bibinfo
  {pages} {124032} (\bibinfo {year} {2009})},\ \Eprint
  {https://arxiv.org/abs/0812.2163} {arXiv:0812.2163 [astro-ph]} \BibitemShut
  {NoStop}%
\bibitem [{\citenamefont {Nakano}(2015)}]{Nakano:2015rda}%
  \BibitemOpen
  \bibfield  {author} {\bibinfo {author} {\bibfnamefont {H.}~\bibnamefont
  {Nakano}},\ }\bibfield  {title} {\bibinfo {title} {{A note on gravitational
  wave extraction from binary simulations}},\ }\href
  {https://doi.org/10.1088/0264-9381/32/17/177002} {\bibfield  {journal}
  {\bibinfo  {journal} {Class. Quant. Grav.}\ }\textbf {\bibinfo {volume}
  {32}},\ \bibinfo {pages} {177002} (\bibinfo {year} {2015})},\ \Eprint
  {https://arxiv.org/abs/1501.02890} {arXiv:1501.02890 [gr-qc]} \BibitemShut
  {NoStop}%
\bibitem [{\citenamefont {Nakano}\ \emph {et~al.}(2015)\citenamefont {Nakano},
  \citenamefont {Healy}, \citenamefont {Lousto},\ and\ \citenamefont
  {Zlochower}}]{Nakano:2015pta}%
  \BibitemOpen
  \bibfield  {author} {\bibinfo {author} {\bibfnamefont {H.}~\bibnamefont
  {Nakano}}, \bibinfo {author} {\bibfnamefont {J.}~\bibnamefont {Healy}},
  \bibinfo {author} {\bibfnamefont {C.~O.}\ \bibnamefont {Lousto}},\ and\
  \bibinfo {author} {\bibfnamefont {Y.}~\bibnamefont {Zlochower}},\ }\bibfield
  {title} {\bibinfo {title} {{Perturbative extraction of gravitational
  waveforms generated with Numerical Relativity}},\ }\href
  {https://doi.org/10.1103/PhysRevD.91.104022} {\bibfield  {journal} {\bibinfo
  {journal} {Phys. Rev. D}\ }\textbf {\bibinfo {volume} {91}},\ \bibinfo
  {pages} {104022} (\bibinfo {year} {2015})},\ \Eprint
  {https://arxiv.org/abs/1503.00718} {arXiv:1503.00718 [gr-qc]} \BibitemShut
  {NoStop}%
\bibitem [{\citenamefont {Arnowitt}\ \emph {et~al.}(2008)\citenamefont
  {Arnowitt}, \citenamefont {Deser},\ and\ \citenamefont
  {Misner}}]{Arnowitt:1962hi}%
  \BibitemOpen
  \bibfield  {author} {\bibinfo {author} {\bibfnamefont {R.~L.}\ \bibnamefont
  {Arnowitt}}, \bibinfo {author} {\bibfnamefont {S.}~\bibnamefont {Deser}},\
  and\ \bibinfo {author} {\bibfnamefont {C.~W.}\ \bibnamefont {Misner}},\
  }\bibfield  {title} {\bibinfo {title} {{The Dynamics of general
  relativity}},\ }\href {https://doi.org/10.1007/s10714-008-0661-1} {\bibfield
  {journal} {\bibinfo  {journal} {Gen. Rel. Grav.}\ }\textbf {\bibinfo {volume}
  {40}},\ \bibinfo {pages} {1997} (\bibinfo {year} {2008})},\ \Eprint
  {https://arxiv.org/abs/gr-qc/0405109} {arXiv:gr-qc/0405109} \BibitemShut
  {NoStop}%
\bibitem [{\citenamefont {Reisswig}\ and\ \citenamefont
  {Pollney}(2011)}]{Reisswig:2010di}%
  \BibitemOpen
  \bibfield  {author} {\bibinfo {author} {\bibfnamefont {C.}~\bibnamefont
  {Reisswig}}\ and\ \bibinfo {author} {\bibfnamefont {D.}~\bibnamefont
  {Pollney}},\ }\bibfield  {title} {\bibinfo {title} {{Notes on the integration
  of numerical relativity waveforms}},\ }\href
  {https://doi.org/10.1088/0264-9381/28/19/195015} {\bibfield  {journal}
  {\bibinfo  {journal} {Class. Quant. Grav.}\ }\textbf {\bibinfo {volume}
  {28}},\ \bibinfo {pages} {195015} (\bibinfo {year} {2011})},\ \Eprint
  {https://arxiv.org/abs/1006.1632} {arXiv:1006.1632 [gr-qc]} \BibitemShut
  {NoStop}%
\bibitem [{\citenamefont {McKechan}\ \emph {et~al.}(2010)\citenamefont
  {McKechan}, \citenamefont {Robinson},\ and\ \citenamefont
  {Sathyaprakash}}]{McKechan:2010kp}%
  \BibitemOpen
  \bibfield  {author} {\bibinfo {author} {\bibfnamefont {D.~J.~A.}\
  \bibnamefont {McKechan}}, \bibinfo {author} {\bibfnamefont {C.}~\bibnamefont
  {Robinson}},\ and\ \bibinfo {author} {\bibfnamefont {B.~S.}\ \bibnamefont
  {Sathyaprakash}},\ }\bibfield  {title} {\bibinfo {title} {{A tapering window
  for time-domain templates and simulated signals in the detection of
  gravitational waves from coalescing compact binaries}},\ }\href
  {https://doi.org/10.1088/0264-9381/27/8/084020} {\bibfield  {journal}
  {\bibinfo  {journal} {Class. Quant. Grav.}\ }\textbf {\bibinfo {volume}
  {27}},\ \bibinfo {pages} {084020} (\bibinfo {year} {2010})},\ \Eprint
  {https://arxiv.org/abs/1003.2939} {arXiv:1003.2939 [gr-qc]} \BibitemShut
  {NoStop}%
\bibitem [{\citenamefont {Harry}\ \emph {et~al.}(2016)\citenamefont {Harry},
  \citenamefont {Privitera}, \citenamefont {Boh\'e},\ and\ \citenamefont
  {Buonanno}}]{Harry:2016ijz}%
  \BibitemOpen
  \bibfield  {author} {\bibinfo {author} {\bibfnamefont {I.}~\bibnamefont
  {Harry}}, \bibinfo {author} {\bibfnamefont {S.}~\bibnamefont {Privitera}},
  \bibinfo {author} {\bibfnamefont {A.}~\bibnamefont {Boh\'e}},\ and\ \bibinfo
  {author} {\bibfnamefont {A.}~\bibnamefont {Buonanno}},\ }\bibfield  {title}
  {\bibinfo {title} {{Searching for Gravitational Waves from Compact Binaries
  with Precessing Spins}},\ }\href {https://doi.org/10.1103/PhysRevD.94.024012}
  {\bibfield  {journal} {\bibinfo  {journal} {Phys. Rev. D}\ }\textbf {\bibinfo
  {volume} {94}},\ \bibinfo {pages} {024012} (\bibinfo {year} {2016})},\
  \Eprint {https://arxiv.org/abs/1603.02444} {arXiv:1603.02444 [gr-qc]}
  \BibitemShut {NoStop}%
\bibitem [{\citenamefont {Venumadhav}\ \emph {et~al.}(2019)\citenamefont
  {Venumadhav}, \citenamefont {Zackay}, \citenamefont {Roulet}, \citenamefont
  {Dai},\ and\ \citenamefont {Zaldarriaga}}]{Venumadhav:2019tad}%
  \BibitemOpen
  \bibfield  {author} {\bibinfo {author} {\bibfnamefont {T.}~\bibnamefont
  {Venumadhav}}, \bibinfo {author} {\bibfnamefont {B.}~\bibnamefont {Zackay}},
  \bibinfo {author} {\bibfnamefont {J.}~\bibnamefont {Roulet}}, \bibinfo
  {author} {\bibfnamefont {L.}~\bibnamefont {Dai}},\ and\ \bibinfo {author}
  {\bibfnamefont {M.}~\bibnamefont {Zaldarriaga}},\ }\bibfield  {title}
  {\bibinfo {title} {{New search pipeline for compact binary mergers: Results
  for binary black holes in the first observing run of Advanced LIGO}},\ }\href
  {https://doi.org/10.1103/PhysRevD.100.023011} {\bibfield  {journal} {\bibinfo
   {journal} {Phys. Rev. D}\ }\textbf {\bibinfo {volume} {100}},\ \bibinfo
  {pages} {023011} (\bibinfo {year} {2019})},\ \Eprint
  {https://arxiv.org/abs/1902.10341} {arXiv:1902.10341 [astro-ph.IM]}
  \BibitemShut {NoStop}%
\bibitem [{\citenamefont {McIsaac}\ \emph {et~al.}(2023)\citenamefont
  {McIsaac}, \citenamefont {Hoy},\ and\ \citenamefont
  {Harry}}]{McIsaac:2023ijd}%
  \BibitemOpen
  \bibfield  {author} {\bibinfo {author} {\bibfnamefont {C.}~\bibnamefont
  {McIsaac}}, \bibinfo {author} {\bibfnamefont {C.}~\bibnamefont {Hoy}},\ and\
  \bibinfo {author} {\bibfnamefont {I.}~\bibnamefont {Harry}},\ }\bibfield
  {title} {\bibinfo {title} {{Search technique to observe precessing compact
  binary mergers in the advanced detector era}},\ }\href
  {https://doi.org/10.1103/PhysRevD.108.123016} {\bibfield  {journal} {\bibinfo
   {journal} {Phys. Rev. D}\ }\textbf {\bibinfo {volume} {108}},\ \bibinfo
  {pages} {123016} (\bibinfo {year} {2023})},\ \Eprint
  {https://arxiv.org/abs/2303.17364} {arXiv:2303.17364 [gr-qc]} \BibitemShut
  {NoStop}%
\bibitem [{\citenamefont {Schmidt}\ \emph {et~al.}(2024)\citenamefont
  {Schmidt}, \citenamefont {Gadre},\ and\ \citenamefont
  {Caudill}}]{Schmidt:2023gzj}%
  \BibitemOpen
  \bibfield  {author} {\bibinfo {author} {\bibfnamefont {S.}~\bibnamefont
  {Schmidt}}, \bibinfo {author} {\bibfnamefont {B.}~\bibnamefont {Gadre}},\
  and\ \bibinfo {author} {\bibfnamefont {S.}~\bibnamefont {Caudill}},\
  }\bibfield  {title} {\bibinfo {title} {{Gravitational-wave template banks for
  novel compact binaries}},\ }\href
  {https://doi.org/10.1103/PhysRevD.109.042005} {\bibfield  {journal} {\bibinfo
   {journal} {Phys. Rev. D}\ }\textbf {\bibinfo {volume} {109}},\ \bibinfo
  {pages} {042005} (\bibinfo {year} {2024})},\ \Eprint
  {https://arxiv.org/abs/2302.00436} {arXiv:2302.00436 [gr-qc]} \BibitemShut
  {NoStop}%
\bibitem [{\citenamefont {Wadekar}\ \emph {et~al.}(2023)\citenamefont
  {Wadekar}, \citenamefont {Venumadhav}, \citenamefont {Mehta}, \citenamefont
  {Roulet}, \citenamefont {Olsen}, \citenamefont {Mushkin}, \citenamefont
  {Zackay},\ and\ \citenamefont {Zaldarriaga}}]{Wadekar:2023kym}%
  \BibitemOpen
  \bibfield  {author} {\bibinfo {author} {\bibfnamefont {D.}~\bibnamefont
  {Wadekar}}, \bibinfo {author} {\bibfnamefont {T.}~\bibnamefont {Venumadhav}},
  \bibinfo {author} {\bibfnamefont {A.~K.}\ \bibnamefont {Mehta}}, \bibinfo
  {author} {\bibfnamefont {J.}~\bibnamefont {Roulet}}, \bibinfo {author}
  {\bibfnamefont {S.}~\bibnamefont {Olsen}}, \bibinfo {author} {\bibfnamefont
  {J.}~\bibnamefont {Mushkin}}, \bibinfo {author} {\bibfnamefont
  {B.}~\bibnamefont {Zackay}},\ and\ \bibinfo {author} {\bibfnamefont
  {M.}~\bibnamefont {Zaldarriaga}},\ }\bibfield  {title} {\bibinfo {title} {{A
  new approach to template banks of gravitational waves with higher harmonics:
  reducing matched-filtering cost by over an order of magnitude}},\ }\href@noop
  {} {\  (\bibinfo {year} {2023})},\ \Eprint {https://arxiv.org/abs/2310.15233}
  {arXiv:2310.15233 [gr-qc]} \BibitemShut {NoStop}%
\bibitem [{\citenamefont {Wadekar}\ \emph {et~al.}(2024)\citenamefont
  {Wadekar}, \citenamefont {Venumadhav}, \citenamefont {Roulet}, \citenamefont
  {Mehta}, \citenamefont {Zackay}, \citenamefont {Mushkin},\ and\ \citenamefont
  {Zaldarriaga}}]{Wadekar:2024zdq}%
  \BibitemOpen
  \bibfield  {author} {\bibinfo {author} {\bibfnamefont {D.}~\bibnamefont
  {Wadekar}}, \bibinfo {author} {\bibfnamefont {T.}~\bibnamefont {Venumadhav}},
  \bibinfo {author} {\bibfnamefont {J.}~\bibnamefont {Roulet}}, \bibinfo
  {author} {\bibfnamefont {A.~K.}\ \bibnamefont {Mehta}}, \bibinfo {author}
  {\bibfnamefont {B.}~\bibnamefont {Zackay}}, \bibinfo {author} {\bibfnamefont
  {J.}~\bibnamefont {Mushkin}},\ and\ \bibinfo {author} {\bibfnamefont
  {M.}~\bibnamefont {Zaldarriaga}},\ }\bibfield  {title} {\bibinfo {title} {{A
  new search pipeline for gravitational waves with higher-order modes using
  mode-by-mode filtering}},\ }\href@noop {} {\  (\bibinfo {year} {2024})},\
  \Eprint {https://arxiv.org/abs/2405.17400} {arXiv:2405.17400 [gr-qc]}
  \BibitemShut {NoStop}%
\bibitem [{\citenamefont {Abbott}\ \emph {et~al.}(2016)\citenamefont {Abbott}
  \emph {et~al.}}]{LIGOScientific:2016vlm}%
  \BibitemOpen
  \bibfield  {author} {\bibinfo {author} {\bibfnamefont {B.~P.}\ \bibnamefont
  {Abbott}} \emph {et~al.} (\bibinfo {collaboration} {LIGO Scientific,
  Virgo}),\ }\bibfield  {title} {\bibinfo {title} {{Properties of the Binary
  Black Hole Merger GW150914}},\ }\href
  {https://doi.org/10.1103/PhysRevLett.116.241102} {\bibfield  {journal}
  {\bibinfo  {journal} {Phys. Rev. Lett.}\ }\textbf {\bibinfo {volume} {116}},\
  \bibinfo {pages} {241102} (\bibinfo {year} {2016})},\ \Eprint
  {https://arxiv.org/abs/1602.03840} {arXiv:1602.03840 [gr-qc]} \BibitemShut
  {NoStop}%
\bibitem [{\citenamefont {Abbott}\ \emph
  {et~al.}(2017{\natexlab{b}})\citenamefont {Abbott} \emph
  {et~al.}}]{LIGOScientific:2017vwq}%
  \BibitemOpen
  \bibfield  {author} {\bibinfo {author} {\bibfnamefont {B.~P.}\ \bibnamefont
  {Abbott}} \emph {et~al.} (\bibinfo {collaboration} {LIGO Scientific,
  Virgo}),\ }\bibfield  {title} {\bibinfo {title} {{GW170817: Observation of
  Gravitational Waves from a Binary Neutron Star Inspiral}},\ }\href
  {https://doi.org/10.1103/PhysRevLett.119.161101} {\bibfield  {journal}
  {\bibinfo  {journal} {Phys. Rev. Lett.}\ }\textbf {\bibinfo {volume} {119}},\
  \bibinfo {pages} {161101} (\bibinfo {year} {2017}{\natexlab{b}})},\ \Eprint
  {https://arxiv.org/abs/1710.05832} {arXiv:1710.05832 [gr-qc]} \BibitemShut
  {NoStop}%
\bibitem [{\citenamefont {Abbott}\ \emph
  {et~al.}(2019{\natexlab{b}})\citenamefont {Abbott} \emph
  {et~al.}}]{LIGOScientific:2018mvr}%
  \BibitemOpen
  \bibfield  {author} {\bibinfo {author} {\bibfnamefont {B.~P.}\ \bibnamefont
  {Abbott}} \emph {et~al.} (\bibinfo {collaboration} {LIGO Scientific,
  Virgo}),\ }\bibfield  {title} {\bibinfo {title} {{GWTC-1: A
  Gravitational-Wave Transient Catalog of Compact Binary Mergers Observed by
  LIGO and Virgo during the First and Second Observing Runs}},\ }\href
  {https://doi.org/10.1103/PhysRevX.9.031040} {\bibfield  {journal} {\bibinfo
  {journal} {Phys. Rev. X}\ }\textbf {\bibinfo {volume} {9}},\ \bibinfo {pages}
  {031040} (\bibinfo {year} {2019}{\natexlab{b}})},\ \Eprint
  {https://arxiv.org/abs/1811.12907} {arXiv:1811.12907 [astro-ph.HE]}
  \BibitemShut {NoStop}%
\bibitem [{\citenamefont {Abbott}\ \emph
  {et~al.}(2021{\natexlab{a}})\citenamefont {Abbott} \emph
  {et~al.}}]{LIGOScientific:2020ibl}%
  \BibitemOpen
  \bibfield  {author} {\bibinfo {author} {\bibfnamefont {R.}~\bibnamefont
  {Abbott}} \emph {et~al.} (\bibinfo {collaboration} {LIGO Scientific,
  Virgo}),\ }\bibfield  {title} {\bibinfo {title} {{GWTC-2: Compact Binary
  Coalescences Observed by LIGO and Virgo During the First Half of the Third
  Observing Run}},\ }\href {https://doi.org/10.1103/PhysRevX.11.021053}
  {\bibfield  {journal} {\bibinfo  {journal} {Phys. Rev. X}\ }\textbf {\bibinfo
  {volume} {11}},\ \bibinfo {pages} {021053} (\bibinfo {year}
  {2021}{\natexlab{a}})},\ \Eprint {https://arxiv.org/abs/2010.14527}
  {arXiv:2010.14527 [gr-qc]} \BibitemShut {NoStop}%
\bibitem [{\citenamefont {Abbott}\ \emph
  {et~al.}(2021{\natexlab{b}})\citenamefont {Abbott} \emph
  {et~al.}}]{LIGOScientific:2021qlt}%
  \BibitemOpen
  \bibfield  {author} {\bibinfo {author} {\bibfnamefont {R.}~\bibnamefont
  {Abbott}} \emph {et~al.} (\bibinfo {collaboration} {LIGO Scientific, KAGRA,
  VIRGO}),\ }\bibfield  {title} {\bibinfo {title} {{Observation of
  Gravitational Waves from Two Neutron Star\textendash{}Black Hole
  Coalescences}},\ }\href {https://doi.org/10.3847/2041-8213/ac082e} {\bibfield
   {journal} {\bibinfo  {journal} {Astrophys. J. Lett.}\ }\textbf {\bibinfo
  {volume} {915}},\ \bibinfo {pages} {L5} (\bibinfo {year}
  {2021}{\natexlab{b}})},\ \Eprint {https://arxiv.org/abs/2106.15163}
  {arXiv:2106.15163 [astro-ph.HE]} \BibitemShut {NoStop}%
\bibitem [{\citenamefont {Abbott}\ \emph {et~al.}(2023)\citenamefont {Abbott}
  \emph {et~al.}}]{KAGRA:2021vkt}%
  \BibitemOpen
  \bibfield  {author} {\bibinfo {author} {\bibfnamefont {R.}~\bibnamefont
  {Abbott}} \emph {et~al.} (\bibinfo {collaboration} {KAGRA, VIRGO, LIGO
  Scientific}),\ }\bibfield  {title} {\bibinfo {title} {{GWTC-3: Compact Binary
  Coalescences Observed by LIGO and Virgo during the Second Part of the Third
  Observing Run}},\ }\href {https://doi.org/10.1103/PhysRevX.13.041039}
  {\bibfield  {journal} {\bibinfo  {journal} {Phys. Rev. X}\ }\textbf {\bibinfo
  {volume} {13}},\ \bibinfo {pages} {041039} (\bibinfo {year} {2023})},\
  \Eprint {https://arxiv.org/abs/2111.03606} {arXiv:2111.03606 [gr-qc]}
  \BibitemShut {NoStop}%
\bibitem [{\citenamefont {Abbott}\ \emph
  {et~al.}(2020{\natexlab{a}})\citenamefont {Abbott} \emph
  {et~al.}}]{LIGOScientific:2020zkf}%
  \BibitemOpen
  \bibfield  {author} {\bibinfo {author} {\bibfnamefont {R.}~\bibnamefont
  {Abbott}} \emph {et~al.} (\bibinfo {collaboration} {LIGO Scientific,
  Virgo}),\ }\bibfield  {title} {\bibinfo {title} {{GW190814: Gravitational
  Waves from the Coalescence of a 23 Solar Mass Black Hole with a 2.6 Solar
  Mass Compact Object}},\ }\href {https://doi.org/10.3847/2041-8213/ab960f}
  {\bibfield  {journal} {\bibinfo  {journal} {Astrophys. J. Lett.}\ }\textbf
  {\bibinfo {volume} {896}},\ \bibinfo {pages} {L44} (\bibinfo {year}
  {2020}{\natexlab{a}})},\ \Eprint {https://arxiv.org/abs/2006.12611}
  {arXiv:2006.12611 [astro-ph.HE]} \BibitemShut {NoStop}%
\bibitem [{\citenamefont {Abbott}\ \emph
  {et~al.}(2020{\natexlab{b}})\citenamefont {Abbott} \emph
  {et~al.}}]{LIGOScientific:2020ufj}%
  \BibitemOpen
  \bibfield  {author} {\bibinfo {author} {\bibfnamefont {R.}~\bibnamefont
  {Abbott}} \emph {et~al.} (\bibinfo {collaboration} {LIGO Scientific,
  Virgo}),\ }\bibfield  {title} {\bibinfo {title} {{Properties and
  Astrophysical Implications of the 150 M$_\odot$ Binary Black Hole Merger
  GW190521}},\ }\href {https://doi.org/10.3847/2041-8213/aba493} {\bibfield
  {journal} {\bibinfo  {journal} {Astrophys. J. Lett.}\ }\textbf {\bibinfo
  {volume} {900}},\ \bibinfo {pages} {L13} (\bibinfo {year}
  {2020}{\natexlab{b}})},\ \Eprint {https://arxiv.org/abs/2009.01190}
  {arXiv:2009.01190 [astro-ph.HE]} \BibitemShut {NoStop}%
\bibitem [{\citenamefont {Abac}\ \emph
  {et~al.}(2024{\natexlab{b}})\citenamefont {Abac} \emph
  {et~al.}}]{LIGOScientific:2024elc}%
  \BibitemOpen
  \bibfield  {author} {\bibinfo {author} {\bibfnamefont {A.~G.}\ \bibnamefont
  {Abac}} \emph {et~al.} (\bibinfo {collaboration} {LIGO Scientific, VIRGO,
  KAGRA}),\ }\bibfield  {title} {\bibinfo {title} {{Observation of
  Gravitational Waves from the Coalescence of a $2.5-4.5~M_\odot$ Compact
  Object and a Neutron Star}},\ }\href@noop {} {\  (\bibinfo {year}
  {2024}{\natexlab{b}})},\ \Eprint {https://arxiv.org/abs/2404.04248}
  {arXiv:2404.04248 [astro-ph.HE]} \BibitemShut {NoStop}%
\bibitem [{\citenamefont {Cabero}\ \emph {et~al.}(2017)\citenamefont {Cabero},
  \citenamefont {Nielsen}, \citenamefont {Lundgren},\ and\ \citenamefont
  {Capano}}]{Cabero:2016ayq}%
  \BibitemOpen
  \bibfield  {author} {\bibinfo {author} {\bibfnamefont {M.}~\bibnamefont
  {Cabero}}, \bibinfo {author} {\bibfnamefont {A.~B.}\ \bibnamefont {Nielsen}},
  \bibinfo {author} {\bibfnamefont {A.~P.}\ \bibnamefont {Lundgren}},\ and\
  \bibinfo {author} {\bibfnamefont {C.~D.}\ \bibnamefont {Capano}},\ }\bibfield
   {title} {\bibinfo {title} {{Minimum energy and the end of the inspiral in
  the post-Newtonian approximation}},\ }\href
  {https://doi.org/10.1103/PhysRevD.95.064016} {\bibfield  {journal} {\bibinfo
  {journal} {Phys. Rev. D}\ }\textbf {\bibinfo {volume} {95}},\ \bibinfo
  {pages} {064016} (\bibinfo {year} {2017})},\ \Eprint
  {https://arxiv.org/abs/1602.03134} {arXiv:1602.03134 [gr-qc]} \BibitemShut
  {NoStop}%
\bibitem [{\citenamefont {Jim\'enez-Forteza}\ \emph {et~al.}(2017)\citenamefont
  {Jim\'enez-Forteza}, \citenamefont {Keitel}, \citenamefont {Husa},
  \citenamefont {Hannam}, \citenamefont {Khan},\ and\ \citenamefont
  {P\"urrer}}]{Jimenez-Forteza:2016oae}%
  \BibitemOpen
  \bibfield  {author} {\bibinfo {author} {\bibfnamefont {X.}~\bibnamefont
  {Jim\'enez-Forteza}}, \bibinfo {author} {\bibfnamefont {D.}~\bibnamefont
  {Keitel}}, \bibinfo {author} {\bibfnamefont {S.}~\bibnamefont {Husa}},
  \bibinfo {author} {\bibfnamefont {M.}~\bibnamefont {Hannam}}, \bibinfo
  {author} {\bibfnamefont {S.}~\bibnamefont {Khan}},\ and\ \bibinfo {author}
  {\bibfnamefont {M.}~\bibnamefont {P\"urrer}},\ }\bibfield  {title} {\bibinfo
  {title} {{Hierarchical data-driven approach to fitting numerical relativity
  data for nonprecessing binary black holes with an application to final spin
  and radiated energy}},\ }\href {https://doi.org/10.1103/PhysRevD.95.064024}
  {\bibfield  {journal} {\bibinfo  {journal} {Phys. Rev. D}\ }\textbf {\bibinfo
  {volume} {95}},\ \bibinfo {pages} {064024} (\bibinfo {year} {2017})},\
  \Eprint {https://arxiv.org/abs/1611.00332} {arXiv:1611.00332 [gr-qc]}
  \BibitemShut {NoStop}%
\bibitem [{\citenamefont {Trefethen}(2019)}]{Trefethen:2019apx}%
  \BibitemOpen
  \bibfield  {author} {\bibinfo {author} {\bibfnamefont {L.~N.}\ \bibnamefont
  {Trefethen}},\ }\href@noop {} {\emph {\bibinfo {title} {Approximation theory
  and approximation practice, extended edition}}}\ (\bibinfo  {publisher}
  {SIAM},\ \bibinfo {year} {2019})\BibitemShut {NoStop}%
\bibitem [{\citenamefont {Marsat}(2015)}]{Marsat:2014xea}%
  \BibitemOpen
  \bibfield  {author} {\bibinfo {author} {\bibfnamefont {S.}~\bibnamefont
  {Marsat}},\ }\bibfield  {title} {\bibinfo {title} {{Cubic order spin effects
  in the dynamics and gravitational wave energy flux of compact object
  binaries}},\ }\href {https://doi.org/10.1088/0264-9381/32/8/085008}
  {\bibfield  {journal} {\bibinfo  {journal} {Class. Quant. Grav.}\ }\textbf
  {\bibinfo {volume} {32}},\ \bibinfo {pages} {085008} (\bibinfo {year}
  {2015})},\ \Eprint {https://arxiv.org/abs/1411.4118} {arXiv:1411.4118
  [gr-qc]} \BibitemShut {NoStop}%
\bibitem [{\citenamefont {Boh\'e}\ \emph {et~al.}(2015)\citenamefont {Boh\'e},
  \citenamefont {Faye}, \citenamefont {Marsat},\ and\ \citenamefont
  {Porter}}]{Bohe:2015ana}%
  \BibitemOpen
  \bibfield  {author} {\bibinfo {author} {\bibfnamefont {A.}~\bibnamefont
  {Boh\'e}}, \bibinfo {author} {\bibfnamefont {G.}~\bibnamefont {Faye}},
  \bibinfo {author} {\bibfnamefont {S.}~\bibnamefont {Marsat}},\ and\ \bibinfo
  {author} {\bibfnamefont {E.~K.}\ \bibnamefont {Porter}},\ }\bibfield  {title}
  {\bibinfo {title} {{Quadratic-in-spin effects in the orbital dynamics and
  gravitational-wave energy flux of compact binaries at the 3PN order}},\
  }\href {https://doi.org/10.1088/0264-9381/32/19/195010} {\bibfield  {journal}
  {\bibinfo  {journal} {Class. Quant. Grav.}\ }\textbf {\bibinfo {volume}
  {32}},\ \bibinfo {pages} {195010} (\bibinfo {year} {2015})},\ \Eprint
  {https://arxiv.org/abs/1501.01529} {arXiv:1501.01529 [gr-qc]} \BibitemShut
  {NoStop}%
\bibitem [{\citenamefont {Poisson}(1998)}]{Poisson:1997ha}%
  \BibitemOpen
  \bibfield  {author} {\bibinfo {author} {\bibfnamefont {E.}~\bibnamefont
  {Poisson}},\ }\bibfield  {title} {\bibinfo {title} {{Gravitational waves from
  inspiraling compact binaries: The Quadrupole moment term}},\ }\href
  {https://doi.org/10.1103/PhysRevD.57.5287} {\bibfield  {journal} {\bibinfo
  {journal} {Phys. Rev. D}\ }\textbf {\bibinfo {volume} {57}},\ \bibinfo
  {pages} {5287} (\bibinfo {year} {1998})},\ \Eprint
  {https://arxiv.org/abs/gr-qc/9709032} {arXiv:gr-qc/9709032} \BibitemShut
  {NoStop}%
\bibitem [{\citenamefont {Porto}\ \emph {et~al.}(2011)\citenamefont {Porto},
  \citenamefont {Ross},\ and\ \citenamefont {Rothstein}}]{Porto:2010zg}%
  \BibitemOpen
  \bibfield  {author} {\bibinfo {author} {\bibfnamefont {R.~A.}\ \bibnamefont
  {Porto}}, \bibinfo {author} {\bibfnamefont {A.}~\bibnamefont {Ross}},\ and\
  \bibinfo {author} {\bibfnamefont {I.~Z.}\ \bibnamefont {Rothstein}},\
  }\bibfield  {title} {\bibinfo {title} {{Spin induced multipole moments for
  the gravitational wave flux from binary inspirals to third Post-Newtonian
  order}},\ }\href {https://doi.org/10.1088/1475-7516/2011/03/009} {\bibfield
  {journal} {\bibinfo  {journal} {JCAP}\ }\textbf {\bibinfo {volume} {03}},\
  \bibinfo {pages} {009}},\ \Eprint {https://arxiv.org/abs/1007.1312}
  {arXiv:1007.1312 [gr-qc]} \BibitemShut {NoStop}%
\bibitem [{\citenamefont {Levi}\ and\ \citenamefont
  {Steinhoff}(2014)}]{Levi:2014sba}%
  \BibitemOpen
  \bibfield  {author} {\bibinfo {author} {\bibfnamefont {M.}~\bibnamefont
  {Levi}}\ and\ \bibinfo {author} {\bibfnamefont {J.}~\bibnamefont
  {Steinhoff}},\ }\bibfield  {title} {\bibinfo {title} {{Equivalence of ADM
  Hamiltonian and Effective Field Theory approaches at next-to-next-to-leading
  order spin1-spin2 coupling of binary inspirals}},\ }\href
  {https://doi.org/10.1088/1475-7516/2014/12/003} {\bibfield  {journal}
  {\bibinfo  {journal} {JCAP}\ }\textbf {\bibinfo {volume} {12}},\ \bibinfo
  {pages} {003}},\ \Eprint {https://arxiv.org/abs/1408.5762} {arXiv:1408.5762
  [gr-qc]} \BibitemShut {NoStop}%
\bibitem [{\citenamefont {Yagi}\ and\ \citenamefont
  {Yunes}(2017{\natexlab{b}})}]{Yagi:2016bkt}%
  \BibitemOpen
  \bibfield  {author} {\bibinfo {author} {\bibfnamefont {K.}~\bibnamefont
  {Yagi}}\ and\ \bibinfo {author} {\bibfnamefont {N.}~\bibnamefont {Yunes}},\
  }\bibfield  {title} {\bibinfo {title} {{Approximate Universal Relations for
  Neutron Stars and Quark Stars}},\ }\href
  {https://doi.org/10.1016/j.physrep.2017.03.002} {\bibfield  {journal}
  {\bibinfo  {journal} {Phys. Rept.}\ }\textbf {\bibinfo {volume} {681}},\
  \bibinfo {pages} {1} (\bibinfo {year} {2017}{\natexlab{b}})},\ \Eprint
  {https://arxiv.org/abs/1608.02582} {arXiv:1608.02582 [gr-qc]} \BibitemShut
  {NoStop}%
\bibitem [{\citenamefont {Abbott}\ \emph
  {et~al.}(2021{\natexlab{c}})\citenamefont {Abbott} \emph
  {et~al.}}]{LIGOScientific:2019lzm}%
  \BibitemOpen
  \bibfield  {author} {\bibinfo {author} {\bibfnamefont {R.}~\bibnamefont
  {Abbott}} \emph {et~al.} (\bibinfo {collaboration} {LIGO Scientific,
  Virgo}),\ }\bibfield  {title} {\bibinfo {title} {{Open data from the first
  and second observing runs of Advanced LIGO and Advanced Virgo}},\ }\href
  {https://doi.org/10.1016/j.softx.2021.100658} {\bibfield  {journal} {\bibinfo
   {journal} {SoftwareX}\ }\textbf {\bibinfo {volume} {13}},\ \bibinfo {pages}
  {100658} (\bibinfo {year} {2021}{\natexlab{c}})},\ \Eprint
  {https://arxiv.org/abs/1912.11716} {arXiv:1912.11716 [gr-qc]} \BibitemShut
  {NoStop}%
\bibitem [{\citenamefont {Ashton}\ \emph {et~al.}(2019)\citenamefont {Ashton}
  \emph {et~al.}}]{Ashton:2018jfp}%
  \BibitemOpen
  \bibfield  {author} {\bibinfo {author} {\bibfnamefont {G.}~\bibnamefont
  {Ashton}} \emph {et~al.},\ }\bibfield  {title} {\bibinfo {title} {{BILBY: A
  user-friendly Bayesian inference library for gravitational-wave astronomy}},\
  }\href {https://doi.org/10.3847/1538-4365/ab06fc} {\bibfield  {journal}
  {\bibinfo  {journal} {Astrophys. J. Suppl.}\ }\textbf {\bibinfo {volume}
  {241}},\ \bibinfo {pages} {27} (\bibinfo {year} {2019})},\ \Eprint
  {https://arxiv.org/abs/1811.02042} {arXiv:1811.02042 [astro-ph.IM]}
  \BibitemShut {NoStop}%
\bibitem [{\citenamefont {Speagle}(2020)}]{Speagle:2020aa}%
  \BibitemOpen
  \bibfield  {author} {\bibinfo {author} {\bibfnamefont {J.~S.}\ \bibnamefont
  {Speagle}},\ }\bibfield  {title} {\bibinfo {title} {{dynesty: a dynamic
  nested sampling package for estimating Bayesian posteriors and evidences}},\
  }\href {https://doi.org/10.1093/mnras/staa278} {\bibfield  {journal}
  {\bibinfo  {journal} {Mon. Not. R. Astron. Soc.}\ }\textbf {\bibinfo {volume}
  {493}},\ \bibinfo {pages} {3132} (\bibinfo {year} {2020})},\ \Eprint
  {https://arxiv.org/abs/1904.02180} {arXiv:1904.02180 [astro-ph.IM]}
  \BibitemShut {NoStop}%
\bibitem [{\citenamefont {Aasi}\ \emph {et~al.}(2015)\citenamefont {Aasi} \emph
  {et~al.}}]{LIGOScientific:2014pky}%
  \BibitemOpen
  \bibfield  {author} {\bibinfo {author} {\bibfnamefont {J.}~\bibnamefont
  {Aasi}} \emph {et~al.} (\bibinfo {collaboration} {LIGO Scientific}),\
  }\bibfield  {title} {\bibinfo {title} {{Advanced LIGO}},\ }\href
  {https://doi.org/10.1088/0264-9381/32/7/074001} {\bibfield  {journal}
  {\bibinfo  {journal} {Class. Quant. Grav.}\ }\textbf {\bibinfo {volume}
  {32}},\ \bibinfo {pages} {074001} (\bibinfo {year} {2015})},\ \Eprint
  {https://arxiv.org/abs/1411.4547} {arXiv:1411.4547 [gr-qc]} \BibitemShut
  {NoStop}%
\bibitem [{\citenamefont {Acernese}\ \emph {et~al.}(2015)\citenamefont
  {Acernese} \emph {et~al.}}]{VIRGO:2014yos}%
  \BibitemOpen
  \bibfield  {author} {\bibinfo {author} {\bibfnamefont {F.}~\bibnamefont
  {Acernese}} \emph {et~al.} (\bibinfo {collaboration} {VIRGO}),\ }\bibfield
  {title} {\bibinfo {title} {{Advanced Virgo: a second-generation
  interferometric gravitational wave detector}},\ }\href
  {https://doi.org/10.1088/0264-9381/32/2/024001} {\bibfield  {journal}
  {\bibinfo  {journal} {Class. Quant. Grav.}\ }\textbf {\bibinfo {volume}
  {32}},\ \bibinfo {pages} {024001} (\bibinfo {year} {2015})},\ \Eprint
  {https://arxiv.org/abs/1408.3978} {arXiv:1408.3978 [gr-qc]} \BibitemShut
  {NoStop}%
\bibitem [{O4P(2022)}]{O4PSDs}%
  \BibitemOpen
  \href@noop {} {\bibinfo {title} {{ Noise curves used for Simulations in the
  update of the Observing Scenarios Paper }}},\ \bibinfo {howpublished}
  {\href{https://dcc.ligo.org/LIGO-T2000012/public}{https://dcc.ligo.org/LIGO-T2000012/public}}
  (\bibinfo {year} {2022})\BibitemShut {NoStop}%
\bibitem [{\citenamefont {Akmal}\ \emph {et~al.}(1998)\citenamefont {Akmal},
  \citenamefont {Pandharipande},\ and\ \citenamefont
  {Ravenhall}}]{Akmal:1998cf}%
  \BibitemOpen
  \bibfield  {author} {\bibinfo {author} {\bibfnamefont {A.}~\bibnamefont
  {Akmal}}, \bibinfo {author} {\bibfnamefont {V.~R.}\ \bibnamefont
  {Pandharipande}},\ and\ \bibinfo {author} {\bibfnamefont {D.~G.}\
  \bibnamefont {Ravenhall}},\ }\bibfield  {title} {\bibinfo {title} {{The
  Equation of state of nucleon matter and neutron star structure}},\ }\href
  {https://doi.org/10.1103/PhysRevC.58.1804} {\bibfield  {journal} {\bibinfo
  {journal} {Phys. Rev.}\ }\textbf {\bibinfo {volume} {C58}},\ \bibinfo {pages}
  {1804} (\bibinfo {year} {1998})},\ \Eprint
  {https://arxiv.org/abs/nucl-th/9804027} {arXiv:nucl-th/9804027 [nucl-th]}
  \BibitemShut {NoStop}%
%%CITATION = NUCL-TH/9804027;%%
\bibitem [{\citenamefont {Abbott}\ \emph
  {et~al.}(2020{\natexlab{c}})\citenamefont {Abbott} \emph
  {et~al.}}]{LIGOScientific:2019eut}%
  \BibitemOpen
  \bibfield  {author} {\bibinfo {author} {\bibfnamefont {B.~P.}\ \bibnamefont
  {Abbott}} \emph {et~al.} (\bibinfo {collaboration} {LIGO Scientific,
  Virgo}),\ }\bibfield  {title} {\bibinfo {title} {{Model comparison from
  LIGO\textendash{}Virgo data on GW170817\textquoteright{}s binary components
  and consequences for the merger remnant}},\ }\href
  {https://doi.org/10.1088/1361-6382/ab5f7c} {\bibfield  {journal} {\bibinfo
  {journal} {Class. Quant. Grav.}\ }\textbf {\bibinfo {volume} {37}},\ \bibinfo
  {pages} {045006} (\bibinfo {year} {2020}{\natexlab{c}})},\ \Eprint
  {https://arxiv.org/abs/1908.01012} {arXiv:1908.01012 [gr-qc]} \BibitemShut
  {NoStop}%
\bibitem [{\citenamefont {Cutler}\ and\ \citenamefont
  {Flanagan}(1994)}]{Cutler:1994ys}%
  \BibitemOpen
  \bibfield  {author} {\bibinfo {author} {\bibfnamefont {C.}~\bibnamefont
  {Cutler}}\ and\ \bibinfo {author} {\bibfnamefont {E.~E.}\ \bibnamefont
  {Flanagan}},\ }\bibfield  {title} {\bibinfo {title} {{Gravitational waves
  from merging compact binaries: How accurately can one extract the binary's
  parameters from the inspiral wave form?}},\ }\href
  {https://doi.org/10.1103/PhysRevD.49.2658} {\bibfield  {journal} {\bibinfo
  {journal} {Phys. Rev. D}\ }\textbf {\bibinfo {volume} {49}},\ \bibinfo
  {pages} {2658} (\bibinfo {year} {1994})},\ \Eprint
  {https://arxiv.org/abs/gr-qc/9402014} {arXiv:gr-qc/9402014} \BibitemShut
  {NoStop}%
\bibitem [{\citenamefont {Poisson}\ and\ \citenamefont
  {Will}(1995)}]{Poisson:1995ef}%
  \BibitemOpen
  \bibfield  {author} {\bibinfo {author} {\bibfnamefont {E.}~\bibnamefont
  {Poisson}}\ and\ \bibinfo {author} {\bibfnamefont {C.~M.}\ \bibnamefont
  {Will}},\ }\bibfield  {title} {\bibinfo {title} {{Gravitational waves from
  inspiraling compact binaries: Parameter estimation using second postNewtonian
  wave forms}},\ }\href {https://doi.org/10.1103/PhysRevD.52.848} {\bibfield
  {journal} {\bibinfo  {journal} {Phys. Rev. D}\ }\textbf {\bibinfo {volume}
  {52}},\ \bibinfo {pages} {848} (\bibinfo {year} {1995})},\ \Eprint
  {https://arxiv.org/abs/gr-qc/9502040} {arXiv:gr-qc/9502040} \BibitemShut
  {NoStop}%
\bibitem [{\citenamefont {Schmidt}\ \emph {et~al.}(2017)\citenamefont
  {Schmidt}, \citenamefont {Harry},\ and\ \citenamefont
  {Pfeiffer}}]{Schmidt:2017btt}%
  \BibitemOpen
  \bibfield  {author} {\bibinfo {author} {\bibfnamefont {P.}~\bibnamefont
  {Schmidt}}, \bibinfo {author} {\bibfnamefont {I.~W.}\ \bibnamefont {Harry}},\
  and\ \bibinfo {author} {\bibfnamefont {H.~P.}\ \bibnamefont {Pfeiffer}},\
  }\bibfield  {title} {\bibinfo {title} {{Numerical Relativity Injection
  Infrastructure}},\ }\href@noop {} {\  (\bibinfo {year} {2017})},\ \Eprint
  {https://arxiv.org/abs/1703.01076} {arXiv:1703.01076 [gr-qc]} \BibitemShut
  {NoStop}%
\bibitem [{\citenamefont {{LIGO Scientific Collaboration}}\ \emph
  {et~al.}(2018)\citenamefont {{LIGO Scientific Collaboration}}, \citenamefont
  {{Virgo Collaboration}},\ and\ \citenamefont {{KAGRA
  Collaboration}}}]{lalsuite}%
  \BibitemOpen
  \bibfield  {author} {\bibinfo {author} {\bibnamefont {{LIGO Scientific
  Collaboration}}}, \bibinfo {author} {\bibnamefont {{Virgo Collaboration}}},\
  and\ \bibinfo {author} {\bibnamefont {{KAGRA Collaboration}}},\ }\href
  {https://doi.org/10.7935/GT1W-FZ16} {\bibinfo {title} {{LVK} {A}lgorithm
  {L}ibrary - {LALS}uite}},\ \bibinfo {howpublished} {Free software (GPL)}
  (\bibinfo {year} {2018})\BibitemShut {NoStop}%
\bibitem [{\citenamefont {Abbott}\ \emph
  {et~al.}(2019{\natexlab{c}})\citenamefont {Abbott} \emph {et~al.}}]{GWTC1}%
  \BibitemOpen
  \bibfield  {author} {\bibinfo {author} {\bibfnamefont {B.~P.}\ \bibnamefont
  {Abbott}} \emph {et~al.} (\bibinfo {collaboration} {LIGO Scientific
  Collaboration and Virgo Collaboration}),\ }\bibfield  {title} {\bibinfo
  {title} {{GWTC-1: A Gravitational-Wave Transient Catalog of Compact Binary
  Mergers Observed by LIGO and Virgo during the First and Second Observing
  Runs}},\ }\href {https://doi.org/10.1103/PhysRevX.9.031040} {\bibfield
  {journal} {\bibinfo  {journal} {Phys. Rev. X}\ }\textbf {\bibinfo {volume}
  {9}},\ \bibinfo {pages} {031040} (\bibinfo {year} {2019}{\natexlab{c}})},\
  \Eprint {https://arxiv.org/abs/1811.12907} {arXiv:1811.12907 [astro-ph.HE]}
  \BibitemShut {NoStop}%
%%CITATION = ARXIV:1811.12907;%%
\bibitem [{\citenamefont {{LIGO Scientific Collaboration, Virgo
  Collaboration}}(2019)}]{GWOSC}%
  \BibitemOpen
  \bibfield  {author} {\bibinfo {author} {\bibnamefont {{LIGO Scientific
  Collaboration, Virgo Collaboration}}},\ }\href@noop {} {\bibinfo {title}
  {{Gravitational Wave Open Science Center}}},\ \bibinfo {howpublished}
  {\href{https://www.gw-openscience.org}{https://www.gw-openscience.org}}
  (\bibinfo {year} {2019})\BibitemShut {NoStop}%
\bibitem [{\citenamefont {Gamba}\ \emph
  {et~al.}(2021{\natexlab{b}})\citenamefont {Gamba}, \citenamefont {Breschi},
  \citenamefont {Bernuzzi}, \citenamefont {Agathos},\ and\ \citenamefont
  {Nagar}}]{Gamba:2020wgg}%
  \BibitemOpen
  \bibfield  {author} {\bibinfo {author} {\bibfnamefont {R.}~\bibnamefont
  {Gamba}}, \bibinfo {author} {\bibfnamefont {M.}~\bibnamefont {Breschi}},
  \bibinfo {author} {\bibfnamefont {S.}~\bibnamefont {Bernuzzi}}, \bibinfo
  {author} {\bibfnamefont {M.}~\bibnamefont {Agathos}},\ and\ \bibinfo {author}
  {\bibfnamefont {A.}~\bibnamefont {Nagar}},\ }\bibfield  {title} {\bibinfo
  {title} {{Waveform systematics in the gravitational-wave inference of tidal
  parameters and equation of state from binary neutron star signals}},\ }\href
  {https://doi.org/10.1103/PhysRevD.103.124015} {\bibfield  {journal} {\bibinfo
   {journal} {Phys. Rev. D}\ }\textbf {\bibinfo {volume} {103}},\ \bibinfo
  {pages} {124015} (\bibinfo {year} {2021}{\natexlab{b}})},\ \Eprint
  {https://arxiv.org/abs/2009.08467} {arXiv:2009.08467 [gr-qc]} \BibitemShut
  {NoStop}%
\bibitem [{\citenamefont {Pratten}\ \emph
  {et~al.}(2020{\natexlab{b}})\citenamefont {Pratten}, \citenamefont
  {Schmidt},\ and\ \citenamefont {Hinderer}}]{Pratten:2019sed}%
  \BibitemOpen
  \bibfield  {author} {\bibinfo {author} {\bibfnamefont {G.}~\bibnamefont
  {Pratten}}, \bibinfo {author} {\bibfnamefont {P.}~\bibnamefont {Schmidt}},\
  and\ \bibinfo {author} {\bibfnamefont {T.}~\bibnamefont {Hinderer}},\
  }\bibfield  {title} {\bibinfo {title} {{Gravitational-Wave Asteroseismology
  with Fundamental Modes from Compact Binary Inspirals}},\ }\href
  {https://doi.org/10.1038/s41467-020-15984-5} {\bibfield  {journal} {\bibinfo
  {journal} {Nature Commun.}\ }\textbf {\bibinfo {volume} {11}},\ \bibinfo
  {pages} {2553} (\bibinfo {year} {2020}{\natexlab{b}})},\ \Eprint
  {https://arxiv.org/abs/1905.00817} {arXiv:1905.00817 [gr-qc]} \BibitemShut
  {NoStop}%
\bibitem [{\citenamefont {Pratten}\ \emph {et~al.}(2022)\citenamefont
  {Pratten}, \citenamefont {Schmidt},\ and\ \citenamefont
  {Williams}}]{Pratten:2021pro}%
  \BibitemOpen
  \bibfield  {author} {\bibinfo {author} {\bibfnamefont {G.}~\bibnamefont
  {Pratten}}, \bibinfo {author} {\bibfnamefont {P.}~\bibnamefont {Schmidt}},\
  and\ \bibinfo {author} {\bibfnamefont {N.}~\bibnamefont {Williams}},\
  }\bibfield  {title} {\bibinfo {title} {{Impact of Dynamical Tides on the
  Reconstruction of the Neutron Star Equation of State}},\ }\href
  {https://doi.org/10.1103/PhysRevLett.129.081102} {\bibfield  {journal}
  {\bibinfo  {journal} {Phys. Rev. Lett.}\ }\textbf {\bibinfo {volume} {129}},\
  \bibinfo {pages} {081102} (\bibinfo {year} {2022})},\ \Eprint
  {https://arxiv.org/abs/2109.07566} {arXiv:2109.07566 [astro-ph.HE]}
  \BibitemShut {NoStop}%
\bibitem [{\citenamefont {Ma}\ \emph {et~al.}(2021)\citenamefont {Ma},
  \citenamefont {Yu},\ and\ \citenamefont {Chen}}]{Ma:2020oni}%
  \BibitemOpen
  \bibfield  {author} {\bibinfo {author} {\bibfnamefont {S.}~\bibnamefont
  {Ma}}, \bibinfo {author} {\bibfnamefont {H.}~\bibnamefont {Yu}},\ and\
  \bibinfo {author} {\bibfnamefont {Y.}~\bibnamefont {Chen}},\ }\bibfield
  {title} {\bibinfo {title} {{Detecting resonant tidal excitations of Rossby
  modes in coalescing neutron-star binaries with third-generation
  gravitational-wave detectors}},\ }\href
  {https://doi.org/10.1103/PhysRevD.103.063020} {\bibfield  {journal} {\bibinfo
   {journal} {Phys. Rev. D}\ }\textbf {\bibinfo {volume} {103}},\ \bibinfo
  {pages} {063020} (\bibinfo {year} {2021})},\ \Eprint
  {https://arxiv.org/abs/2010.03066} {arXiv:2010.03066 [gr-qc]} \BibitemShut
  {NoStop}%
\bibitem [{\citenamefont {Poisson}(2020)}]{Poisson:2020eki}%
  \BibitemOpen
  \bibfield  {author} {\bibinfo {author} {\bibfnamefont {E.}~\bibnamefont
  {Poisson}},\ }\bibfield  {title} {\bibinfo {title} {{Gravitomagnetic tidal
  resonance in neutron-star binary inspirals}},\ }\href
  {https://doi.org/10.1103/PhysRevD.101.104028} {\bibfield  {journal} {\bibinfo
   {journal} {Phys. Rev. D}\ }\textbf {\bibinfo {volume} {101}},\ \bibinfo
  {pages} {104028} (\bibinfo {year} {2020})},\ \Eprint
  {https://arxiv.org/abs/2003.10427} {arXiv:2003.10427 [gr-qc]} \BibitemShut
  {NoStop}%
\bibitem [{\citenamefont {Ho}\ and\ \citenamefont
  {Andersson}(2023)}]{Ho:2023shr}%
  \BibitemOpen
  \bibfield  {author} {\bibinfo {author} {\bibfnamefont {W.~C.~G.}\
  \bibnamefont {Ho}}\ and\ \bibinfo {author} {\bibfnamefont {N.}~\bibnamefont
  {Andersson}},\ }\bibfield  {title} {\bibinfo {title} {{New dynamical tide
  constraints from current and future gravitational wave detections of
  inspiralling neutron stars}},\ }\href
  {https://doi.org/10.1103/PhysRevD.108.043003} {\bibfield  {journal} {\bibinfo
   {journal} {Phys. Rev. D}\ }\textbf {\bibinfo {volume} {108}},\ \bibinfo
  {pages} {043003} (\bibinfo {year} {2023})},\ \Eprint
  {https://arxiv.org/abs/2307.10721} {arXiv:2307.10721 [astro-ph.HE]}
  \BibitemShut {NoStop}%
\bibitem [{\citenamefont {Chirenti}\ \emph {et~al.}(2017)\citenamefont
  {Chirenti}, \citenamefont {Gold},\ and\ \citenamefont
  {Miller}}]{Chirenti:2016xys}%
  \BibitemOpen
  \bibfield  {author} {\bibinfo {author} {\bibfnamefont {C.}~\bibnamefont
  {Chirenti}}, \bibinfo {author} {\bibfnamefont {R.}~\bibnamefont {Gold}},\
  and\ \bibinfo {author} {\bibfnamefont {M.~C.}\ \bibnamefont {Miller}},\
  }\bibfield  {title} {\bibinfo {title} {{Gravitational waves from f-modes
  excited by the inspiral of highly eccentric neutron star binaries}},\ }\href
  {https://doi.org/10.3847/1538-4357/aa5ebb} {\bibfield  {journal} {\bibinfo
  {journal} {Astrophys. J.}\ }\textbf {\bibinfo {volume} {837}},\ \bibinfo
  {pages} {67} (\bibinfo {year} {2017})},\ \Eprint
  {https://arxiv.org/abs/1612.07097} {arXiv:1612.07097 [astro-ph.HE]}
  \BibitemShut {NoStop}%
\bibitem [{\citenamefont {Dutta~Roy}\ and\ \citenamefont
  {Saini}(2024)}]{DuttaRoy:2024aew}%
  \BibitemOpen
  \bibfield  {author} {\bibinfo {author} {\bibfnamefont {P.}~\bibnamefont
  {Dutta~Roy}}\ and\ \bibinfo {author} {\bibfnamefont {P.}~\bibnamefont
  {Saini}},\ }\bibfield  {title} {\bibinfo {title} {{Impact of unmodeled
  eccentricity on the tidal deformability measurement and implications for
  gravitational wave physics inference}},\ }\href@noop {} {\  (\bibinfo {year}
  {2024})},\ \Eprint {https://arxiv.org/abs/2403.02404} {arXiv:2403.02404
  [astro-ph.HE]} \BibitemShut {NoStop}%
\bibitem [{\citenamefont {Lorimer}(2008)}]{Lorimer:2008se}%
  \BibitemOpen
  \bibfield  {author} {\bibinfo {author} {\bibfnamefont {D.~R.}\ \bibnamefont
  {Lorimer}},\ }\bibfield  {title} {\bibinfo {title} {{Binary and Millisecond
  Pulsars}},\ }\href {https://doi.org/10.12942/lrr-2008-8} {\bibfield
  {journal} {\bibinfo  {journal} {Living Rev. Rel.}\ }\textbf {\bibinfo
  {volume} {11}},\ \bibinfo {pages} {8} (\bibinfo {year} {2008})},\ \Eprint
  {https://arxiv.org/abs/0811.0762} {arXiv:0811.0762 [astro-ph]} \BibitemShut
  {NoStop}%
\bibitem [{\citenamefont {Schmidt}\ \emph {et~al.}(2015)\citenamefont
  {Schmidt}, \citenamefont {Ohme},\ and\ \citenamefont
  {Hannam}}]{Schmidt:2014iyl}%
  \BibitemOpen
  \bibfield  {author} {\bibinfo {author} {\bibfnamefont {P.}~\bibnamefont
  {Schmidt}}, \bibinfo {author} {\bibfnamefont {F.}~\bibnamefont {Ohme}},\ and\
  \bibinfo {author} {\bibfnamefont {M.}~\bibnamefont {Hannam}},\ }\bibfield
  {title} {\bibinfo {title} {{Towards models of gravitational waveforms from
  generic binaries II: Modelling precession effects with a single effective
  precession parameter}},\ }\href {https://doi.org/10.1103/PhysRevD.91.024043}
  {\bibfield  {journal} {\bibinfo  {journal} {Phys. Rev. D}\ }\textbf {\bibinfo
  {volume} {91}},\ \bibinfo {pages} {024043} (\bibinfo {year} {2015})},\
  \Eprint {https://arxiv.org/abs/1408.1810} {arXiv:1408.1810 [gr-qc]}
  \BibitemShut {NoStop}%
\bibitem [{\citenamefont {Colleoni}\ \emph {et~al.}(2023)\citenamefont
  {Colleoni}, \citenamefont {Vidal}, \citenamefont {Johnson-McDaniel},
  \citenamefont {Dietrich}, \citenamefont {Haney},\ and\ \citenamefont
  {Pratten}}]{Colleoni:2023czp}%
  \BibitemOpen
  \bibfield  {author} {\bibinfo {author} {\bibfnamefont {M.}~\bibnamefont
  {Colleoni}}, \bibinfo {author} {\bibfnamefont {F.~A.~R.}\ \bibnamefont
  {Vidal}}, \bibinfo {author} {\bibfnamefont {N.~K.}\ \bibnamefont
  {Johnson-McDaniel}}, \bibinfo {author} {\bibfnamefont {T.}~\bibnamefont
  {Dietrich}}, \bibinfo {author} {\bibfnamefont {M.}~\bibnamefont {Haney}},\
  and\ \bibinfo {author} {\bibfnamefont {G.}~\bibnamefont {Pratten}},\
  }\bibfield  {title} {\bibinfo {title} {{IMRPhenomXP\_NRTidalv2: An improved
  frequency-domain precessing binary neutron star waveform model}},\
  }\href@noop {} {\  (\bibinfo {year} {2023})},\ \Eprint
  {https://arxiv.org/abs/2311.15978} {arXiv:2311.15978 [gr-qc]} \BibitemShut
  {NoStop}%
\bibitem [{\citenamefont {Chatziioannou}\ \emph {et~al.}(2015)\citenamefont
  {Chatziioannou}, \citenamefont {Cornish}, \citenamefont {Klein},\ and\
  \citenamefont {Yunes}}]{Chatziioannou:2014coa}%
  \BibitemOpen
  \bibfield  {author} {\bibinfo {author} {\bibfnamefont {K.}~\bibnamefont
  {Chatziioannou}}, \bibinfo {author} {\bibfnamefont {N.}~\bibnamefont
  {Cornish}}, \bibinfo {author} {\bibfnamefont {A.}~\bibnamefont {Klein}},\
  and\ \bibinfo {author} {\bibfnamefont {N.}~\bibnamefont {Yunes}},\ }\bibfield
   {title} {\bibinfo {title} {{Spin-Precession: Breaking the Black
  Hole--Neutron Star Degeneracy}},\ }\href
  {https://doi.org/10.1088/2041-8205/798/1/L17} {\bibfield  {journal} {\bibinfo
   {journal} {Astrophys. J. Lett.}\ }\textbf {\bibinfo {volume} {798}},\
  \bibinfo {pages} {L17} (\bibinfo {year} {2015})},\ \Eprint
  {https://arxiv.org/abs/1402.3581} {arXiv:1402.3581 [gr-qc]} \BibitemShut
  {NoStop}%
\bibitem [{\citenamefont {Pratten}\ \emph
  {et~al.}(2020{\natexlab{c}})\citenamefont {Pratten}, \citenamefont {Schmidt},
  \citenamefont {Buscicchio},\ and\ \citenamefont {Thomas}}]{Pratten:2020igi}%
  \BibitemOpen
  \bibfield  {author} {\bibinfo {author} {\bibfnamefont {G.}~\bibnamefont
  {Pratten}}, \bibinfo {author} {\bibfnamefont {P.}~\bibnamefont {Schmidt}},
  \bibinfo {author} {\bibfnamefont {R.}~\bibnamefont {Buscicchio}},\ and\
  \bibinfo {author} {\bibfnamefont {L.~M.}\ \bibnamefont {Thomas}},\ }\bibfield
   {title} {\bibinfo {title} {{Measuring precession in asymmetric compact
  binaries}},\ }\href {https://doi.org/10.1103/PhysRevResearch.2.043096}
  {\bibfield  {journal} {\bibinfo  {journal} {Phys. Rev. Res.}\ }\textbf
  {\bibinfo {volume} {2}},\ \bibinfo {pages} {043096} (\bibinfo {year}
  {2020}{\natexlab{c}})},\ \Eprint {https://arxiv.org/abs/2006.16153}
  {arXiv:2006.16153 [gr-qc]} \BibitemShut {NoStop}%
\bibitem [{\citenamefont {Inc.}()}]{Mathematica}%
  \BibitemOpen
  \bibfield  {author} {\bibinfo {author} {\bibfnamefont {W.~R.}\ \bibnamefont
  {Inc.}},\ }\href {https://www.wolfram.com/mathematica} {\bibinfo {title}
  {Mathematica, {V}ersion 14.0}},\ \bibinfo {note} {{C}hampaign, {IL},
  2024}\BibitemShut {NoStop}%
\bibitem [{\citenamefont {Hunter}(2007)}]{Hunter:2007}%
  \BibitemOpen
  \bibfield  {author} {\bibinfo {author} {\bibfnamefont {J.~D.}\ \bibnamefont
  {Hunter}},\ }\bibfield  {title} {\bibinfo {title} {Matplotlib: A 2d graphics
  environment},\ }\href@noop {} {\bibfield  {journal} {\bibinfo  {journal}
  {Computing In Science \& Engineering}\ }\textbf {\bibinfo {volume} {9}},\
  \bibinfo {pages} {90} (\bibinfo {year} {2007})}\BibitemShut {NoStop}%
\bibitem [{\citenamefont {Bernuzzi}\ \emph
  {et~al.}(2024{\natexlab{b}})\citenamefont {Bernuzzi}, \citenamefont {Nagar}
  \emph {et~al.}}]{TEOBgit}%
  \BibitemOpen
  \bibfield  {author} {\bibinfo {author} {\bibfnamefont {S.}~\bibnamefont
  {Bernuzzi}}, \bibinfo {author} {\bibfnamefont {A.}~\bibnamefont {Nagar}},
  \emph {et~al.},\ }\href
  {https://bitbucket.org/eob_ihes/teobresums/commits/branch/GIOTTO} {\bibinfo
  {title} {{TEOBR}esum{S} {B}itbucket {R}espository}} (\bibinfo {year}
  {2024}{\natexlab{b}})\BibitemShut {NoStop}%
\end{thebibliography}%
%%%%%%%%%%%%%%%%%%%%%%%%%%%%%

%%%%%%%%%%%%%%%%%%%%%%%%%
\end{document}